**Ice Melt, Sea Level Rise and Superstorms: Evidence from Paleoclimate Data, Climate Modeling, and Modern Observations that 2°C Global Warming is Dangerous**

James Hansen[1], Makiko Sato[1], Paul Hearty[2], Reto Ruedy[3,4], Maxwell Kelley[3,4], Valerie Masson-Delmotte[5], Gary Russell[4], George Tselioudis[4], Junji Cao[6], Eric Rignot[7,8], Isabella Velicogna[8,7], Blair Tormey[9], Bailey Donovan[10], Evgeniya Kandiano[11], Karina von Schuckmann[12], Pushker Kharecha[1,4], Allegra N. Legrande[4], Michael Bauer[13,4], Kwak-Wai Lo[3,4]

**Abstract.** We use numerical climate simulations, paleoclimate data, and modern observations to study the effect of growing ice melt from Antarctica and Greenland. Meltwater tends to stabilize the ocean column, inducing amplifying feedbacks that increase subsurface ocean warming and ice shelf melting. Cold meltwater and induced dynamical effects cause ocean surface cooling in the Southern Ocean and North Atlantic, thus increasing Earth's energy imbalance and heat flux into most of the global ocean's surface. Southern Ocean surface cooling, while lower latitudes are warming, increases precipitation on the Southern Ocean, increasing ocean stratification, slowing deepwater formation, and increasing ice sheet mass loss. These feedbacks make ice sheets in contact with the ocean vulnerable to accelerating disintegration. We hypothesize that ice mass loss from the most vulnerable ice, sufficient to raise sea level several meters, is better approximated as exponential than by a more linear response. Doubling times of 10, 20 or 40 years yield multi-meter sea level rise in about 50, 100 or 200 years. Recent ice melt doubling times are near the lower end of the 10-40 year range, but the record is too short to confirm the nature of the response. The feedbacks, including subsurface ocean warming, help explain paleoclimate data and point to a dominant Southern Ocean role in controlling atmospheric $CO_2$, which in turn exercised tight control on global temperature and sea level. The millennial (500-2000 year) time scale of deep ocean ventilation affects the time scale for natural $CO_2$ change and thus the time scale for paleo global climate, ice sheet, and sea level changes, but this paleo millennial time scale should not be misinterpreted as the time scale for ice sheet response to a rapid large human-made climate forcing. These climate feedbacks aid interpretation of events late in the prior interglacial, when sea level rose to +6-9 meters with evidence of extreme storms while Earth was less than 1°C warmer than today. Ice melt cooling of the North Atlantic and Southern Oceans, increases atmospheric temperature gradients, eddy kinetic energy and baroclinicity, thus driving more powerful storms. The modeling, paleoclimate evidence, and ongoing observations together imply that 2°C global warming above the preindustrial level would be dangerous. Continued high fossil fuel emissions this century are predicted to yield: (1) cooling of the Southern Ocean, especially in the Western Hemisphere, (2) slowing of the Southern Ocean overturning circulation, warming of the ice shelves, and growing ice sheet mass loss, (3) slowdown and eventual shutdown of the Atlantic overturning circulation with cooling of the North Atlantic region, (4) increasingly powerful storms, and (5) nonlinearly growing sea

[1] Climate Science, Awareness and Solutions, Columbia University Earth Institute, New York, NY 10115, USA

[2] Department of Environmental Studies, University of North Carolina at Wilmington, North Carolina 28403, USA

[3] Trinnovium LLC, New York, NY 10025, USA

[4] NASA Goddard Institute for Space Studies, 2880 Broadway, New York, NY 10025, USA

[5] Institut Pierre Simon Laplace, Laboratoire des Sciences du Climat et de l'Environnement (CEA-CNRS-UVSQ), Gif-sur-Yvette, France

[6] Key Lab of Aerosol Chemistry & Physics, Institute of Earth Environment, Chinese Academy of Sciences, Xi'an 710075, China

[7] Jet Propulsion Laboratory, California Institute of Technology, Pasadena, California, 91109, USA

[8] Department of Earth System Science, University of California, Irvine, California, 92697, USA

[9] Program for the Study of Developed Shorelines, Western Carolina University, Cullowhee, NC 28723, USA

[10] Department of Geological Sciences, East Carolina University, Greenville, NC 27858, USA

[11] GEOMAR, Helmholtz Centre for Ocean Research, Wischhofstrasse 1-3, Kiel 24148, Germany

[12] Mediterranean Institut of Oceanography, University of Toulon, La Garde, France

[13] Department of Applied Physics and Applied Mathematics, Columbia University, New York, NY, 10027, USA

level rise, reaching several meters over a time scale of 50-150 years. These predictions, especially the cooling in the Southern Ocean and North Atlantic with markedly reduced warming or even cooling in Europe, differ fundamentally from existing climate change assessments. We discuss observations and modeling studies needed to refute or clarify these assertions.

## 1 Introduction

Humanity is rapidly extracting and burning fossil fuels without full understanding of the consequences. Current assessments place emphasis on practical effects such as increasing extremes of heat waves, droughts, heavy rainfall, floods, and encroaching seas (IPCC, 2014; USNCA, 2014). These assessments and our recent study (Hansen et al., 2013a) conclude that there is an urgency to slow carbon dioxide ($CO_2$) emissions, because the longevity of the carbon in the climate system (Archer, 2005) and persistence of the induced warming (Solomon et al., 2010) may lock in unavoidable highly undesirable consequences.

Despite these warnings, fossil fuels remain the world's primary energy source and global $CO_2$ emissions continue at a high level, perhaps with an expectation that humanity can adapt to climate change and find ways to minimize effects via advanced technologies. We suggest that this viewpoint fails to appreciate the nature of the threat posed by ice sheet instability and sea level rise. If the ocean continues to accumulate heat and increase melting of marine-terminating ice shelves of Antarctica and Greenland, a point will be reached at which it is impossible to avoid large scale ice sheet disintegration with sea level rise of at least several meters. The economic and social cost of losing functionality of all coastal cities is practically incalculable. We suggest that a strategy relying on adaptation to such consequences will be unacceptable to most of humanity, so it is important to understand this threat as soon as possible.

We investigate the climate threat using a combination of atmosphere-ocean modeling, information from paleoclimate data, and observations of ongoing climate change. Each of these has limitations: modeling is an imperfect representation of the climate system, paleo data consist mainly of proxy climate information usually with substantial ambiguities, and modern observations are limited in scope and accuracy. However, with the help of a large body of research by the scientific community, it is possible to draw meaningful conclusions.

## 2 Background information and organization of paper

Our study germinated a decade ago. Hansen (2005, 2007) argued that the modest 21[st] century sea level rise projected by IPCC (2001), less than a meter, was inconsistent with presumed climate forcings, which were larger than paleoclimate forcings associated with sea level rise of many meters. His argument about the potential rate of sea level rise was necessarily heuristic, because ice sheet models are at an early stage of development, depending sensitively on many processes that are poorly understood. This uncertainty is illustrated by Pollard et al. (2015), who found that addition of hydro-fracturing and cliff failure into their ice sheet model increased simulated sea level rise from 2 m to 17 m, in response to only 2°C ocean warming and accelerated the time for substantial change from several centuries to several decades.

The focus for our paper developed in 2007, when the first author (JH) read several papers by co-author P. Hearty. Hearty used geologic field data to make a persuasive case for rapid sea level rise late in the prior interglacial period to a height +6-9 m relative to today, and he presented evidence of strong storms in the Bahamas and Bermuda at that time. Hearty's data suggested violent climate behavior on a planet only slightly warmer than today.

Our study was designed to shed light on, or at least raise questions about, physical processes that could help account for the paleoclimate data and have relevance to ongoing and future climate change. Our assumption was that extraction of significant information on these processes would require use of and analysis of (1) climate modeling, (2) paleoclimate data, and

(3) modern observations. It is the combination of all of these that helps us interpret the intricate paleoclimate data and extract implications about future sea level and storms.

Our approach is to postulate existence of feedbacks that can rapidly accelerate ice melt, impose such rapidly growing freshwater injection on a climate model, and look for a climate response that supports such acceleration. Our imposed ice melt grows nonlinearly in time, specifically exponentially, so the rate is characterized by a doubling time. Total amounts of freshwater injection are chosen in the range 1-5 m of sea level, amounts that can be provided by vulnerable ice masses in contact with the ocean. We find significant impact of meltwater on global climate and feedbacks that support ice melt acceleration. We obtain this information without use of ice sheet models, which are still at an early stage of development, in contrast to global general circulation models that were developed over more than half a century and do a capable job of simulating atmosphere and ocean circulation.

Our principal finding concerns the effect of meltwater on stratification of the high latitude ocean and resulting ocean heat sequestration that leads to melting of ice shelves and catastrophic ice sheet collapse. Stratification contrasts with homogenization. Winter conditions on parts of the North Atlantic Ocean and around the edges of Antarctica normally produce cold salty water that is dense enough to sink to the deep ocean, thus stirring and tending to homogenize the water column. Injection of fresh meltwater reduces the density of the upper ocean wind-stirred mixed layer, thus reducing the rate at which cold surface water sinks in winter at high latitudes. Vertical mixing normally brings warmer water to the surface, where heat is released to the atmosphere and space. Thus the increased stratification due to freshwater injection causes heat to be retained at ocean depth, where it is available to melt ice shelves. Despite improvements that we make in our ocean model, which allow Antarctic Bottom Water to be formed at proper locations, we suggest that excessive mixing in many climate models, ours included, limits this stratification effect. Thus human impact on ice sheets and sea level may be even more imminent than in our model, suggesting a need for confirmatory observations.

Our paper published in Atmospheric Chemistry and Physics Discussion was organized in the chronological order of our investigation. Here we reorganize the work to make the science easier to follow. First we describe our climate simulations with specified growing freshwater sources in the North Atlantic and Southern Oceans. Second we analyze paleoclimate data for evidence of these processes and possible implications for the future. Third we examine modern data for evidence that the simulated climate changes are already occurring.

We use paleoclimate data to find support for and deeper understanding of these processes, focusing especially on events in the last interglacial period warmer than today, called Marine Isotope Stage (MIS) 5e in studies of ocean sediment cores, Eemian in European climate studies, and sometimes Sangamonian in American literature (see Sec. 4.2 for timescale diagram of Marine Isotope Stages). Accurately known changes of Earth's astronomical configuration altered the seasonal and geographical distribution of incoming radiation during the Eemian. Resulting global warming was due to feedbacks that amplified the orbital forcing. While the Eemian is not an analog of future warming, it is useful for investigating climate feedbacks, including the interplay between ice melt at high latitudes and ocean circulation.

## 3  Simulations of 1850-2300 climate change

We make simulations for 1850-2300 with radiative forcings that were used in CMIP (Climate Model Intercomparison Project) simulations reported by IPCC (2007, 2013). This allows comparison of our present simulations with prior studies. First, for the sake of later raising and discussing fundamental questions about ocean mixing and climate response time, we define climate forcings and the relation of forcings to Earth's energy imbalance and global temperature.

### 3.1 Climate forcing, Earth's energy imbalance, and climate response function

A climate forcing is an imposed perturbation of Earth's energy balance, such as change of solar irradiance or a radiatively effective constituent of the atmosphere or surface. Non-radiative climate forcings are possible, e.g., change of Earth's surface roughness or rotation rate, but these are small and radiative feedbacks likely dominate global climate response even in such cases. The net forcing driving climate change in our simulations (Fig. S16) is almost 2 W/m$^2$ at present and increases to 5-6 W/m$^2$ at the end of this century, depending on how much the (negative) aerosol forcing is assumed to reduce the greenhouse gas (GHG) forcing. The GHG forcing is based on IPCC scenario A1B. "Orbital" forcings, i.e., changes in the seasonal and geographical distribution of insolation on millennial time scales caused by changes of Earth's orbit and spin axis tilt, are near zero on global average, but they spur "slow feedbacks" of several W/m$^2$, mainly change of surface reflectivity and GHGs..

When a climate forcing changes, say solar irradiance increases or atmospheric $CO_2$ increases, Earth is temporarily out of energy balance, more energy coming in than going out in these cases, so Earth's temperature will increase until energy balance is restored. Earth's energy imbalance is a result of the climate system's inertia, the slowness of the surface temperature to respond to changing global climate forcing. Earth's energy imbalance is a function of ocean mixing, as well as climate forcing and climate sensitivity, the latter being the equilibrium global temperature response to a specified climate forcing. Earth's present energy imbalance, +0.5-1 W/m$^2$ (von Schuckmann et al., 2016), provides an indication of how much additional global warming is still "in the pipeline" if climate forcings remain unchanged. However, climate change generated by today's energy imbalance, especially the rate at which it occurs, is quite different than climate change in response to a new forcing of equal magnitude. Understanding this difference is relevant to issues raised in this paper.

The different effect of old and new climate forcings is implicit in the shape of the climate response function, R(t), where R is the fraction of the equilibrium global temperature change achieved as a function of time following imposition of a forcing. Global climate models find that a large fraction of the equilibrium response is obtained quickly, about half of the response occurring within several years, but the remainder is "recalcitrant" (Held et al., 2010), requiring many decades or even centuries for nearly complete response. Hansen (2008) showed that once a climate model's response function R is known, based on simulations for an instant forcing, global temperature change, T(t), in response to any climate forcing history, F(t), can be accurately obtained from a simple (Green's function) integration of R over time

$$T(t) \; = \; \int R(t) \, [dF/dt] \, dt \tag{1}$$

dF/dt is the annual increment of the net forcing and the integration begins before human-made climate forcing became substantial.

We use these concepts in discussing evidence that most ocean models, ours included, are too diffusive. Such excessive mixing causes the Southern and North Atlantic Oceans in the models to have unrealistically slow response to surface meltwater injection. Implications include more imminent threat of slowdowns of Antarctic Bottom Water and North Atlantic Deep Water formation than present models suggest, with regional and global climate impacts.

### 3.1 Climate model

Simulations are made with an improved version of a coarse-resolution model that allows long runs at low cost, GISS (Goddard Institute for Space Studies) model E-R. The atmosphere model is the documented modelE (Schmidt et al., 2006). The ocean is based on the Russell et al. (1995) model that conserves water and salt mass, has a free surface with divergent flow, uses a linear upstream scheme for advection, allows flow through 12 sub-resolution straits, has background diffusivity 0.3 cm²/s, 4°×5° resolution and 13 layers that increase in thickness with depth.

However, the ocean model includes simple but significant changes, compared with the version documented in simulations by Miller et al. (2014). First, an error in the calculation of neutral surfaces in the Gent-McWilliams (GM, Gent and McWilliams, 1990) mesoscale eddy parameterization was corrected; the resulting increased slope of neutral surfaces provides proper leverage to the restratification process and correctly orients eddy stirring along those surfaces.

Second, the calculation of eddy diffusivity $K_{meso}$ for GM following Visbeck et al. (1997) was simplified to use a length scale independent of the density structure (J. Marshall, pers. comm.):

$$K_{meso} = C/[T_{eady} \times f(latitude)] \hspace{4cm} (1)$$

where $C = (27.9 \text{ km})^2$, Eady growth rate $1/T_{eady} = \{|S \times N|\}$, S is the neutral surface slope, N the Brunt-Vaisala frequency, { } signifies averaging over the upper D meters of ocean depth, D = min(max(depth, 400 m), 1000 m), and $f(latitude) = \max(0.1, \sin(|latitude|))^{14}$ to qualitatively mimic the larger values of the Rossby radius of deformation at low latitudes. These choices for $K_{meso}$, whose simplicity is congruent with the use of a depth-independent eddy diffusivity and the use of $1/T_{eady}$ as a metric of eddy energy, result in the zonal average diffusivity shown in Fig. 1.

Third, the so-called nonlocal terms in the KPP mixing parameterization (Large et al., 1994) were activated. All of these modifications tend to increase the ocean stratification, and in particular the Southern Ocean state is fundamentally improved. For example, we show in Sec. 3.7.5 that our current model produces Antarctic Bottom Water on the Antarctic coastline, as observed, rather than in the middle of the Southern Ocean as occurs in many models, including the GISS-ER model documented in CMIP5. However, although overall realism of the ocean circulation is much improved, significant model deficiencies remain, as we will describe.

The simulated Atlantic Meridional Overturning Circulation (AMOC) has maximum flux that varies within the range ~14-18 Sv in the model control run (Figs. 2 and 3). AMOC strength in recent observations is 17.5 ± 1.6 Sv (Baringer et al., 2013; Srokosz et al., 2012), based on eight years (2004-2011) data for an in situ mooring array (Rayner et al., 2011; Johns et al., 2011).

Ocean model control run initial conditions are climatology for temperature and salinity (Levitus and Boyer, 1994; Levitus et al., 1994); atmospheric composition is that of 1880 (Hansen et al., 2011). Overall model drift from control run initial conditions is moderate (see Fig. S1 for planetary energy imbalance and global temperature), but there is drift in the North Atlantic circulation. The AMOC circulation cell initially is confined to the upper 3 km at all latitudes (1st century in Figs. 2 and 3), but by the 5th century the cell reaches deeper at high latitudes.

Atmospheric and surface climate in the present model is similar to the documented modelE-R, but because of changes to the ocean model we provide several diagnostics in the Supplement. A notable flaw in the simulated surface climate is the unrealistic double precipitation maximum in the tropical Pacific (Fig. S2). This double inter-tropical convergence zone (ITCZ) occurs in many models and may be related to cloud and radiation biases over the Southern Ocean (Hwang and Frierson, 2013) or deficient low level clouds in the tropical Pacific (de Szoeke and Xie, 2008). Another flaw is unrealistic hemispheric sea ice, with too much sea ice in the Northern Hemisphere and too little in the Southern Hemisphere (Figs. S3 and S4). Excessive Northern Hemisphere sea ice might be caused by deficient poleward heat transport in the Atlantic Ocean (Fig. S5). However, the AMOC has realistic strength and Atlantic meridional heat transport is only slightly below observations at high latitudes (Fig. S5). Thus we suspect that the problem may lie in sea ice parameterizations or deficient dynamical transport of ice out of the Arctic. The deficient Southern Hemisphere sea ice, at least in part, is likely related to excessive poleward (southward) transport of heat by the simulated global ocean (Fig. S5), which is related to deficient northward transport of heat in the modeled Atlantic Ocean (Fig. S5).

---

[14] Where ocean depth exceeds 1000 m, these conditions yield D = 1000 m, thus excluding any first-order abyssal bathymetric imprint on upper ocean eddy energy, consistent with theory and observations. The other objective of the stated condition is to limit release of potential energy in the few ocean gridboxes with ocean depth less than 400 m, because shallow depths limit the ability of baroclinic eddies to release potential energy via vertical motion.

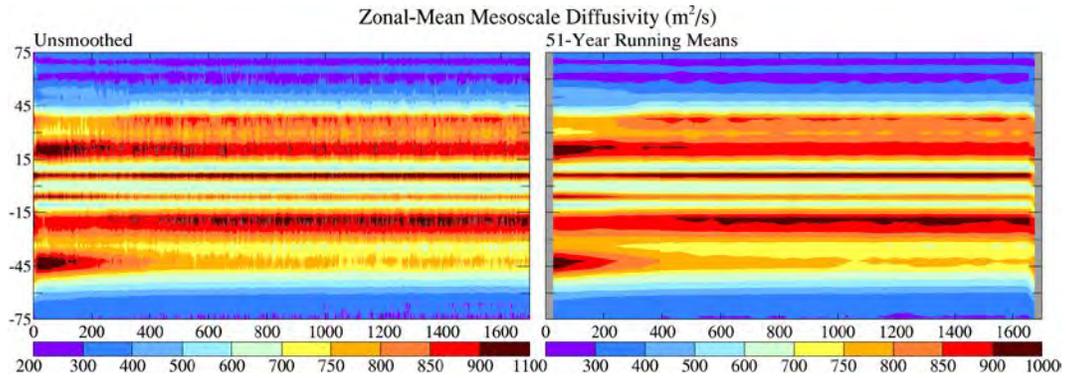

**Fig. 1.** Zonal-mean mesoscale diffusivity (m²/s) versus time in control run.

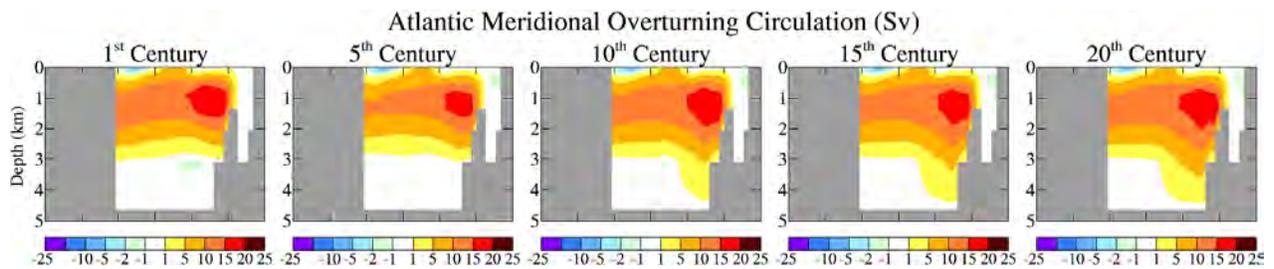

**Fig. 2.** AMOC (Sv) in the 1st, 5th, 10th, 15th and 20th centuries of the control run.

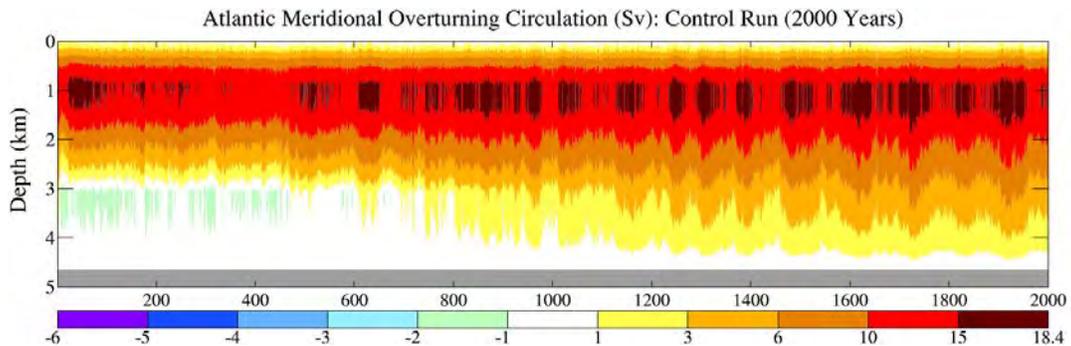

**Fig. 3.** Annual mean AMOC (Sv) at 28°N in the model control run.

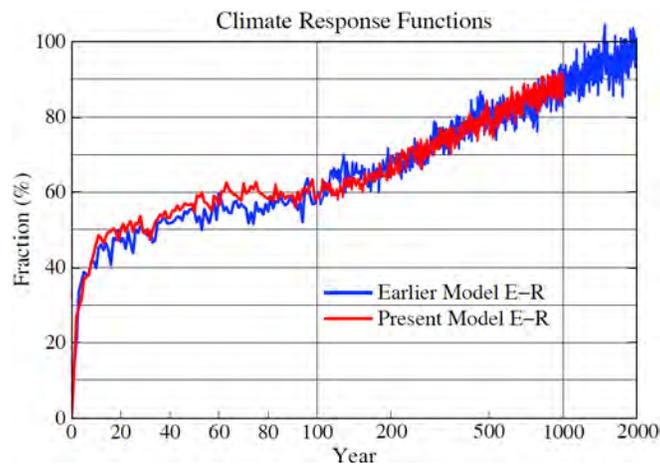

**Fig. 4.** Climate response function, R(t), i.e., the fraction (%) of equilibrium surface temperature response for GISS model E-R based on a 2000 year control run (Hansen et al., 2007a). Forcing was instant $CO_2$ doubling with fixed ice sheets, vegetation distribution, and other long-lived GHGs.

A key characteristic of the model and the real world is the response time: how fast does the surface temperature adjust to a climate forcing? ModelE-R response is about 40% in five years (Fig. 4) and 60% in 100 years, with the remainder requiring many centuries. Hansen et al. (2011) concluded that most ocean models, including modelE-R, mix a surface temperature perturbation downward too efficiently and thus have a slower surface response than the real world. The basis for this conclusion was empirical analysis using climate response functions, with 50%, 75% and 90% response at year 100 for climate simulations (Hansen et al., 2011). Earth's measured energy imbalance in recent years and global temperature change in the past century revealed that the response function with 75% response in 100 years provided a much better fit with observations than the other choices. Durack et al. (2012) compared observations of how rapidly surface salinity changes are mixed into the deeper ocean with the large number of global models in CMIP3, reaching a similar conclusion, that the models mix too rapidly.

Our present ocean model has a faster response on 10-75 year time scales than the old model (Fig. 4), but the change is small. Although the response time in our model is similar to that in many other ocean models (Hansen et al., 2011), we believe that it is likely slower than the real world response on time scales of a few decades and longer. A too slow surface response could result from excessive small scale mixing. We will argue, after the studies below, that excessive mixing likely has other consequences, e.g., causing the effect of freshwater stratification on slowing Antarctic Bottom Water (AABW) formation and growth of Antarctic sea ice cover to occur 1-2 decades later than in the real world. Similarly, excessive mixing probably makes the AMOC in the model less sensitive to freshwater forcing than the real world AMOC.

## 3.2 Experiment definition: exponentially increasing fresh water

Freshwater injection is 360 Gt/yr (1 mm sea level) in 2003-2015, then grows with 5, 10 or 20 year doubling time (Fig. 5) and terminates when global sea level reaches 1m or 5m. Doubling times of 10, 20 and 40 years, reaching meter scale sea level rise in 50, 100, and 200 years may be a more realistic range of time scales, but 40 years yields little effect this century, the time of most interest, so we learn more with less computing time using the 5, 10 and 20 year doubling times. Observed ice sheet mass loss doubling rates, although records are short, are ~10 years (Sec. 5.1). Our sharp cut-off of melt aids separation of immediate forcing effects and feedbacks.

We argue that such rapid increase of meltwater is plausible if GHGs keep growing rapidly. Greenland and Antarctica have outlet glaciers in canyons with bedrock below sea level well back into the ice sheet (Fretwell et al., 2013; Morlighem et al., 2014; Pollard et al., 2015). Feedbacks, including ice sheet darkening due to surface melt (Hansen et al., 2007b; Robinson et al., 2012; Tedesco et al., 2012; Box et al., 2012) and lowering and thus warming of the near-coastal ice sheet surface, make increasing ice melt likely. Paleoclimate data reveal sea level rise of several meters in a century (Fairbanks, 1989; Deschamps et al., 2012). Those cases involved ice sheets at lower latitudes, but 21st century climate forcing is larger and increasing much more rapidly.

Radiative forcings (Fig. S16a,b) are from Hansen et al. (2007c) through 2003 and IPCC scenario A1B for later GHGs. A1B is an intermediate IPCC scenario over the century, but on the high side early this century (Fig. 2, Hansen et al., 2007c). We add freshwater to the North Atlantic (ocean area within 52°N-72°N and 15°E-65°N) or Southern Ocean (ocean south of 60°S), or equally divided between the two oceans. Ice sheet discharge (icebergs plus meltwater) is mixed as fresh water with mean temperature −15°C into the top three ocean layers (Fig. S6).

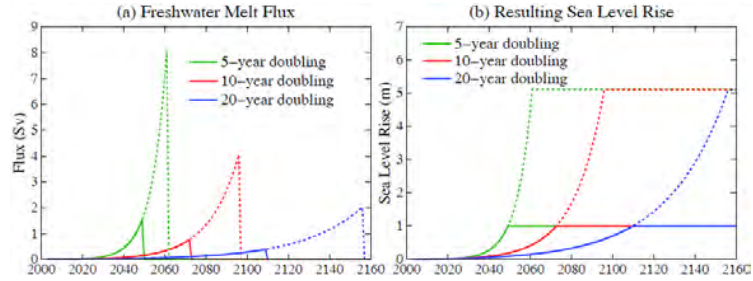

**Fig. 5.** (a) total fresh water flux added in North Atlantic and Southern Oceans, (b) resulting sea level rise. Solid lines for 1 m sea level rise, dotted for 5 m. One Sverdrup (Sv) is $10^6$ m³/s, which is ~3×$10^4$ Gt/year.

Annual Mean Surface Air Temperature Change (°C)

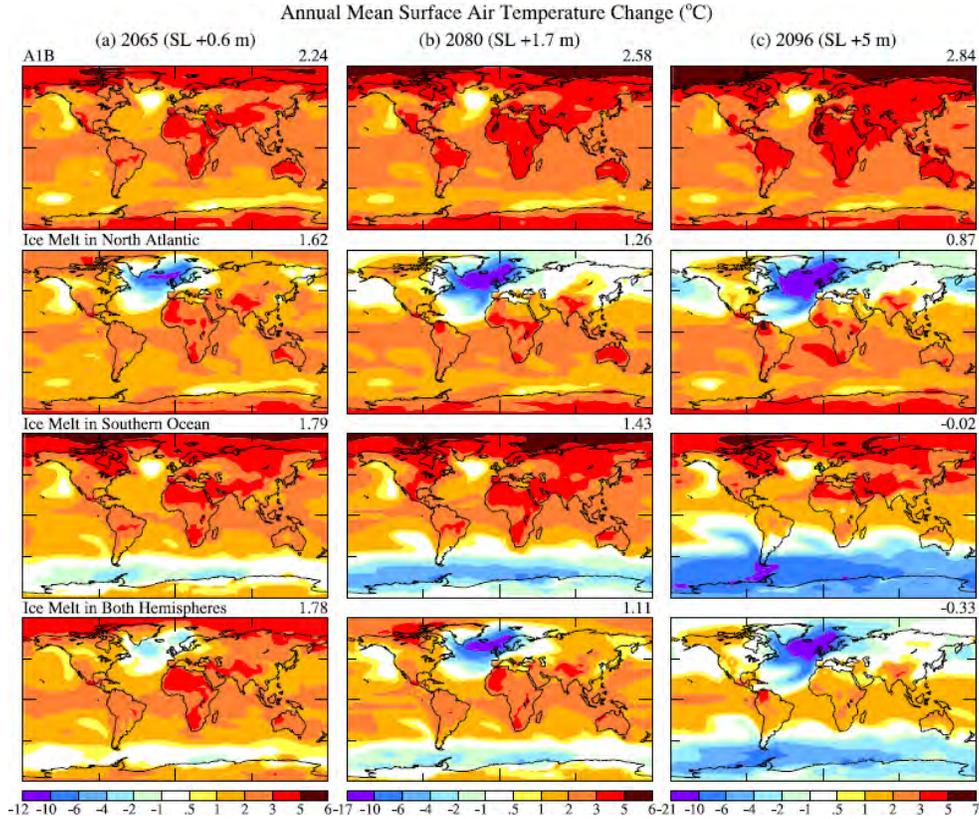

**Fig. 6.** Surface air temperature (°C) relative to 1880-1920 in (a) 2065, (b) 2080, and (c) 2096. Top row is IPCC scenario A1B. Ice melt with 10-year doubling is added in other scenarios.

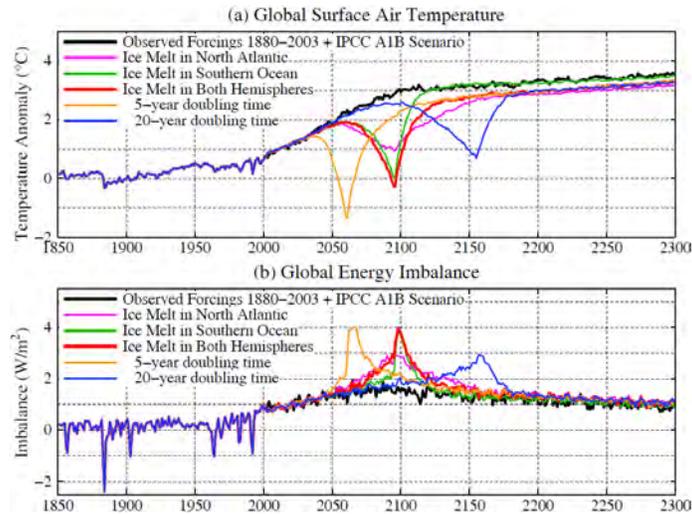

**Fig. 7.** (a) Surface air temperature (°C) relative to 1880-1920 for several scenarios. (b) Global energy imbalance (W/m²) for the same scenarios.

### 3.3 Simulated surface temperature and energy balance

We present surface temperature and planetary energy balance first, thus providing a global overview. Then we examine changes in ocean circulation and compare results with prior studies.

Temperature change in 2065, 2080 and 2096 for 10-year doubling time (Fig. 6) should be thought of as results when sea level rise reaches 0.6, 1.7 and 5 m, because the dates depend on initial freshwater flux. Actual current freshwater flux may be about a factor of four higher than assumed in these initial runs, as we will discuss, and thus effects may occur ~20 years earlier. A sea level rise of 5 m in a century is about the most extreme in the paleo record (Fairbanks, 1989; Deschamps et al., 2012), but the assumed 21st century climate forcing is also more rapidly growing than any known natural forcing.

Meltwater injected into the North Atlantic has larger initial impact, but Southern Hemisphere ice melt has a greater global effect for larger melt as the effectiveness of more meltwater in the North Atlantic begins to decline. The global effect is large long before sea level rise of 5 m is reached. Meltwater reduces global warming about half by the time sea level rise reaches 1.7 m. Cooling due to ice melt more than eliminates A1B warming in large areas of the globe.

The large cooling effect of ice melt does not decrease much as the ice melting rate varies between doubling times of 5, 10 or 20 years (Fig. 7a). In other words, the cumulative ice sheet melt, rather than the rate of ice melt, largely determines the climate impact for the range of melt rates covered by 5, 10 and 20 year doubling times. Thus if ice sheet loss occurs even to an extent of 1.7 m sea level rise (Fig. 7b), a large impact on climate and climate change is predicted.

Greater global cooling occurs for freshwater injected on the Southern Ocean, but the cooling lasts much longer for North Atlantic injection (Fig. 7a). That persistent cooling, mainly at Northern Hemisphere middle and high latitudes (Fig. S7), is a consequence of the sensitivity, hysteresis effects, and long recovery time of the AMOC (Stocker and Wright, 1991; Rahmstorf, 1995 and earlier studies referenced therein). AMOC changes are described below.

When freshwater injection on the Southern Ocean is halted, global temperature jumps back within two decades to the value it would have had without any freshwater addition (Fig. 7a). Quick recovery is consistent with the Southern Ocean-centric picture of the global overturning circulation (Fig. 4, Talley, 2013), as the Southern Meridional Overturning Circulation (SMOC), driven by AABW formation, responds to change of the vertical stability of the ocean column near Antarctica (Sec. 3.6) and the ocean mixed layer and sea ice have limited thermal inertia.

Cooling from ice melt is largely regional, temporary, and does not alleviate concerns about global warming. Southern Hemisphere cooling is mainly in uninhabited regions. Northern Hemisphere cooling increases temperature gradients that will drive stronger storms (Sec. 3.8).

Global cooling due to ice melt causes a large increase in Earth's energy imbalance (Fig. 7b), adding about +2 W/m$^2$, which is larger than the imbalance caused by increasing GHGs. Thus, although the cold fresh water from ice sheet disintegration provides a negative feedback on regional and global surface temperature, it increases the planet's energy imbalance, thus providing more energy for ice melt (Hansen, 2005). This added energy is pumped into the ocean.

Increased downward energy flux at the top of the atmosphere is not located in the regions cooled by ice melt. On the contrary, those regions suffer a large reduction of net incoming energy (Fig. 8a). The regional energy reduction is a consequence of increased cloud cover (Fig. 8b) in response to the colder ocean surface. However, the colder ocean surface reduces upward radiative, sensible and latent heat fluxes, thus causing a large (~50 W/m$^2$) increase of energy into the North Atlantic and a substantial but smaller flux into the Southern Ocean (Fig. 8c).

Below we conclude that the principal mechanism by which this ocean heat increases ice melt is via its effect on ice shelves. Discussion requires examination of how the freshwater injections alter the ocean circulation and internal ocean temperature.

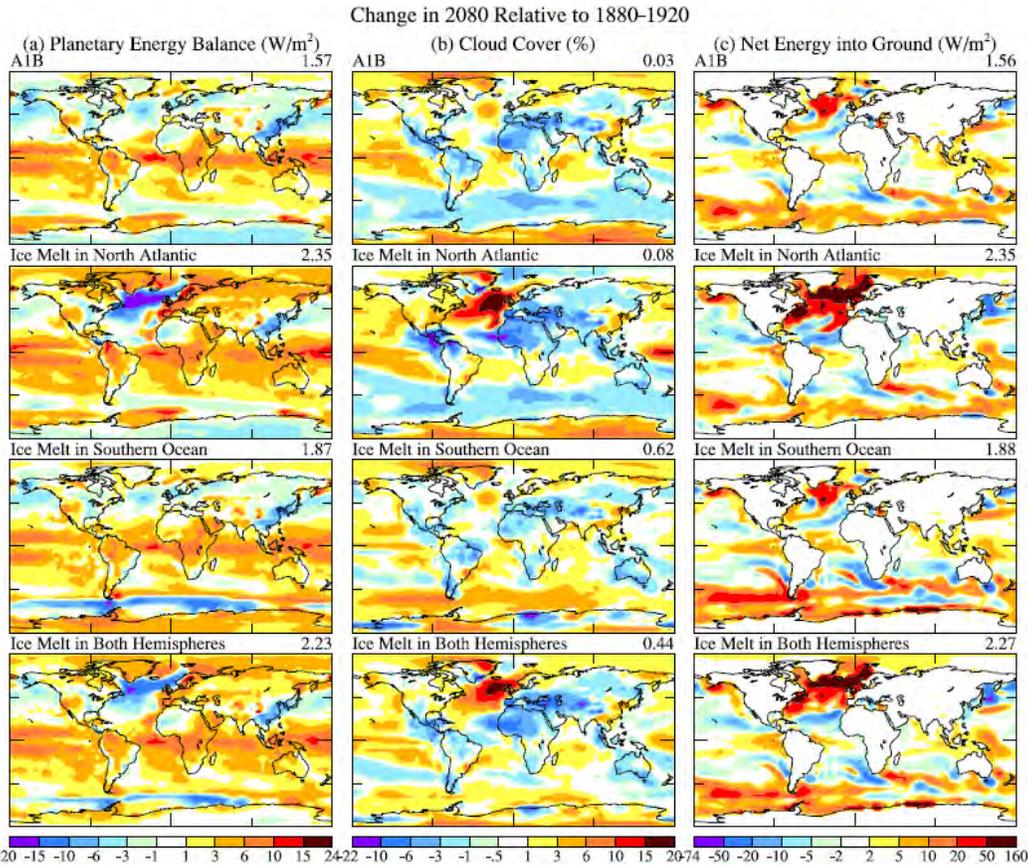

**Fig. 8.** Change in 2080 (mean of 2078-2082), relative to 1880-1920, of annual mean **(a)** planetary energy balance (W/m$^2$), **(b)** cloud cover (%), and **(c)** net energy into ground (W/m$^2$) for same scenarios as Fig. 6.

### 3.4 Simulated Atlantic Meridional Overturning Circulation (AMOC)

Broecker's articulation of likely effects of freshwater outbursts in the North Atlantic on ocean circulation and global climate (Broecker, 1990; Broecker et al., 1990) spurred quantitative studies with idealized ocean models (Stocker and Wright, 1991) and global atmosphere-ocean models (Manabe and Stouffer, 1995; Rahmstorf 1995, 1996). Scores of modeling studies have since been carried out, many reviewed by Barreiro et al. (2008), and observing systems are being developed to monitor modern changes in the AMOC (Carton and Hakkinen, 2011).

Our climate simulations in this section are 5-member ensembles of runs initiated at 25-year intervals at years 901-1001 of the control run. We chose this part of the control run because the planet is then in energy balance (Fig. S1), although by that time model drift had altered the slow deep ocean circulation. Some model drift away from initial climatological conditions is inevitable, as all models are imperfect, and we carry out the experiments with cognizance of model limitations. However, there is strong incentive to seek basic improvements in representation of physical processes to reduce drift in future versions of the model.

GHGs alone (scenario A1B) slow AMOC by the early 21$^{st}$ century (Fig. 9), but variability among individual runs (Fig. S8) would make definitive detection difficult at present. Freshwater injected onto the North Atlantic or in both hemispheres shuts down the AMOC (Fig. 9 right side). GHG amounts are fixed after 2100 and ice melt is zero, but after two centuries of stable climate forcing the AMOC has not recovered to its earlier state. This slow recovery was found in the earliest simulations by Manabe and Stouffer (1994) and Rahmstorf (1995, 1996).

Freshwater injection already has a large impact when ice melt is a fraction of 1 m of sea level. By the time sea level rise reaches 59 cm (2065 in the present scenarios), when fresh water flux is 0.48 Sv, the impact on AMOC is already large, consistent with the substantial surface cooling in the North Atlantic (Fig. 6).

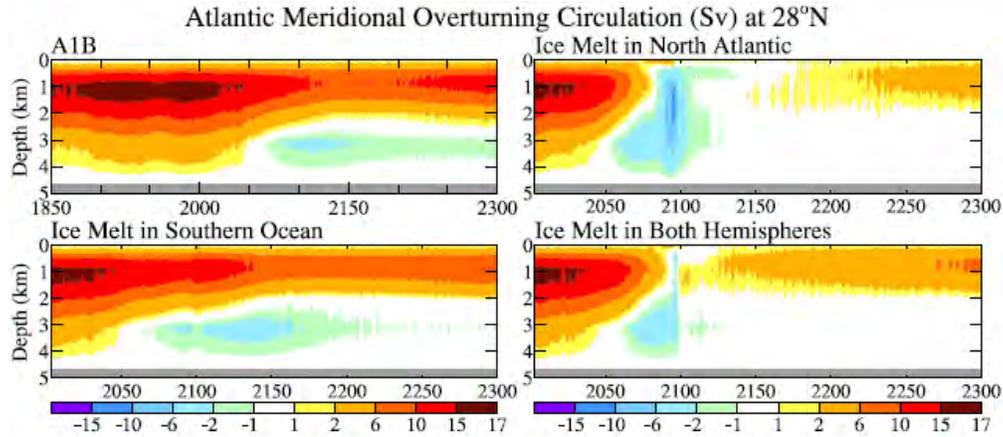

**Fig. 9.** Ensemble-mean AMOC (Sv) at 28N versus time for the same four scenarios as in Fig. 6, with ice melt reaching 5 m at the end of the 21$^{st}$ century in the three experiments with ice melt.

### 3.5 Comparison with prior simulations

AMOC sensitivity to GHG forcing has been examined extensively based on CMIP studies. Schmittner et al. (2005) found that AMOC weakened 25±25% by the end of the 21$^{st}$ century in 28 simulations of 9 different models forced by the A1B emission scenario. Gregory et al. (2005) found 10-50% AMOC weakening in 11 models for $CO_2$ quadrupling (1%/year increase for 140 years), with largest decreases in models with strong AMOCs. Weaver et al. (2007) found a 15-31% AMOC weakening for $CO_2$ quadrupling in a single model for 17 climate states differing in initial GHG amount. AMOC in our model weakens 30% between 1990-2000 and 2090-2100, the period used by Schmittner et al (2005), for A1B forcing (Fig. S8). Thus our model is more sensitive than the average, but within the range of other models, a conclusion that continues to be valid in comparison with 10 CMIP5 models (Cheng et al., 2013).

AMOC sensitivity to freshwater forcing has not been compared as systematically among models. Several studies find little impact of Greenland melt on AMOC (Huybrechts et al., 2002; Jungclaus et al., 2006; Vizcaino et al., 2008) while others find substantial North Atlantic cooling (Fichefet et al., 2003; Swingedouw et al., 2007; Hu et al., 2009, 2011). Studies with little impact calculated or assumed small ice sheet melt rates, e.g., Greenland contributed only 4 cm of sea level rise in the 21$^{st}$ century in the ice sheet model of Huybrechts et al. (2002). Fichefet et al. (2003), using nearly the same atmosphere-ocean model as Huybrechts et al. (2002) but a more responsive ice sheet model, found AMOC weakening from 20 to 13 Sv late in the 21$^{st}$ century, but separate contributions of ice melt and GHGs to AMOC slowdown were not defined.

Hu et al. (2009, 2011) use the A1B scenario and freshwater from Greenland starting at 1 mm sea level per year increasing 7%/year, similar to our 10-year doubling case. Hu et al. keep the melt rate constant after it reaches 0.3 Sv (in 2050), yielding 1.65 m sea level rise in 2100 and 4.2 m in 2200. Global warming found by Hu et al. for scenario A1B resembles our result but is 20-30% smaller [compare Fig. 2b of Hu et al. (2009) to our Fig. 6], and cooling they obtain from the freshwater flux is moderately less than that in our model. AMOC is slowed about one-third by the latter 21$^{st}$ century in the Hu et al. (2011) 7%/year experiment, comparable to our result.

General consistency holds for other quantities, such as changes of precipitation. Our model yields southward shifting of the Inter-Tropical Convergence Zone (ITCZ) and intensification of the subtropical dry region with increasing GHGs (Fig. S9), as has been reported in modeling studies of Swingedouw et al. (2007, 2009). These effects are intensified by ice melt and cooling in the North Atlantic region (Fig. S9).

A recent 5-model study (Swingedouw et al., 2014) finds a small effect on AMOC for 0.1 Sv Greenland freshwater flux added in 2050 to simulations with a strong GHG forcing. Our larger response is likely due, at least in part, to our freshwater flux reaching several tenths of a Sv.

### 3.6 Pure freshwater experiments

We assumed, in discussing the relevance of these experiments to Eemian climate, that effects of freshwater injection dominate over changing GHG amount, as seems likely because of the large freshwater effect on SSTs and sea level pressure. However, Eemian $CO_2$ was actually almost constant at ~275 ppm (Luthi et al., 2008). Thus, to isolate effects better, we now carry out simulations with fixed GHG amount, which helps clarify important feedback processes.

Our pure freshwater experiments are 5-member ensembles starting at years 1001, 1101, 1201, 1301, and 1401 of the control run. Each experiment ran 300 years. Freshwater flux in the initial decade averaged 180 km$^3$/year (0.5 mm sea level) in the hemisphere with ice melt and increased with a 10-year doubling time. Freshwater input is terminated when it reaches 0.5 m sea level rise per hemisphere for three 5-member ensembles: two ensembles with injection in the individual hemispheres and one ensemble with input in both hemispheres (1 m total sea level rise). Three additional ensembles were obtained by continuing freshwater injection until hemispheric sea level contributions reached 2.5 m. Here we provide a few model diagnostics central to discussions that follow. Additional results are provided in Figs. S10-S12.

The AMOC shuts down for Northern Hemisphere freshwater input yielding 2.5 m sea level rise (Fig. 10). By year 300, more than 200 years after cessation of all freshwater input, AMOC is still far from full recovery for this large freshwater input. On the other hand, freshwater input of 0.5 m does not cause full shutdown, and AMOC recovery occurs in less than a century.

Global temperature change (Fig. 11) reflects the fundamentally different impact of freshwater forcings of 0.5 m and 2.5 m. The response also differs greatly depending on the hemisphere of the freshwater input. The case with freshwater forcing in both hemispheres is shown only in the Supplement because, to a good approximation, the response is simply the sum of the responses to the individual hemispheric forcings (see Figs. S10-S12). The sum of responses to hemispheric forcings moderately exceeds the response to global forcing.

Global cooling continues for centuries for the case with freshwater forcing sufficient to shut down the AMOC (Fig. 11). If the forcing is only 0.5 m of sea level, the temperature recovers in a few decades. However, the freshwater forcing required to reach the tipping point of AMOC shutdown may be less in the real world than in our model, as discussed below. Global cooling due to freshwater input on the Southern Ocean disappears in a few years after freshwater input ceases (Fig. 11), for both the smaller (0.5 m of sea level) and larger (2.5 m) freshwater forcings.

Injection of a large amount of surface freshwater in either hemisphere has a notable impact on heat uptake by the ocean and the internal ocean heat distribution (Fig. 12). Despite continuous injection of a large amount of very cold (-15°C) water in these pure freshwater experiments, substantial portions of the ocean interior become warmer. Tropical and Southern Hemisphere warming is the well-known effect of reduced heat transport to northern latitudes in response to the AMOC shutdown (Rahmstorf, 1996; Barreiro et al., 2008).

However, deep warming in the Southern Ocean may have greater consequences. Warming is maximum at grounding line depths (~1-2 km) of Antarctic ice shelves (Rignot and Jacobs, 2002). Ice shelves near their grounding lines (Fig. 13 of Jenkins and Doake, 1991) are sensitive to temperature of the proximate ocean, with ice shelf melting increasing 1 meter per year for each 0.1°C temperature increase (Rignot and Jacobs, 2002). The foot of an ice shelf provides most of the restraining force that ice shelves exert on landward ice (Fig. 14 of Jenkins and Doake, 1991), making ice near the grounding line the buttress of the buttress. Pritchard et al. (2012) deduce from satellite altimetry that ice shelf melt has primary control of Antarctic ice sheet mass loss.

Thus we examine our simulations in more detail (Fig. 13). The pure freshwater experiments add 5 mm sea level in the first decade (requiring an initial 0.346 mm/year for 10-year doubling), 10 mm in the second decade, and so on (Fig. 13a). Cumulative freshwater injection reaches 0.5 m in year 68 and 2.5 m in year 90.

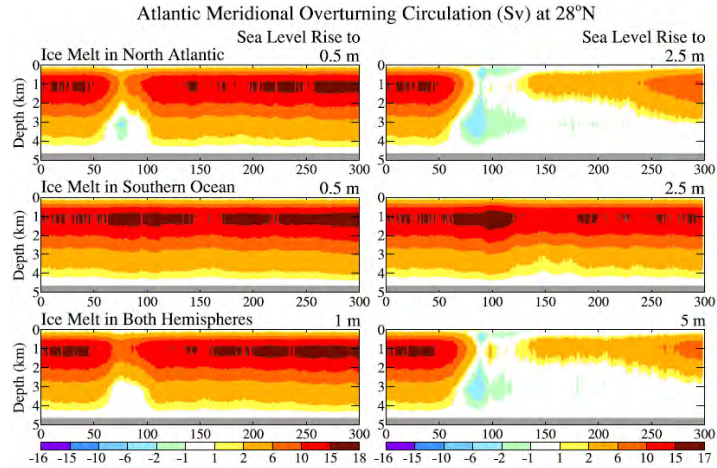

**Fig. 10.** Ensemble-mean AMOC (Sv) at 28°N versus time for six pure freshwater forcing experiments.

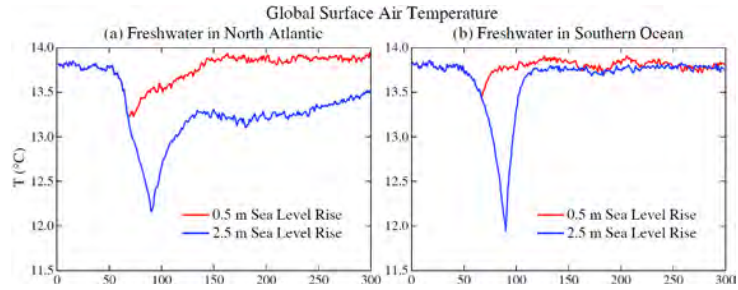

**Fig. 11.** Ensemble-mean global surface air temperature (°C) for experiments (years on x-axis) with freshwater forcing in either the North Atlantic Ocean (left) or the Southern Ocean (right).

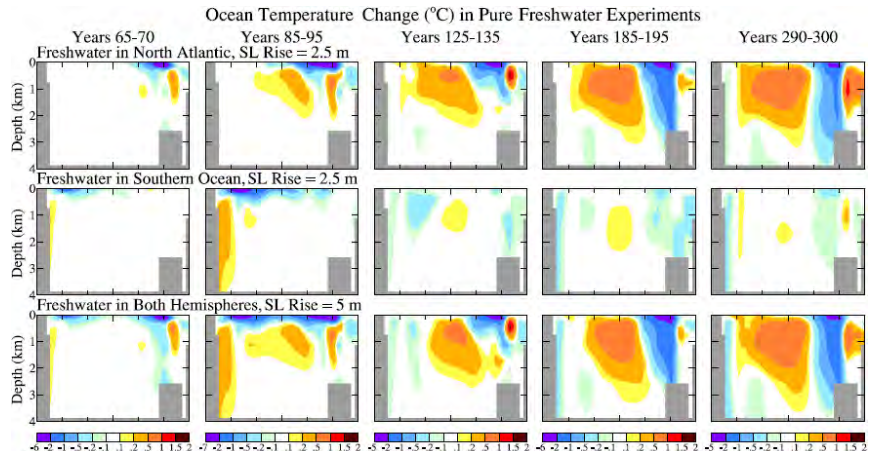

**Fig. 12.** Change of ocean temperature (°C) relative to control run due to freshwater input that reaches 2.5 m of global sea level in a hemisphere (thus 5 m sea level rise in the bottom row).

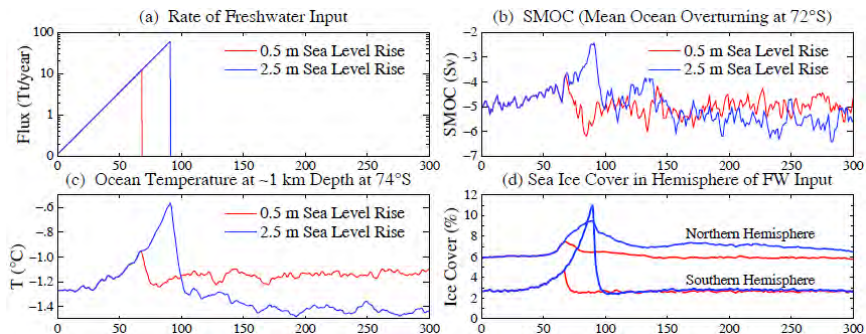

**Fig. 13.** (a) Freshwater input (Tt/year) to Southern Ocean (1 Tt = 1000 km³). (b, c, d) Simulated overturning strength (Sv) of AABW cell at 72°S, temperature (°C) at depth 1.13 km at 74S, and sea ice cover (%).

Antarctic Bottom Water (AABW) formation is reduced ~20% by year 68 and ~50% by year 90 (Fig. 13b). When freshwater injection ceases, AABW formation rapidly regains full strength, in contrast to the long delay in reestablishing North Atlantic Deep Water (NADW) formation after AMOC shutdown. The Southern Ocean mixed layer response time dictates the recovery time for AABW formation. Thus rapid recovery also applies to ocean temperature at depths of ice shelf grounding lines (Fig. 13c). The rapid response of the Southern Meridional Circulation (SMOC) implies that the rate of freshwater addition to the mixed layer is the driving factor.

Freshwater flux has little effect on simulated Northern Hemisphere sea ice until the 7[th] decade of freshwater growth (Fig. 13d), but Southern Hemisphere sea ice is more sensitive, with substantial response in the 5[th] decade and large response in the 6[th] decade. Below we show that "5[th] decade" freshwater flux (2880 Gt/year) is already relevant to the Southern Ocean today.

## 3.7 Simulations to 2100 with modified (more realistic) forcings

Recent data show that current ice melt is larger than assumed in our 1850-2300 simulations. Thus we make one more simulation and include minor improvements in the radiative forcing.

### 3.7.1 Advanced (earlier) freshwater injection

Atmosphere-ocean climate models, including ours, commonly include a fixed freshwater flux from the Greenland and Antarctic ice sheets to the ocean. This flux is chosen to balance snow accumulation in the model's control run, with the rationale that approximate balance is expected between net accumulation and mass loss including icebergs and ice shelf melting. Global warming creates a mass imbalance that we want to investigate. Ice sheet models can calculate the imbalance, but it is unclear how reliably ice sheet models simulate ice sheet disintegration. We forgo ice sheet modeling, instead adding a growing freshwater amount to polar oceans with alternative growth rates and initial freshwater amount estimated from available data.

Change of freshwater flux into the ocean in a warming world with shrinking ice sheets consists of two terms, Term 1 being net ice melt and Term 2 being change of P-E (precipitation minus evaporation) over the relevant ocean. Term 1 includes land based ice mass loss, which can be detected by satellite gravity measurements, loss of ice shelves, and net sea ice mass change. Term 2 is calculated in a climate model forced by changing atmospheric composition, but it is not included in our pure freshwater experiments that have no global warming.

IPCC (Vaughan et al., 2013) estimated land ice loss in Antarctica that increased from 30 Gt/year in 1992-2001 to 147 Gt/year in 2002-2011 and in Greenland from 34 Gt/year to 215 Gt/year, with uncertainties discussed by Vaughan et al. (2013). Gravity satellite data suggest Greenland ice sheet mass loss ~300-400 Gt/year in the past few years (Barletta et al., 2013). A newer analysis of gravity data for 2003-2013 (Velicogna et al., 2014), discussed in more detail in Sec. 5.1, finds a Greenland mass loss 280 ± 58 Gt/yr and Antarctic mass loss 67 ± 44 Gt/year.

One estimate of net ice loss from Antarctica, including ice shelves, is obtained by surveying and adding the mass flux from all ice shelves and comparing this freshwater mass loss with the freshwater mass gain from the continental surface mass budget. Rignot et al. (2013) and Depoorter et al. (2013) independently assessed the freshwater mass fluxes from Antarctic ice shelves. Their respective estimates for the basal melt are 1500 ±237 Gt/year and 1454 ±174 Gt/year. Their respective estimates for calving are 1265 ± 139 Gt/year and 1321 ± 144 Gt/year.

This estimated freshwater loss via the ice shelves (~2800 Gt/year) is larger than freshwater gain by Antarctica. Vaughan et al. (1999) estimated net surface mass balance of the continent as +1811 Gt/year and +2288 Gt/year including precipitation on ice shelves. Vaughan et al. (2013) estimates the net Antarctic surface mass balance as +1983 ± 122 Gt/year excluding ice shelves. Thus comparison of continental freshwater input with ice shelf output suggests a net export of freshwater to the Southern Ocean of several hundred Gt/year in recent years. However, substantial uncertainty exists in the difference between these two large numbers.

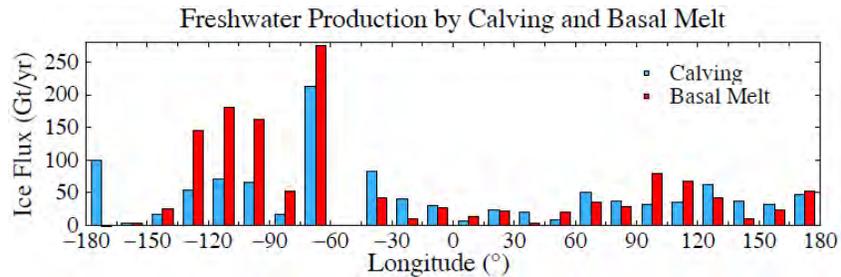

**Fig. 14.** Freshwater flux (Gt/year) from Antarctic ice shelves based on data of Rignot et al. (2013), integrated here into intervals of 15° of longitude. Depoorter et al. (2013) data yield a similar distribution.

An independent evaluation has recently been achieved by Rye et al. (2014) using satellite measured changes of sea level around Antarctica in the period 1992-2011. Sea level along the Antarctic coast rose 2 mm/year faster than the regional mean sea level rise in the Southern Ocean south of 50°S, an effect that they conclude is almost entirely a steric adjustment caused by accelerating freshwater discharge from Antarctica. They conclude that an excess freshwater input of 430 ± 230 Gt/year, above the rate needed to maintain a steady ocean salinity, is required. Rye et al. (2014) note that these values constitute a lower bound for the actual excess discharge above a "steady salinity" rate, because numerous in situ data, discussed below, indicate that freshening began earlier than 1992.

Term 2, change of P-E over the Southern Ocean relative to its pre-industrial amount, is large in our climate simulations. In our ensemble of runs (using observed GHGs for 1850-2003 and scenario A1B thereafter) the increase of P-E in the decade 2011-2020, relative to the control run, was in the range 3500 to 4000 Gt/year, as mean precipitation over the Southern Ocean increased ~35 mm/year and evaporation decreased ~3 mm/year.

Increasing ice melt and increasing P-E are climate feedbacks, their growth in recent decades driven by global warming. Our pure freshwater simulations indicate that their sum, at least 4000 Gt/year, is sufficient to affect ocean circulation, sea ice cover, and surface temperature, which can spur other climate feedbacks. We investigate these feedbacks via climate simulations using improved estimates of freshwater flux from ice melt. P-E is computed by the model.

We take freshwater injection to be 720 Gt/year from Antarctica and 360 Gt/year in the North Atlantic in 2011, with injection rates at earlier and later times defined by assumption of a 10-year doubling time. Resulting mean freshwater injection around Antarctica in 1992-2011 is ~400 Gt/year, similar to the estimate of Rye et al. (2014). A recent estimate of 310 ±74 km³ volume loss of floating Antarctic ice shelves in 2003-2012 (Paolo et al., 2015) is not inconsistent, as the radar altimeter data employed for ice shelves does not include contributions from the ice sheet or fast ice tongues at the ice shelf grounding line. Greenland ice sheet mass loss provides most of the assumed 360 Gt/year freshwater, and this would be supplemented by shrinking ice shelves (Rignot and Steffen, 2008) and small ice caps in the North Atlantic and west of Greenland (Ohmura, 2009) that are losing mass (Abdalati et al., 2004; Bahr et al., 2009).

We add freshwater around Antarctica at coastal grid boxes (Fig. S13) guided by the data of Rignot et al. (2013) and Depoorter et al. (2013). Injection in the western hemisphere, especially from the Weddell Sea to the Ross Sea, is more than twice that in the other hemisphere (Fig. 14). Specified freshwater flux around Greenland is similar on the east and west coasts, and small along the north coast (Fig. S13).

### 3.7.2 Modified radiative forcings

Actual GHG forcing is less than scenario A1B, because CH₄ and minor gas growth declined after IPCC scenarios were defined (Fig. 5, Hansen et al., 2013c, update at http://www.columbia .edu/~mhs119/GHGs/). As a simple improvement we decreased the A1B CH₄ scenario during 2003-2013 so that subsequent CH₄ is reduced 100 ppb, decreasing radiative forcing ~0.05 W/m².

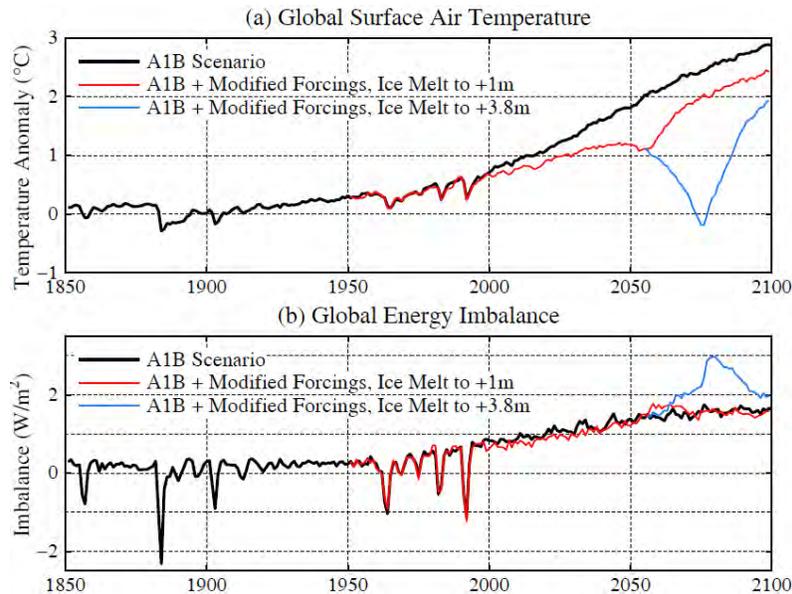

**Fig. 15.** (a) Surface air temperature (°C) change relative to 1880-1920 and (b) global energy imbalance (W/m²) for the modified forcing scenario including cases with global ice melt reaching 1 m and 3.8 m.

Stratospheric aerosol forcing to 2014 uses the data set of Sato et al. (1993) as updated at http://www.columbia.edu/~mhs119/StratAer/. Future years have constant aerosol optical depth 0.0052 yielding effective forcing -0.12 W/m², implemented by using fixed 1997 aerosol data. Tropospheric aerosol growth is assumed to slow smoothly, leveling out at -2 W/m² in 2100. Future solar forcing is assumed to have an 11-year cycle with amplitude 0.25 W/m². Net forcing exceeds 5 W/m² by the end of the 21$^{st}$ century, about three times the current forcing (Fig. S16).

### 3.7.3 Climate simulations with modified forcings

Global temperature has a maximum at +1.2°C in the 2040s for the modified forcings (Fig. 15). Ice melt cooling is advanced as global ice melt reaches 1 m of sea level in 2060, 1/3 from Greenland and 2/3 from Antarctica. Global temperature rise resumes in the 2060s after cessation of freshwater injection.

Global temperature becomes an unreliable diagnostic of planetary condition as the ice melt rate increases. Global energy imbalance (Fig. 15b) is a more meaningful measure of planetary status as well as an estimate of the climate forcing change required to stabilize climate. Our calculated present energy imbalance of ~0.8 W/m² (Fig. 15b) is larger than the observed 0.58 ± 0.15 W/m² during 2005-2010 (Hansen et al., 2011). The discrepancy is likely accounted for by excessive ocean heat uptake at low latitudes in our model, a problem related to the model's slow surface response time (Fig. 4) that may be caused by excessive small scale ocean mixing.

Large scale regional cooling occurs in the North Atlantic and Southern Oceans by mid-century (Fig. 16) for 10-year doubling of freshwater injection. A 20-year doubling places similar cooling near the end of this century, 40 years earlier than in our prior simulations (Fig. 7), as the factor of four increase of current freshwater from Antarctica is a 40-year advance.

Cumulative North Atlantic freshwater forcing in Sverdrup years (Sv years) is 0.2 Sv years in 2014, 2.4 Sv years in 2050, and 3.4 Sv years (its maximum) prior to 2060 (Fig. S14). The critical issue is whether human-spurred ice sheet mass loss can be approximated as an exponential process during the next few decades. Such nonlinear behavior depends upon amplifying feedbacks, which, indeed, our climate simulations reveal in the Southern Ocean.

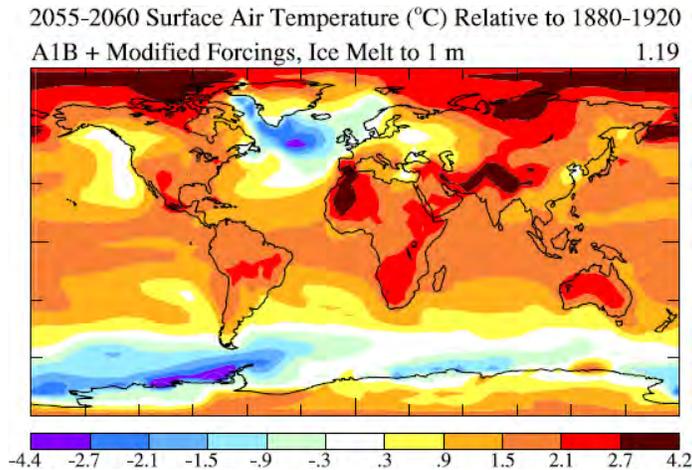

**Fig. 16.** Surface air temperature (°C) change relative to 1880-1920 in 2055-2060 for modified forcings.

### 3.7.4 Southern Ocean feedbacks

Amplifying feedbacks in the Southern Ocean and atmosphere contribute to dramatic climate change in our simulations (Fig. 16). We first summarize the feedbacks to identify processes that must be simulated well to draw valid conclusions. While recognizing the complexity of the global ocean circulation (Lozier, 2012; Lumpkin and Speer, 2007; Marshall and Speer, 2012; Munk and Wunsch, 1998; Orsi et al., 1999; Sheen et al., 2014; Talley, 2013; Wunsch and Ferrari, 2004), we use a simple two-dimensional representation to discuss the feedbacks.

Climate change includes slowdown of AABW formation, indeed shutdown by midcentury if freshwater injection increases with a doubling time as short as 10 years (Fig. 17). Implications of AABW shutdown are so great that we must ask whether the mechanisms are simulated with sufficient realism in our climate model, which has coarse resolution and relevant deficiencies that we have noted. After discussing the feedbacks here, we examine how well the processes are included in our model (Sec. 3.7.5). Paleoclimate data (Sec. 4) provides much insight about these processes and modern observations (Sec. 5) suggest that these feedbacks are already underway.

Large-scale climate processes affecting ice sheets are sketched in Fig. 18. The role of the ocean circulation in the global energy and carbon cycles is captured to a useful extent by the two-dimensional (zonal mean) overturning circulation featuring deep water (NADW) and bottom water (AABW) formation in the polar regions. Marshall and Speer (2012) discuss the circulation based in part on tracer data and analyses by Lumpkin and Speer (2007). Talley (2013) extends the discussion with diagrams clarifying the role of the Pacific and Indian Oceans.

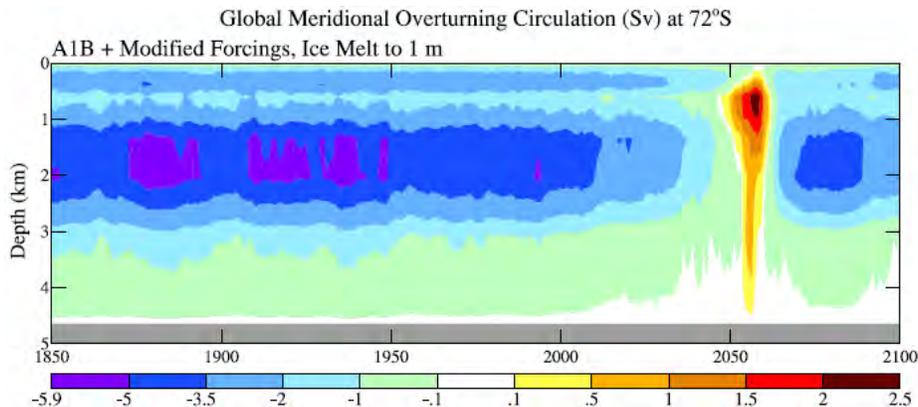

**Fig. 17.** SMOC, ocean overturning strength (Sv) at 72°S, including only the mean (Eulerian) transport. This is the average of a 5-member model ensemble for the modified forcing including advanced ice melt (720 Gt/year from Antarctica in 2011) and 10-year doubling.

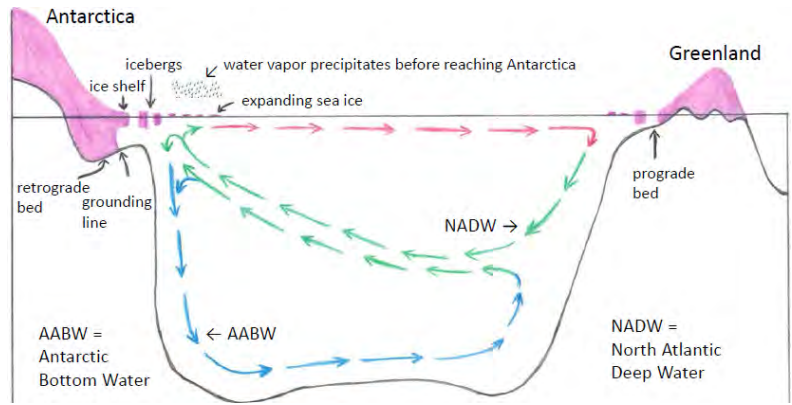

**Fig. 18.** Schematic of stratification and precipitation amplifying feedbacks. Stratification: increased freshwater flux reduces surface water density, thus reducing AABW formation, trapping NADW heat, and increasing ice shelf melt. Precipitation: increased freshwater flux cools ocean mixed layer, increases sea ice area, causing precipitation to fall before it reaches Antarctica, reducing ice sheet growth and increasing ocean surface freshening. Ice in West Antarctica and the Wilkes Basin, East Antarctica is most vulnerable because of the instability of retrograde beds.

Wunsch (2002) emphasizes that the ocean circulation is driven primarily by atmospheric winds and secondarily by tidal stirring. Strong circumpolar westerly winds provide energy drawing deep water toward the surface in the Southern Ocean. Ocean circulation also depends on processes maintaining the ocean's vertical density stratification. Winter cooling of the North Atlantic surface produces water dense enough to sink (Fig. 18), forming North Atlantic Deep Water (NADW). However, because North Atlantic water is relatively fresh, compared to the average ocean, NADW does not sink all the way to the global ocean bottom. Bottom water is formed instead in the winter around the Antarctic coast, where very salty cold water (AABW) can sink to the ocean floor. This ocean circulation (Fig. 18) is altered by natural and human-made forcings, including freshwater from ice sheets, engendering powerful feedback processes.

A key Southern Ocean feedback is meltwater stratification effect, which reduces ventilation of ocean heat to the atmosphere and space. Our "pure freshwater" experiments show that the low density lid causes deep ocean warming, especially at depths of ice shelf grounding lines that provide most of the restraining force limiting ice sheet discharge (Fig. 14 of Jenkins and Doake, 1991). West Antarctica and Wilkes Basin in East Antarctica have potential to cause rapid sea level rise, because much of their ice sits on retrograde beds (beds sloping inland), a situation that can lead to unstable grounding line retreat and ice sheet disintegration (Mercer, 1978).

Another feedback occurs via the effect of surface and atmospheric cooling on precipitation and evaporation over the Southern Ocean. CMIP5 climate simulations, which do not include increasing freshwater injection in the Southern Ocean, find snowfall increases on Antarctica in the 21st century, thus providing a negative term to sea level change. Frieler et al. (2015) note that 35 climate models are consistent in showing that warming climate yields increasing snow accumulation in accord with paleo data for warmer climates, but the paleo data refer to slowly changing climate in quasi-equilibrium with ocean boundary conditions. In our experiments with growing freshwater injection, the increasing sea ice cover and cooling of the Southern Ocean surface and atmosphere cause the increased precipitation to occur over the Southern Ocean, rather than over Antarctica. This feedback not only reduces any increase of snowfall over Antarctica, it also provides a large freshening term to the surface of the Southern Ocean, thus magnifying the direct freshening effect from increasing ice sheet melt.

North Atlantic meltwater stratification effects are also important, but different. Meltwater from Greenland can slow or shutdown NADW formation, cooling the North Atlantic, with global impacts even in the Southern Ocean, as we will discuss later. One important difference is that the North Atlantic can take centuries to recover from NADW shutdown, while the Southern Ocean recovers within 1-2 decades after freshwater injection stops (Sec. 3.6).

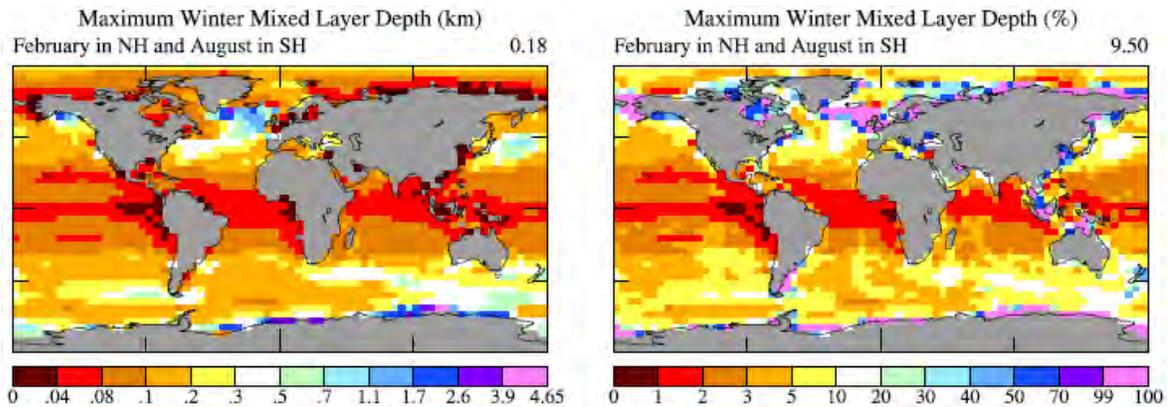

**Fig. 19.** Maximum mixed layer depth (in km, left, and % of ocean depth, right) in February (Northern Hemisphere) and August (Southern Hemisphere) using the mixed layer definition of Heuze et al. (2013).

### 3.7.5 Model's ability to simulate these feedbacks

Realistic representation of these feedbacks places requirements on both the atmosphere and ocean components of our climate model. We discuss first the atmosphere, then the ocean.

There are two main requirements on the atmospheric model. First, it must simulate well P-E, because of its importance for ocean circulation and the amplifying feedback in the Southern Ocean. Second, it must simulate well winds, because these drive the ocean.

Simulated P-E (Fig. S15b) agrees well with meteorological reanalysis (Fig. 3.4b of Rhein et al., 2013). Resulting sea surface salinity (SSS) patterns in the model (Fig. S15a) agree well with global ocean surface salinity patterns (Antonov et al., 2010 and Fig. 3.4a of Rhein et al., 2013). SSS trends in our simulation (Fig. S15c), with the Pacific on average becoming fresher while most of the Atlantic and the subtropics in the Southern Hemisphere become saltier, are consistent with observed salinity trends (Durack and Wijffels, 2010). Recent freshening of the Southern Ocean in our simulation is somewhat less than in observed data (Figs. 3.4c,d of Rhein et al., 2013), implying that the amplifying feedback may be *under*estimated in our simulation. A likely reason for that is discussed below in conjunction with observed sea ice change.

Obtaining accurate winds requires the model to simulate well atmospheric pressure patterns and their change in response to climate forcings. A test is provided by observed changes of the Southern Annular Mode (SAM), with a decrease of surface pressure near Antarctica and a small increase at mid-latitudes (Marshall, 2003) that D. Thompson et al. (2011) relate to stratospheric ozone loss and increasing GHGs. Our climate forcing (Fig. S16) includes ozone change (Fig. 2, Hansen et al., 2007a) with stratospheric ozone depletion in 1979-1997 and constant ozone thereafter. Our model produces a trend toward the high index polarity of SAM (Fig. S17) similar to observations, although perhaps a slightly smaller change than observed (compare Fig. S17 with Fig. 3 of Marshall, 2003). SAM continues to increase in our model after ozone stabilizes (Fig. S17), suggesting that GHGs may provide a larger portion of the SAM response in our model than in the model study of D. Thompson et al. (2011). It would not be surprising if the stratospheric dynamical response to ozone change were weak in our model, given the coarse resolution and simplified representation of atmospheric drag and dynamical effects in the stratosphere (Hansen et al., 2007a), but that is not a major concern for our present purposes.

The ocean model must be able to simulate realistically the ocean's overturning circulation and its response to forcings including freshwater additions. Heuze et al. (2013, 2015) point out that simulated deep convection in the Southern Ocean is unrealistic in most models, with AABW formation occurring in the open ocean where it rarely occurs in nature. Our present ocean model contains significant improvements (see Sec. 3.1) compared to the GISS E2-R model that Heuze et al. include in their comparisons. Thus we show (Fig. 19) the maximum mixed layer depth in winter (February in the Northern Hemisphere and August in the Southern Hemisphere) using the same criterion as Heuze et al. to define the mixed layer depth, i.e., the layers with a density difference from the ocean surface layer less than 0.03 kg/m$^3$.

Southern Ocean mixing in the model reaches a depth of ~500 m in a wide belt near 60°S stretching west from the southern tip of South America, with similar depths south of Australia. These open ocean mixed layer depths compare favorably with observations shown in Fig. 2a of Heuze et al. (2015), based on data of de Boyer Montegut et al. (2004). There is no open ocean deep convection in our model.

Deep convection occurs only along the coast of Antarctica (Fig. 19). Coastal grid boxes on the continental shelf are a realistic location for AABW formation. Orsi et al. (1999) suggest that most AABW is formed on shelves around the Weddell-Enderby Basin (60%) and shelves of the Adelie-Wilkes Coast and Ross Sea (40%). Our model produces mixing down to the shelf in those locations (Fig. 19b), and also on the Amery Ice Shelf near the location where Ohshima et al. (2013) identified AABW production, which they term Cape Darnley Bottom Water.

With our coarse 4° stair step to the ocean bottom, AABW cannot readily slide down the slope to the ocean floor. Thus dense shelf water mixes into the open ocean grid boxes, making our modeled Southern Ocean less stratified than the real world (cf. temporal drift of Southern Ocean salinity in Fig. S18), because the denser water must move several degrees of latitude horizontally before it can move deeper. Nevertheless, our Southern Ocean is sufficiently stratified to avoid the unrealistic open ocean convection that infects many models (Heuze et al., 2013, 2015).

Orsi et al. (1999) estimate the AABW formation rate in several ways, obtaining values in the range 8-12 Sv, larger than our modeled 5-6 Sv (Fig. 17). However, as in most models (Heuze et al., 2015) our SMOC diagnostic (Fig. 17) is the mean (Eulerian) circulation, i.e., excluding eddy-induced transport. Rerun of a 20-year segment of our control run to save eddy-induced changes reveals an increase of SMOC at 72°S by 1-2 Sv, with negligible change at middle and low latitudes, making our simulated transport close to the range estimated by Orsi et al. (1999).

We conclude that the model may simulate Southern Ocean feedbacks that magnify the effect of freshwater injected onto the Southern Ocean: the P-E feedback that wrings global-warming-enhanced water vapor from the air before it reaches Antarctica and the AABW slowdown that traps deep ocean heat, leaving that heat at levels where it accelerates ice shelf melting. Indeed, we will argue that both of these feedbacks are probably underestimated in our current model.

The model seems less capable in Northern Hemisphere polar regions. Deep convection today is believed to occur mainly in the Greenland-Iceland-Norwegian (GIN) Sea and at the southern end of Baffin Bay (Fig. 2b, Heuze et al., 2015). In our model, perhaps because of excessive sea ice in those regions, open ocean deep convection occurs to the southeast of the southern tip of Greenland and at less deep grid boxes between that location and the United Kingdom (Fig. 19). Mixing reaching the ocean floor on the Siberian coast in our model (Fig. 19) may be realistic, as coastal polynya are observed on the Siberian continental shelf (D. Bauch et al., 2012). However, the winter mixed layer on the Alaska south coast is unrealistically deep (Fig. 19). These model limitations must be kept in mind in interpreting simulated Northern Hemisphere climate change.

## 3.8 Impact of ice melt on storms

Our inferences about potential storm changes from continued high growth of atmospheric GHGs are fundamentally different than modeling results described in IPCC (2013, 2014), where the latter are based on CMIP5 climate model results without substantial ice sheet melt. Lehmann et al. (2014) note ambiguous results for storm changes from prior model studies and describe implications of the CMIP5 ensemble of coupled climate models. Storm changes are moderate in nature, with even a weakening of storms in some locations and seasons. This is not surprising, because warming is greater at high latitudes, reducing meridional temperature gradients.

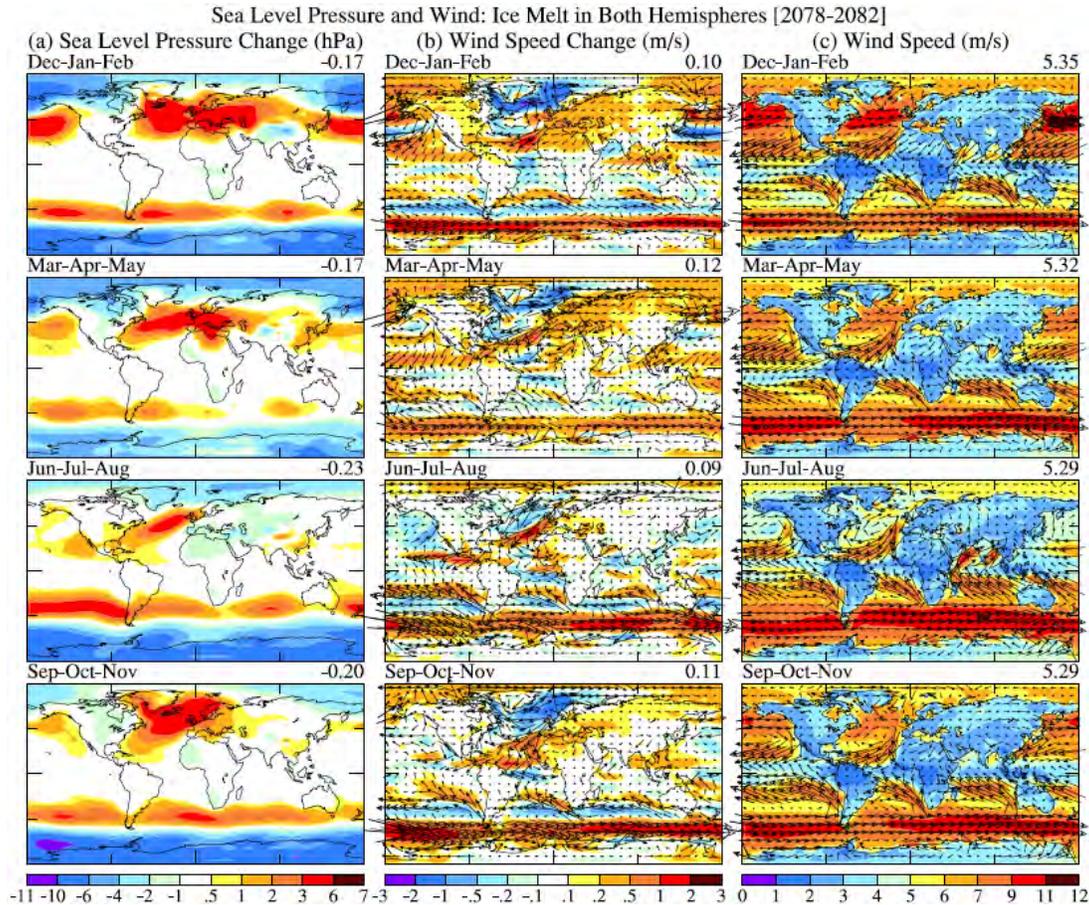

**Fig. 20.** Change of seasonal mean **(a)** sea level pressure (hPa), **(b)** wind speed (m/s) in 2078-2082 relative to 1880-1920, and **(c)** the wind speed (m/s) itself, all for the scenario with ice melt in both hemispheres.

Before describing our model results, we note the model limitations for study of storms, including its coarse resolution (4°×5°), which may contribute to slight misplacement of the Bermuda high pressure system for today's climate (Fig. S2). Excessive Northern Hemisphere sea ice may cause a bias in location of deepwater formation toward lower latitudes. Simulated effects also depend on the location chosen for freshwater injection; in model results shown here (Fig. 20) freshwater was spread uniformly over all longitudes in the North Atlantic between 65°W and 15°E. It would be useful to carry out similar studies with higher resolution models including the most realistic possible distribution of meltwater.

Despite these caveats, we have shown that the model realistically simulates meridional changes of sea level pressure in response to climate forcings (Sec. 3.7.5). Specifically, the model yields a realistic trend to the positive phase of the Southern Annular Mode (SAM) in response to decrease of stratospheric ozone and increase of other GHGs (Fig. S17). We also note that the modeled response of atmospheric pressure to the cooling effect of ice melt is large scale, tending to be of a meridional nature that should be handled by our model resolution.

Today's climate, not Eemian climate, is the base climate state upon which we inject polar freshwater. However, the simulated climate effects of the freshwater are so large that they should also be relevant to freshwater injection in the Eemian period.

### 3.8.1 Modeling insights on Eemian storms

Ice melt in the North Atlantic increases simulated sea level pressure in that region in all seasons (Fig. 20). In summer the increased surface pressure strengthens and moves northward the Bermuda high pressure system (Fig. S2). Circulation around the high pressure creates stronger prevailing northeasterly winds at latitudes of Bermuda and the Bahamas. A1B climate forcing alone (Fig. S21 top row) has only a small impact on the winds, but cold meltwater in the North Atlantic causes a strengthening and poleward shift of the high pressure.

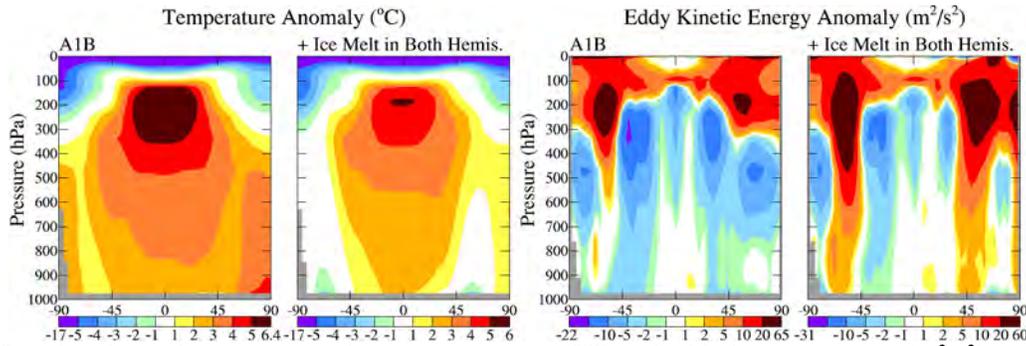

**Fig. 21.** Simulated zonal mean atmospheric temperature (°C) and eddy kinetic energy (m²/s²) in 2078-2082 relative to 1880-1920 for A1B scenario and A1B plus 2.5 m ice melt in each hemisphere.

The high pressure in the model is located further east than needed to produce the fastest possible winds at the Bahamas. Our coarse resolution (4°×5°) model may be partly responsible for the displacement. However, the location of high pressure also depends on meltwater placement, which we spread uniformly over all longitudes in the North Atlantic between 65°W and 15°E and on specific location of ocean currents and surface temperature during the Eemian.

North Atlantic cooling from AMOC shutdown creates faster winds in our simulations, with . seasonal-mean increment as much as 10-20%. Such a percentage translates into an increase of storm power dissipation by a factor ~1.4-2, because dissipation is proportional to the cube of wind speed (Emanuel, 1987, 2005). Our simulated changes refer to mean winds over large grid-boxes, not individual storms, for which the change of the most extreme cases might be larger.

Increased North Atlantic high pressure strengthens prevailing northeasterly winds blowing onto the Bahamas in the direction of Eemian wave-formed deposits (Sec. 4.1.2). Consistent increase of these winds would contribute to creation of long wavelength, deep ocean waves that scour the ocean floor as they reach the shallow near-shore region. However, extreme events may require the combined effect of increased prevailing winds and tropical storms guided by the strengthened blocking high pressure and nurtured by the unusually warm late-Eemian tropical sea surface temperatures (Cortijo et al., 1999), which would favor more powerful tropical storms (Emanuel, 1987). This enhanced meridional temperature gradient, warmer tropics and cooler high latitudes, was enhanced by low obliquity of Earth's spin axis in the late Eemian.

### 3.8.2  21st Century storms

If GHGs continue to increase rapidly and ice melt grows, our simulations yield shutdown or major slowdown of the AMOC in the 21st century, implying an increase of severe weather. This is shown by zonal mean temperature and eddy kinetic energy changes in simulations of Sec. 3.2 with and without ice melt (Fig. 21). Without ice melt, surface warming is largest in the Arctic (Fig. 21, left), resulting in a decrease of lower tropospheric eddy energy. However, the surface cooling from ice melt increases surface and lower tropospheric temperature gradients, and in stark contrast to the case without ice melt, there is a large increase of mid-latitude eddy energy throughout the midlatitude troposphere. The increase of zonal-mean midlatitude baroclinicity (Fig. 21) is in agreement with the localized, North Atlantic-centered increases in baroclinicity found in the higher resolution simulations of Jackson et al. (2015) and Brayshaw et al. (2009).

Increased baroclinicity produced by a stronger temperature gradient provides energy for more severe weather events. Many of the most memorable and devastating storms in eastern North

America and western Europe, popularly known as superstorms, have been winter cyclonic storms, though sometimes occurring in late fall or early spring, that generate near-hurricane force winds and often large amounts of snowfall (Chapter 11, Hansen, 2009). Continued warming of low latitude oceans in coming decades will provide a larger water vapor repository that can strengthen such storms. If this tropical warming is combined with a cooler North Atlantic Ocean from AMOC slowdown and an increase in midlatitude eddy energy (Fig. 21), we can anticipate more severe baroclinic storms. Increased high pressure due to cooler high latitude ocean (Fig. 20) can make blocking situations more extreme, with a steeper pressure gradient between the storm's low pressure center and the blocking high, thus driving stronger North Atlantic storms.

Freshwater injection on the North Atlantic and Southern Oceans increases sea level pressure at middle latitudes and decreases it at polar latitudes (Figs. 20, S22), but the impact is different in the North Atlantic than in the Southern Ocean. In the Southern Ocean the increased meridional temperature gradient increases the strength of westerlies in all seasons at all longitudes. In the North Atlantic Ocean the increase of sea level pressure in winter slows the westerlies (Fig. 20). Thus instead of a strong zonal wind that keeps cold polar air locked in the Arctic, there is a tendency for a less zonal flow and thus more cold air outbreaks to middle latitudes.

## 4 Earth's climate history

Earth's climate history is our richest source of information about climate processes. We first examine the Eemian or MIS 5e period, the last time Earth was as warm as today, because it is especially relevant to the issue of rapid sea level rise and storms when ice sheets existed only on Greenland and Antarctica. A fuller interpretation of late-Eemian climate events, as well as projection of climate change in the Anthropocene, requires understanding mechanisms involved in Earth's millennial climate oscillations, which we discuss in the following subsection.

### 4.1 Eemian interglacial period (marine isotope substage MIS 5e)

We first discuss Eemian sea level (4.1.1), especially evidence for rapid sea level rise late in the Eemian to +6-9 m relative to today's sea level, and then evidence for strong late-Eemian storms (4.1.2). We provide in the Supplement more detailed geologic analysis of data on Eemian sea level, because the rapid late-Eemian sea level rise relates to our expectation of likely near-future events if rapid global warming continues. In Sec. 4.1.3 we present evidence from ocean sediment cores for strong late-Eemian cooling in the North Atlantic associated with shutdown of the Atlantic Meridional Overturning Circulation (AMOC), and in Sec. 4.1.4 we show that Earth orbital parameters in the late Eemian were consistent with cooling in the North Atlantic and global sea level rise from Antarctic ice sheet collapse.

#### 4.1.1 Eemian sea level

Eemian sea level is of special interest because Eemian climate was little warmer than today. Masson-Delmotte et al. (2013) conclude, based on multiple data and model sources, that peak Eemian temperature probably was only a few tenths of a degree warmer than today. Yet Eemian sea level reached heights several meters above today's level (Land et al., 1967; Chen et al., 1991; Neumann and Hearty, 1996; Hearty et al., 2007; Kopp et al., 2009; Dutton and Lambeck, 2012; O'Leary et al., 2013; Dutton et al., 2015).

Change of sea level within the Eemian period is particularly relevant to concerns about ice sheet stability and the potential for rapid sea level rise. Hearty et al. (2007) used data from 15 sites around the world to construct an Eemian sea level curve that had sea level rising in the early Eemian to +2-3 m ("+" indicates above today's sea level), possibly falling in mid-Eemian to near today's sea level, rapidly rising in late-Eemian to +6-9 m, and then plummeting as Earth moved from the Eemian into the 100,000 year glacial period preceding the Holocene. Evidence from a variety of sources supports this interpretation, as discussed in the Supplement.

The most comprehensive analyses of sea level and paleoclimate storms are obtained by combining information from different geologic sources, each with strengths and weaknesses. Coral reefs, for example, allow absolute U/Th dating with age uncertainty as small as 1-2 ky, but inferred sea levels are highly uncertain because coral grows below sea level at variable depths as great as several meters. Carbonate platforms such as Bermuda and the Bahamas, in contrast, have few coral reefs for absolute dating, but the ability of carbonate sediments to cement rapidly preserves rock evidence of short-lived events such as rapid sea level rise and storms.

The important conclusion, that sea level rose rapidly in late-Eemian by several meters, to +6-9 m, is supported by records preserved in both the limestone platforms and coral reefs. Figure 6 of Hearty and Kindler (1995), for example, based on Bermuda and Bahamas geological data from marine and eolian limestone, reveals the rapid late-Eemian sea level rise and fall. Based on the limited size of the notches cut in Bahamian shore during the rapid late Eemian level rise and crest, Neumann and Hearty (1996) inferred that this period was at most a few hundred years. Independently, Blanchon et al. (2009) used coral reef "back-stepping" on Yucatan peninsula, i.e., movement of coral reef building shoreward as sea level rises, to conclude that sea level in late-Eemian jumped 2-3 m within an "ecological" period, i.e., within several decades.

Despite general consistency among these studies, considerable uncertainty remains about absolute Eemian sea level elevation and exact timing of end-Eemian events. Uncertainties include effects of local tectonics and glacio-isostatic adjustment (GIA) of Earth's crust. Models of GIA of Earth's crust to ice sheet loading and unloading are increasingly used to improve assessments. O'Leary et al. (2013) use over 100 corals from reefs at 28 sites along the 1400 km west coast of Australia, incorporating minor GIA corrections, to conclude that sea level in most of the Eemian was relatively stable at +3-4 m, followed by a rapid late-Eemian sea level rise to about +9 m. U-series dating of the corals has peak sea level at 118.1 ± 1.4 ky b2k.

A more complete discussion of data on Eemian sea level is provided in the Supplement.

Late-Eemian sea level rise may seem a paradox, because orbital forcing then favored growth of Northern Hemisphere ice sheets. We will find evidence, however, that the sea level rise and increased storminess are consistent, and likely related to events in the Southern Ocean.

### 4.1.2 Evidence of end-Eemian storms in Bahamas and Bermuda

Geologic data indicate that the rapid end-Eemian sea level oscillation was accompanied by increased temperature gradients and storminess in the North Atlantic region. We summarize several interconnected lines of evidence for end-Eemian storminess, based on geological studies in Bermuda and the Bahamas referenced below. It is important to consider *all* the physical evidence of storminess rather than exclusively the transport mechanism of the boulders, indeed, it is essential to integrate data from obviously wave-produced runup and chevron deposits that exist a few km distant in North Eleuthera, Bahamas as well as across the Bahama Platform.

The Bahama Banks are flat, low-lying carbonate platforms that are exposed as massive islands during glacials and largely inundated during interglacial high stands. From a tectonic perspective, the platforms are relatively stable, as indicated by near horizontal +2-3 m elevation of Eemian reef crests across the archipelago (Hearty and Neumann, 2001). During MIS 5e sea level high stands, an enormous volume of aragonitic oolitic grains blanketed the shallow, high-energy banks. Sea level shifts and storms formed shoals, ridges, and dunes. Oolitic sediments indurated rapidly ($\sim 10^1$ to $\sim 10^2$ yr) once stabilized, preserving detailed and delicate lithic evidence of these brief, high-energy events. This shifting sedimentary substrate across the banks was inimical to coral growth, which partially explains the rarity of reefs during late MIS 5e.

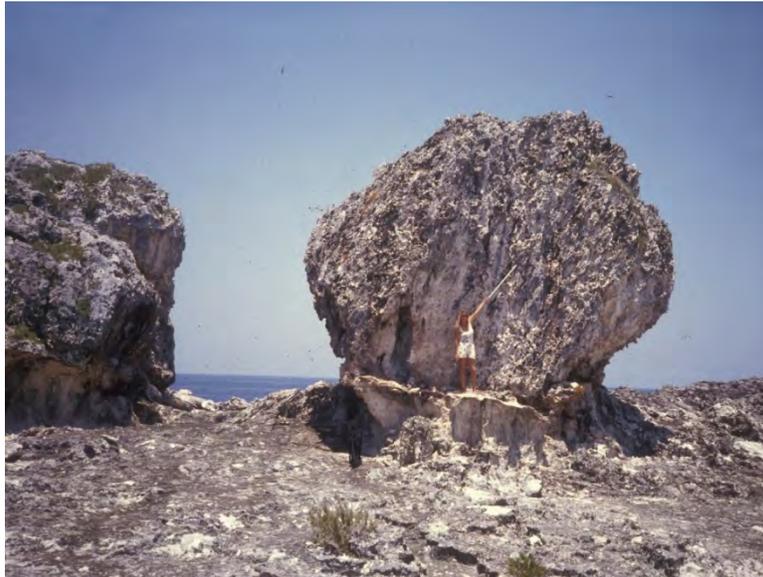

**Fig. 22.** Megaboulders #1 (left) and #2 resting on MIS 5e eolianite at the crest of a 20 m high ridge with person (1.7 m) showing scale and orientation of bedding planes in the middle Pleistocene limestone. The greater age compared to underlying strata and disorientation of the primary bedding beyond natural *in situ* angles indicates that the boulders were wave transported.

The preserved regional stratigraphic, sedimentary and geomorphic features attest to a turbulent end-Eemian transition in the North Atlantic. As outlined below, a coastal gradient of sedimentological features corresponds with coastal morphology, distance from the coast, and increasing elevation, reflecting the attenuating force and inland 'reach' of large waves, riding on high late-Eemian sea levels. On rocky, steep coasts, giant limestone boulders were detached and catapulted onto and over the coastal ridge by ocean waves. On higher, Atlantic-facing built up dune ridges, waves ran up to over 40 m elevation, leaving meter-thick sequences of fenestral beds, pebble lenses, and scour structures. Across kilometers of low-lying tidal inlets and flats, "nested" chevron clusters were formed as stacked, multi-meter thick, tabular fenestral beds.

The complexity of geomorphology and stratigraphy of these features are temporal measures of sustained sea level and storm events, encompassing perhaps hundreds of years. These features exclude a single wave cluster from a local point-source tsunami. Here we present data showing the connections among the megaboulders, runup deposits, and chevron ridges.

### 4.1.2.1 Megaboulders

In North Eleuthera enormous boulders were plucked from seaward middle Pleistocene outcrops and washed onto a younger Pleistocene landscape (Hearty and Neumann, 2001). The average 1000-ton megaclasts provide a metric of powerful waves at the end of MIS 5e. Evidence of transport by waves includes: 1) They are composed of recrystallized oolitic-peloidal limestone of MIS 9 or 11 age (300 - 400 ky; Kindler and Hearty, 1996) and hammer-ringing hardness; 2) They rest on oolitic sediments typical of early to mid MIS 5e that are soft and punky under hammer blows; 3) *Cerion* land snail fossils beneath boulder #4 (Hearty, 1997) correlate with the last interglacial period (Garrett and Gould, 1984; Hearty and Kaufman, 2009); 4) Calibrated amino acid racemization (AAR) ratios (Hearty, 1997; Hearty et al., 1998; Hearty and Kaufman, 2000; 2009) confirm the last interglacial age of the deposits as well as the stratigraphic reversal; 5) Dips of bedding planes in boulders between 50° and 75° (Fig. 22) far exceed natural angles; and 6) Some of the largest boulders are located on MIS 5e deposits at the crest of the island's ridge, proving that they are not karstic relics of an ancient landscape (Mylroie, 2008).

The ability of storm waves to transport large boulders is demonstrated. Storms in the North Atlantic tossed boulders as large as 80 tons to a height +11 m on the shore on Ireland's Aran Islands (Cox et al., 2012), this specific storm on 5 January 1991 being driven by a low pressure system that recorded a minimum 946 mb, producing wind gusts to 80 knots and sustained winds of 40 knots for 5 hours (Cox et al., 2012). Typhoon Haiyan (8 November 2013) in the Philippines produced longshore transport of a 180 ton block and lifted boulders of up to ~24 tons to elevations as high as 10 m (May et al., 2015). May et al. (2015) conclude that these observed facts "…*demand a careful re-evaluation of storm-related transport where it, based on the boulder's sheer size, has previously been ascribed to tsunamis*."

The situation of the North Eleuthera megaboulders is special in two ways. First, all the large boulders are located at the apex of a horseshoe-shaped bay that would funnel energy of storm waves coming from the northeast, the direction of prevailing winds. Second, the boulders are above a vertical cliff at right angles to the incoming waves, a situation that allows constructive interference of reflected and incoming waves (Cox et al., 2012). The ability of waves hitting that cliff to produce large near-vertical splash is shown by a photograph in the Supplement taken on 31 October 1991 when a storm in the North Atlantic produced large waves impacting Eleuthera.

It is generally accepted that the boulders were wave transported in the late Eemian. The boulders were deposited near complex chevron ridges and widespread runup deposits, which must be considered in analyzing wave-generating mechanisms. Lower elevation areas such as tidal inlets would have been flooded and scoured by the same waves, forming chevron ridges, and such large waves would also wash up onto higher, older ridges.

#### 4.1.2.2 Runup deposits

Across several hundred kilometers of the Bahama Islands, older built up dune ridges are mantled with wave runup deposits that reach heights over +40 m (Fig. 23). They are generally 1-5 m thick, fenestrae-filled, and seaward-sloping tabular beds (Wanless and Dravis, 1989; Chen et al., 1991; Neumann and Hearty, 1996; Tormey and Donovan, 2015). These stratigraphically youngest Eemian deposits mantle older MIS 5e dune deposits on the shore-parallel ridges, and are the upland correlative to wave-generated boulders and chevron formations.

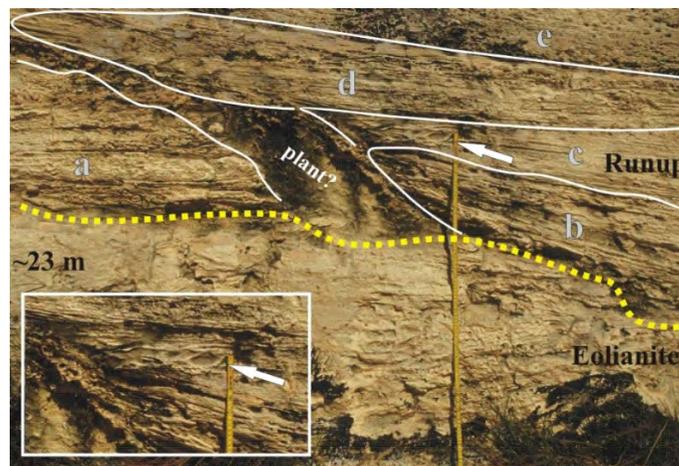

**Fig. 23.** Photograph of runup deposits in a road cutting above +23 m (1 m scale in photo) on Old Land Road, Great Exuma Island, situated deep in Exuma Sound ~200 km south of north Eleuthera. The older built-up eolianite forms the lower half of the image; the upper half has multiple "packages" of planar, fenestrae-filled beach sets. The upper progression of sedimentary packages (labelled a-e) clearly shows an onlapping, rising sequence of beds, indicating increasing wave energy and degree of runup. Further, the individual laminae of scour structures (arrows and inset image) display the same onlapping, upward climbing succession. It would be impossible to achieve such bedding if rain-saturated sediments were sloughing downhill on low angle slopes under the influence of gravity, especially near the crest of a ridge.

If these are deposits of powerful storms driven by an unusually warm tropical ocean and strong temperature gradients in the North Atlantic, as opposed to a tsunami, should there not be evidence of comparable end-Eemian storms in Bermuda?  Indeed, along several kilometers of the north coast of Bermuda (Land et al., 1967; Vacher and Rowe, 1997; Hearty et al., 1998) there are seaward sloping planar beds rising to about +20 m.  Although interpretations of these beds vary, they are filled with beach fenestrae and stratigraphically of latest MIS 5e carbonate sediments equivalent to runup in the Bahamas.  These planar beds contrast with older MIS 5e sedimentary (dune) structures that underlie them (Hearty et al., 1998).  Massive subtidal cross beds comprise the seaward facies of the elevated beach beds, pointing to an exceptional energy anomaly on the normally tranquil, shallow, broad and protected north shore platform of Bermuda.

### 4.1.2.3 Chevrons

In the Bahama Islands, extensive oolitic sand ridges with a distinctive landward-pointing V-shape are common, standing ~5-15 m high across several kilometers on broad, low lying platforms or ramps throughout the Atlantic-facing, deep-water margins of the Bahamas (Hearty et al., 1998). Hearty et al. (1998) examined 35 areas with chevron ridges across the Bahamas, which all point generally in a southwest direction (S65°W) with no apparent relation to the variable aspect of the coastline, nor to a point source event as would be expected for a tsunami generated by flank margin collapse.

These chevron formations are the lowland correlative to the wave-generated rocky coast boulder deposits. The chevron ridges often occur in nested groups of several ridges (e.g., North Eleuthera and Great Exuma; Hearty et al., 1998) and show multiple complex sets and subsets of fenestrae-filled beds, indicating the passage of a sustained interval of time late in the interglacial. Their definitive and complex characteristics preclude formation during a single tsunami event.

The character of fenestral beds in both the Eemian chevron ridges and runup deposits change with increasing elevation and distance from shore, as does the abundance and geometry of fenestral pores (Tormey and Donovan, 2015): 1) at low elevations and in proximal locations, the chevron ridges are dominated by multiple truncated, thick, tabular, fenestrae-rich beds (Fig. 24A,B); 2) at moderate elevations and further inland, fenestrae are concentrated in discrete packages within eolianites, often associated with scour (Fig. 23) and rip-up clasts; and 3) in the highest and most distal eolian ridges, only rare, thin, discontinuous fenestrae beds can be found (Fig. 24C; Tormey and Donovan, 2015). This spatial transition is improbable if torrential rain was falling across the area during a storm as asserted by Bain and Kindler (1994); rather this is exactly the pattern expected as waves attenuate with greater distance and elevation inland.

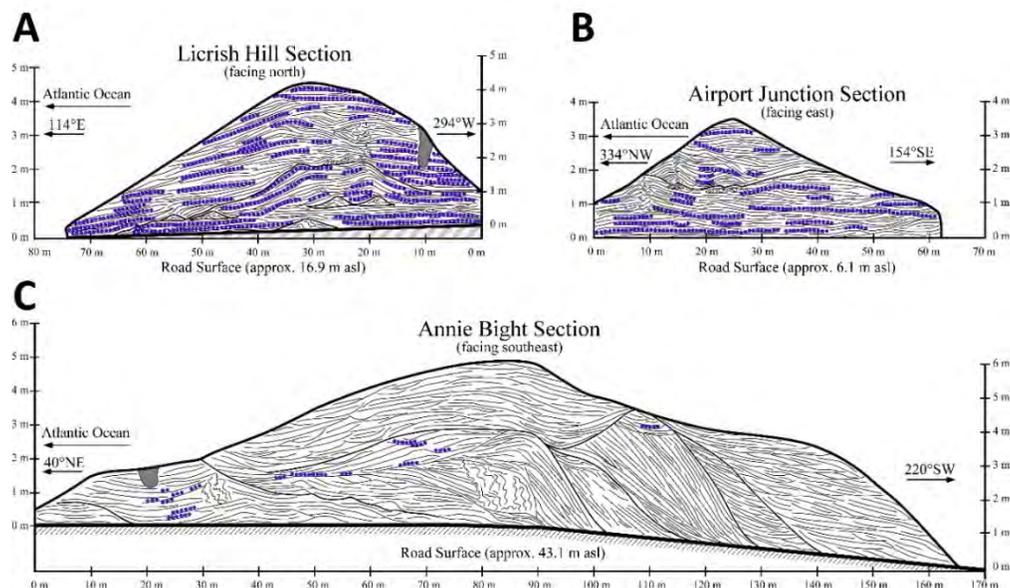

**Fig. 24.** Cross-section diagrams (Tormey, 1999) of Eemian chevron and dune deposits in North Eleuthera (A,B ~10 km west of megaboulders) showing geometry of bedding, fenestral porosity (lines of blue dots), and fossil roots (vertical wavy lines).  A) Chevron ridge exposure at Licrish Hill characterized by rising sequences of thick, tabular fenestral beds.  B) Chevron ridge exposure at Airport Junction characterized by rising sequences of thick, tabular fenestral beds.  C) Eolian ridge exposure at a higher elevation road cutting at Annie Bight (6 km south of megaboulders) characterized by dominantly backset and topset bedding with scattered, thin, wispy fenestrae beds.

Presence of a few eolian structures (Engel et al., 2015) does not imply that the chevron ridges are parabolic dunes; it suggests the deposits were sub-aerially exposed and wind blew during periods of relative quiescence (as commonly observed on today's beaches after a storm). Unlike parabolic dunes, the chevron ridges are dominated by thick, low-angle (<10°) seaward-dipping, aggradational oolitic bedding (Hearty et al., 1998; Tormey, 1999; Fig. 24A-C). Foreset beds, diagnostic of migrating parabolic dunes, are rare or absent from many chevron ridges, supporting formation primarily by waves (Hearty et al., 2002). Furthering the distinction, fenestral porosity in low-angle bedding is prevalent throughout chevron ridges, occurring in repeated cycles of cm-thick beds that onlap the underlying strata, and often comprise meter-thick fenestrae-rich packages that can be followed in outcrop for tens of meters (Fig. 24A,B).

### 4.1.2.4 Summary of evidence

Alternative interpretations of the geologic data have been made (Bain and Kindler, 1994; Kindler and Strasser, 2000, 2002; Engel et al., 2015), specifically: 1) the boulders were thrown by a tsunami caused by flank margin collapse in North Eleuthera; 2) beach fenestrae in runup and chevron beds were caused by heavy rainfall; and 3) the chevron beach ridges are parabolic dunes. These views are challenged by Hearty et al. (2002) and again here for these reasons: 1) Extensive research in the Bahamas has revealed no geologic evidence of a point-source tsunami radiating from North Eleuthera. A slow speed margin failure is possible, without a tsunami, and indeed such a flank margin collapse could have been initiated by massive storm waves impacting an over-steepened margin. 2) If heavy rainfall was a significant process in the formation of fenestrae in dunes, they should commonly occur in all dunes of all ages, which is not the case. 3) Carbonate dunes, particularly oolitic ones, generally do not migrate unless exposed to extremely arid climates, which contradicts #2), and chevrons lack the most diagnostic feature of migration – foreset bedding.

In is too random and chronologically coincidental to argue that the trilogy of evidence – boulders, runup deposits, and chevron ridges – were caused by unconnected processes. If large, long-period waves lifted 1000-ton boulders onto and over the coastal ridge, as is generally agreed, the same waves must have also impacted large areas of the eastern Bahamas, for which there is abundant documentation. An attenuating gradient of wave energy, as from a tsunami generated by bank margin collapse, should yield predictable changes in bedforms and landforms with greater distance and elevation from the coastline, which are not observed. Absence of evidence for tsunamis on the United States East Coast refutes the possibility of a large remote tsunami source.

Our interpretation of these features is the most parsimonious, and we have argued that it is most consistent with the data. A common, synchronous, and non-random set of super-storm-related processes best explains boulder transport by waves, emplacement of runup deposits on older built up ridges, and the formation of complex chevron deposits over time across lower areas of the Bahamas. Indeed, given the geologic evidence of high seas and storminess from Bermuda and the Bahamas, Hearty and Neumann (2001) suggested "Steeper pressure, temperature, and moisture gradients adjacent to warm tropical waters could presumably spawn larger and more frequent cyclonic storms in the North Atlantic than those seen today."

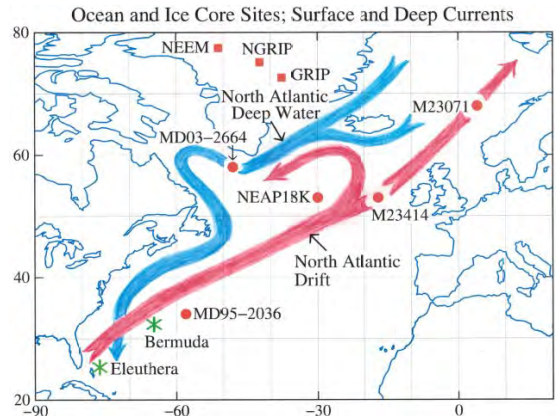

**Fig. 25.** Ocean and ice core sites and simplified sketch of upper ocean North Atlantic Current and North Atlantic Deep Water return flow. In interglacial periods the North Atlantic Current extends further north, allowing the Greenland-Iceland-Norwegian Sea to become an important source of deepwater formation.

We now seek evidence about end-Eemian climate change to help clarify how North Atlantic storms could have dispersed such strong long-period, well-organized waves to the southwest.

### 4.1.3 End-Eemian cold event: evidence from North Atlantic sediment cores

We present here evidence for rapid end-Eemian cooling in the North Atlantic at a time with the tropics warmer than today. The cooling marked initial descent from interglacial conditions toward global ice age conditions, occurring at ~118 ky b2k in ocean cores with uncertainty ~2 ky. It is identified by Chapman and Shackleton (1999) as cold event C26 in Greenland ice cores.

This section discusses ocean core data, but we first note the relation with ice core data and dating uncertainties. Ice cores have become of great value for climate studies, partly because the relative timing of events in ice cores at different locations can be determined very accurately via marker events such as volcanic eruptions and $CH_4$ fluctuations, even though the absolute dating error in ice cores is comparable to the dating uncertainty in ocean cores.

C26 is the cold phase of Dansgaard-Oeschger climate oscillation D-O 26 in the NGRIP (North Greenland Ice Core Project) ice core (NGRIP, 2004). C26 begins with a sharp cooling at 119.14 ky b2k on the GICC05modelext time scale (Rasmussen et al., 2014). The GICC05 time scale is based on annual layer counting in Greenland ice cores for the last 60 ky and an ice flow-model extension for earlier times. An alternative time scale is provided by Antarctic ice core chronology AICC2012 (Bazin et al., 2013; Veres et al., 2013) on which Greenland ice core records are synchronized via global markers, mainly oscillations of atmospheric $CH_4$ amount, which is globally well-mixed. C26 on Greenland is at 116.72 ky b2k on the AICC2012 time scale. Fig. S19 shows the difference between GICC05 and AICC2012 times scales versus time.

This age uncertainty for C26 is consistent with the ice core 2σ error estimate of 3.2 ky at Eemian time (Bazin et al., 2013). Despite this absolute age uncertainty, we can use Greenland data synchronized to the AICC2012 time scale to determine the relative timing of Greenland and Antarctic climate changes (Sec. 4.2.1) to an accuracy of a few decades (Bazin et al., 2013).

Sediment cores from multiple locations provide information not only on ocean temperature and circulation (Fig. 25), but also on ice sheet via information inferred from ice rafted debris. Comparison of data from different sites is affected by inaccuracy in absolute dating and use of different age models. Dating of sediments is usually based on tuning to the time scale of Earth orbital variations (Martinson et al., 1987) or "wiggle-matching" to another record (Sirocko et al., 2005), which limits accuracy to several ky. Temporal resolution is limited by bioturbation of sediments; thus resolution varies with core location and climate (Keigwin and Jones, 1994). For example, high deposition rates during ice ages at the Bermuda Rise yield a resolution of a few decades, but low sedimentation rates during the Eemian yield a resolution of a few centuries (Lehman et al., 2002). Lateral transport of sedimentary material prior to deposition complicates data interpretation and can introduce uncertainty, as argued specifically regarding data from the Bermuda Rise (Ohkouchi et al., 2002; Engelbrecht and Sachs, 2005).

Adkins et al. (1997) analyzed sediment core (MD95-2036, 34°N,58°W) from the Bermuda Rise using an age model based on Martinson et al. (1987) orbital tuning with the MIS stage 5/6 transition set at 131 ky b2k and the stage 5d/5e transition at 114 ky b2k. They found that oxygen isotope $\delta^{18}O$ of planktonic (near-surface dwelling) foraminifera and benthic (deep ocean) foraminifera both attain full interglacial values at ~128 ky b2k and remain nearly constant for ~10 ky (their Fig. 2). Adkins et al. (1997) infer that: "Late within isotope stage 5e (~118 ky b2k), there is a rapid shift in oceanic conditions in the western North Atlantic…" They find in the sediments at that point an abrupt increase of clays indicative of enhanced land-based glacier melt and an increase of high nutrient "southern source waters". The latter change implies a shutdown or diminution of NADW formation that allows Antarctic Bottom Water (AABW) to push into the deep North Atlantic Ocean (Duplessy et al., 1988; Govin et al., 2009). Adkins et al. (1997) continue: "The rapid deep and surface hydrographic changes found in this core mark the end of the peak interglacial and the beginning of climate deterioration towards the semi-glacial stage 5d. Before and immediately after this event, signaling the impeding end of stage 5e, deep-water chemistry is similar to modern NADW." This last sentence refers to a temporary rebound to near interglacial conditions. In Sec. 4.2.4 we use accurately synchronized Greenland and Antarctic ice cores, which also reveal this temporary end-Eemian climate rebound, to interpret the glacial inception and its relation to ice melt and late-Eemian sea level rise.

Ice rafted debris (IRD) found in ocean cores provides a useful climate diagnostic tool (Heinrich, 1988; Hemming, 2004). Massive ice rafting ("Heinrich") events are often associated with decreased NADW production and shutdown or slowdown of the Atlantic Meridional Overturning Circulation (AMOC) (Broecker, 2002; Barreiro et al., 2008; Srokosz et al., 2012). However, ice rafting occurs on a continuum of scales, and significant IRD is found in the cold phase of all the 24 Dansgaard-Oeschger (D-O) climate oscillations first identified in Greenland ice cores (Dansgaard et al., 1993). D-O events exhibit rapid warming on Greenland of at least several degrees within a few decades or less, followed by cooling over a longer period. Chapman and Shackleton (1999) found IRD events in the NEAP18K core for all D-O events (C19-C24) within the core interval that they studied, and they also labeled two additional events (C25 and C26). C26 did not produce identifiable IRD at the NEAP18K site, but it was added to the series because of its strong surface cooling.

Lehman et al. (2002) quantify the C26 cooling event using the same Bermuda Rise core (MD95-2036) and age model as Adkins et al. (1997). Based on the alkenone paleo-temperature technique (Sachs and Lehman, 1999), Lehman et al. (2002) find a sharp sea surface temperature (SST) decrease of ~3°C (their Fig. 1) at ~118 ky BP, coinciding with the end-Eemian shoulder of the benthic $\delta^{18}O$ plateau that defines stage 5e in the deep ocean. The SST partially recovered after several centuries, but C26 marked the start of a long slide into the depths of stage 5d cold, as ice sheets grew and sea level fell ~50 m in 10 ky (Lambeck and Chappell, 2001; Rohling et al., 2009). Lehman et al. (2002) wiggle-match the MD95-2036 and NEAP18K cores, finding a simple adjustment to the age model of Chapman and Shackleton (1999) that maximizes correlation of the benthic $\delta^{18}O$ records with the Adkins et al. (1997) $\delta^{18}O$ record. Specifically, they adjust the NEAP time scale by +4 ky before the MIS 5b $\delta^{18}O$ minimum and by +2 ky after it, which places C26 cooling at 118 ky b2k in both records. They give preference to the Adkins et al. (1997) age scale because it employs a $^{230}$Th-based time scale between 100 and 130 ky b2k.

We do not assert that the end-Eemian C-26 cooling was necessarily at 118 ky b2k, but we suggest that the strong rapid cooling observed in several sediment cores in this region of the subtropical and midlatitude North Atlantic Drift at about this time were all probably the same event. Such a large cooling lasting for centuries would not likely be confined to a small region. The dating models in several other studies place the date of the end-Eemian shoulder of the deep ocean $\delta^{18}O$ and an accompanying surface cooling event in the range 116-118 ky b2k.

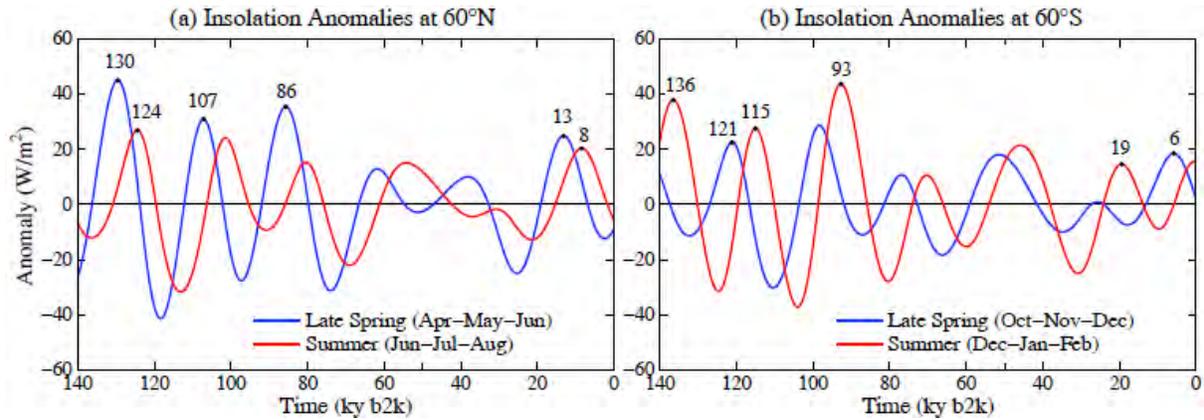

**Fig. 26.** Summer (Jun-Jul-Aug) and late spring (Apr-May-Jun) insolation anomalies (W/m²) at 60ºN and summer (Dec-Jan-Feb) and late spring (Oct-Nov-Dec) anomalies at 60ºS.

Kandiano et al. (2004) and Bauch and Kandiano (2007) analyze core M23414 (53°N,17°W), west of Ireland, finding a major SST end-Eemian cooling that they identify as C26 and place at 117 ky b2k. The 1 ky change in the timing of this event compared with Lehman et al. (2002), is due to a minor change in the age model, specifically, Bauch and Kandiano say: "The original age model of MD95-2036 (Lehman et al., 2002) has been adjusted to our core M23414 by alignment of the 4 per mil level in the benthic $\delta^{18}O$ records (at 130 ka in M23414) and the prominent C24 event in both cores." Bauch and Erlenkeuser (2008) and H. Bauch et al. (2012) examine ocean cores along the North Atlantic Current including its continuation into the Nordic seas. They find that in the Greenland-Iceland-Norwegian (GIN) Sea, unlike middle latitudes, the Eemian was warmest near the end of the interglacial period. The age model employed by Bauch and Erlenkeuser (2008) has the Eemian about 2 ky younger than the Adkins et al. (1997) age model, Bauch and Erlenkeuser (2008) having the benthic $\delta^{18}O$ plateau at ~ 116-124 ky BP (their Fig. 6). Rapid cooling they illustrate there at ~ 116.6 ky BP for core M23071 on the Voring Plateau (67°N,3°E) likely corresponds to the C26 end-Eemian cooling event.

Identification of end-Eemian cooling in ocean cores is hampered by the fact that Eemian North Atlantic climate was more variable than in the Holocene (Fronval and Jansen, 1996). There were at least three cooling events within the Eemian, each with minor increases in IRD, which are labeled C27, C27a and C27b by Oppo et al. (2006); see their Fig. 2 for core site ODP-980 in the eastern North Atlantic (55°N,15°W) near Ireland. High (sub-centennial) resolution cores in the Eirik drift region (MD03-2664, 57°N,49°W) near the southern tip of Greenland reveal an event with rapid cooling accompanied by reduction in NADW production (Irvali et al., 2012; Galaasen et al., 2014), which they place at ~117 ky b2k. However, their age scale has the benthic $\delta^{18}O$ shoulder at ~115 ky b2k (Fig. S1, Galaasen et al., 2014), so that event may have been C27b with C26 being stronger cooling that occurred thereafter.

### 4.1.4 Eemian timing consistency with insolation anomalies

Glacial-interglacial climate cycles are related to insolation change, as shown persuasively by Hays et al. (1976). Each "termination" (Broecker, 1984) of glacial conditions in the past several hundred thousand years coincided with a large positive warm-season insolation anomaly at the latitude of North American and Eurasian ice sheets (Raymo, 1997; Paillard, 2001). The explanation is that positive summer insolation anomalies (negative in winter) favor increased summer melting and reduced winter snowfall, thus shrinking ice sheets.

Termination timing is predicted better by high Northern Hemisphere late spring (April-May-June) insolation than by summer anomalies. For example, Raymo (1997) places Terminations I and II (preceding the Holocene and Eemian) midpoints at 13.5 and 128-131 ky b2k. Late spring insolation maxima are at 13.2 and 129.5 ky b2k (Fig. 26a). The AICC2012 ice core chronology (Bazin et al., 2013) places Termination II at 128.5 ky b2k, with 2σ uncertainty 3.2 ky. Late spring irradiance maximizes warm-season ice melt by producing the earliest feasible warm-season ice sheet darkening via snow melt and snow recrystallization (Hansen et al., 2007b).

Summer insolation anomalies are also shown in Fig. 26, because interglacial periods can be expected to continue as long as summer insolation is large enough to prevent ice sheet genesis. Summer insolation anomalies at 60°N became negative at ~118 ky b2k (Fig. 26a). Dating of insolation anomalies has high absolute accuracy, unlike ocean and ice core dating, as orbital anomalies are based on well-known planetary orbital mechanics (Berger, 1978).

Late Eemian sea level rise might appear to be a paradox, because glacial-interglacial sea level change is mainly a result of the growth and decay of Northern Hemisphere ice sheets. Northern warm-season insolation anomalies were declining rapidly in the latter part of the Eemian (Fig. 26a), so Northern Hemisphere ice should have been just beginning to grow. We suggest that the explanation for a late-Eemian sea level maximum is a late-Eemian collapse of Antarctic ice facilitated by the positive warm-season insolation anomaly on Antarctica and the Southern Ocean during the late Eemian (Fig. 26b) and possibly aided by an AMOC shutdown, which would increase warming of the Southern Ocean.

Persuasive evidence for this interpretation is provided by detailed paleoclimate data discussed in the next section, and is supported by modeling of relevant climate mechanisms. We will show that these mechanisms in turn help to explain ongoing climate change today, with implications for continuing climate change this century.

## 4.2 Millennial climate oscillations

Paleoclimate data are essential for understanding the major climate feedbacks. Processes of special importance are: (1) the role of the Southern Ocean in ventilating the deep ocean, affecting $CO_2$ control of global temperature, and (2) the role of subsurface ocean warming in ice shelf melt, affecting ice sheet disintegration and sea level rise. An understanding of time scales imparted by the ocean and the carbon cycle onto climate change is important, so that slow paleo ice sheet changes are not ascribed to ice physics, when the time scale is actually set elsewhere.

Major glacial-interglacial climate oscillations are spurred by periodic variation of seasonal and geographical insolation (Hays et al., 1976). Insolation anomalies are due to slow changes of the eccentricity of Earth's orbit, tilt of Earth's spin axis, and precession of the equinoxes, thus the day of year at which Earth is closest to the Sun, with dominant periodicities near 100,000, 40,000 and 20,000 years (Berger, 1978). These periods emerge in long climate records, yet a large fraction of climate variability at any site is stochastic (Wunsch, 2004; Lisiecki and Raymo, 2005). Such behavior is expected for a weakly-forced system characterized by amplifying feedbacks, complex dynamics, and multiple sources of inertia with a range of time scales.

Large glacial-interglacial climate change and stochastic variability are a result of two strong amplifying feedbacks, surface albedo and atmospheric $CO_2$. Orbit-induced insolation anomalies, per se, cause a direct climate forcing, i.e., an imposed Earth energy imbalance, only of order 0.1 W/m$^2$, but the persistent regional insolation anomalies spur changes of ice sheet size and GHGs. The albedo and GHG changes arise as slow climate feedbacks, but they are the forcings that maintain a quasi-equilibrium climate state nearly in global radiative balance. Glacial-interglacial albedo and greenhouse forcings are each ~3 W/m$^2$ (Fig. 27e,f)[15]. These forcings fully account for glacial-interglacial global temperature change with a climate sensitivity 0.5-1°C per W/m$^2$ (Hansen et al., 2008; Masson-Delmotte et al., 2010; Palaeosens, 2012).

The insolation anomaly peaking at 129.5 ky b2k (Fig. 27a) succeeded in removing ice sheets from North America and Eurasia and in driving atmospheric $CO_2$ up to ~285 ppm, as discussed below. However, smaller climate oscillations within the last glacial cycle are also instructive about ice feedbacks. Some of these oscillations are related to weak insolation anomalies and all are affected by predominately amplifying climate feedbacks.

Insolation anomalies peaking at 107 and 86 ky b2k (Fig. 27a) led to ~40 m sea level rises at rates exceeding 1 m/century (Stirling et al., 1998; Cutler et al., 2003) in early MIS 5c and 5a (Fig. 27f), but $CO_2$ did not rise above 250 ppm and interglacial status (with large ice sheets only on Greenland and Antarctica) was not achieved. $CO_2$ then continued on a 100 ky decline until ~18 ky b2k. Sea level continued its long decline, in concert with $CO_2$, reaching a minimum at least 120 m below today's sea level (Peltier and Fairbanks, 2006; Lambeck et al., 2014).

Progress achieved by the paleoclimate and oceanographic research communities allows interpretation of the role of the Southern Ocean in the tight relationship between $CO_2$ and temperature, as well as discussion of the role of subsurface ocean warming in sea level rise. Both topics are needed to interpret end-Eemian climate change and ongoing climate change.

### 4.2.1 Southern Ocean and atmospheric $CO_2$

There is ample evidence that reduced atmospheric $CO_2$ in glacial times, at least in substantial part, results from increased stratification of the Southern Ocean that reduces ventilation of the deep ocean (Toggweiler, 1999; Anderson et al., 2009; Skinner et al., 2010; Tschumi et al., 2011; Burke and Robinson, 2012; Schmitt et al., 2012; Marcott et al., 2014). Today the average "age" of deep water, i.e., the time since it left the ocean surface, is ~1000 years (DeVries and Primeau, 2011), but it was more than twice that old during the last glacial maximum (Skinner et al., 2010). The Southern Ocean dominates exchange between the deep ocean and atmosphere because ~80% of deep water resurfaces in the Southern Ocean (Lumpkin and Speer, 2007), as westerly circumpolar winds and surface flow draw up deep water (Talley, 2013).

Mechanisms causing more rapid deep ocean ventilation during interglacials include warmer Antarctic climate that increases heat flux into the ocean and buoyancy mixing that supports upwelling (Watson and Garabato, 2006), poleward shift of the westerlies (Toggweiler et al., 2006), and reduced sea ice (Keeling and Stephens, 2001). Fischer et al. (2010) question whether the latitudinal shift of westerlies is an important contributor; however, the basic point is the empirical fact that a warmer interglacial Southern Ocean produces faster ventilation of the deep ocean via a combination of mechanisms.

Although a complete quantitative understanding is lacking for mechanisms to produce the large glacial-interglacial swings of atmospheric $CO_2$, we can safely assume that deep ocean ventilation acts to oppose the sequestration of carbon in the ocean by the "pumps" that move carbon from the surface to ocean depths, and changes in the ventilation rate have a significant effect on atmospheric $CO_2$. Ridgwell and Arndt (2015) describe several conceptual pumps: (1) the organic carbon pump, in which the sinking material also controls nutrient cycling by the ocean, (2) the carbonate pump, with biological precipitation of mainly calcium carbonate, a fraction of which escapes dissolution to form a new geological carbon reservoir, (3) the simple solubility pump, as $CO_2$ is more soluble in the cold polar waters where deep water forms, and (4) a microbial carbon pump that seems capable of altering deep ocean dissolved organic carbon.

---

[15] Other parts of Fig. 27 are discussed later, but they are most informative if aligned together. In interpreting Fig. 27, note that long-lived greenhouse gas amounts in ice cores have global relevance, but ice core temperatures are local to Greenland and Antarctica. Also, because our analysis does not depend on absolute temperature, we do not need to convert the temperature proxy, $\delta^{18}O$, into an estimated absolute temperature. We include $CH_4$ and $N_2O$ in the total GHG climate forcing, but we do not discuss the reasons for $CH_4$ and $N_2O$ variability (see Schilt et al., 2010), because $CO_2$ provides ~80% of the GHG forcing.

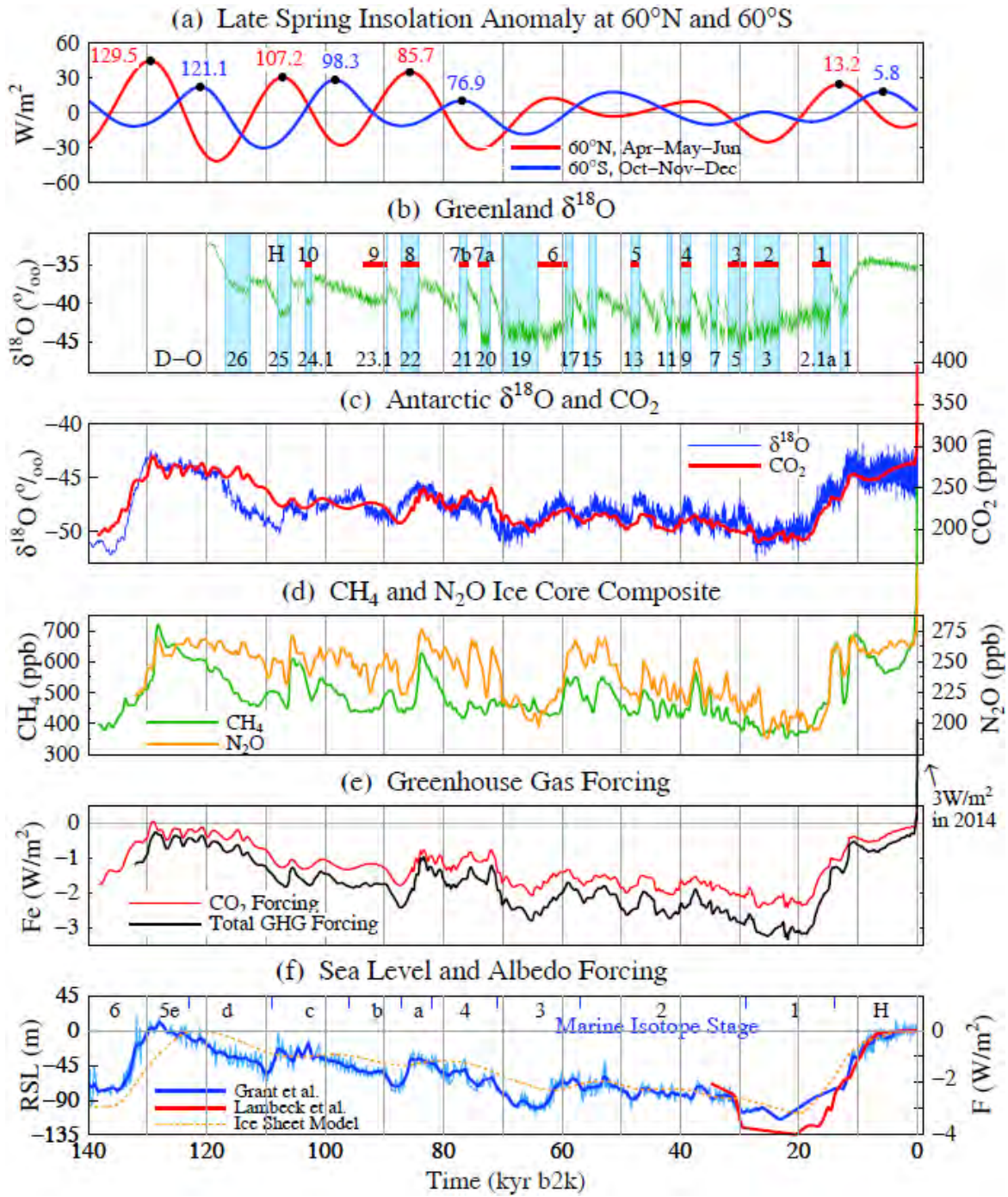

**Fig. 27.** (a) Late spring insolation anomalies relative to the mean for the past million years, (b) $\delta^{18}O_{ice}$ of composite Greenland ice cores (Rassmussen et al., 2014) with Heinrich events of Guillevic et al. (2014), (c, d) $\delta^{18}O_{ice}$ of EDML Antarctic ice core (Ruth et al., 2007), multi-ice core $CO_2$, $CH_4$, and $N_2O$ based on spline fit with 1000-year cut-off (Schilt et al., 2010), scales are such that $CO_2$ and $\delta^{18}O$ means coincide and standard deviations have the same magnitude, (e) GHG forcings from equations in Table 1 of Hansen et al. (2000), but with the $CO_2$, $CH_4$, and $N_2O$ forcings multiplied by factors 1.024, 1.60, and 1.074, respectively, to account for each forcing's "efficacy" (Hansen et al., 2005a), with $CH_4$ including factor 1.4 to account for indirect effect on ozone and stratospheric water vapor, (f) sea level data from Grant et al. (2012) and Lambeck et al. (2014) and ice sheet model results from de Boer et al. (2010). Marine isotope stage boundaries from Lisiecki and Raymo (2005). (b-e) are on AICC2012 time scale (Bazin et al., 2013),

No doubt the terrestrial biosphere also contributes to glacial-interglacial atmospheric $CO_2$ change (Archer et al., 2000; Sigman and Boyle, 2000; Kohler et al., 2005; Menviel et al., 2012; Fischer et al., 2015). Also the efficacy of the ocean pumps depends on terrestrial conditions, e.g., dust-borne iron fertilization of the biological pump (Martin and Fitzwater, 1988) contributes to millennial and full glacial $CO_2$ drawdown (Martinez-Garcia et al., 2014). And the Southern Ocean is not the only conduit to the deep ocean, e.g., AMOC changes are associated with at least two rapid $CO_2$ increases of about 10 ppm, as revealed by a high resolution West Antarctic ice core (Marcott et al., 2014). Nevertheless, it is reasonable to hypothesize that sequestration of $CO_2$ in the glacial ocean is the largest cause of glacial-interglacial $CO_2$ change, and it is known that ocean ventilation occurs mainly via the Southern Ocean.

Southern Ocean ventilation, as the dominant cause of atmospheric $CO_2$ change, helps explain temperature-$CO_2$ leads and lags. Temperature and $CO_2$ rises are almost congruent at ice age terminations (Masson-Delmotte et al., 2010; Pedro et al., 2012; Parrenin et al., 2013). Southern Ocean temperature is expected to lead, spurring deep ocean ventilation and atmospheric $CO_2$ increase, with global temperature following. Termination I is dated best and Shakun et al. (2012) have reconstructed global temperature then, finding evidence for this expected order of events.

Correlation of $\delta^{18}O$ and $CO_2$ over the past 140 ky (Fig. 27c) is 84.4% with $CO_2$ lagging by 760 years. For the period 100-20 ky b2k, which excludes the two terminations, the correlation is 77.5% with $CO_2$ lagging by 1040 years. Briefer lag for the longer period and longer lag during glacial inception are consistent with the rapid deep ocean ventilation that occurs at terminations.

### 4.2.2 $CO_2$ as climate control knob

$CO_2$ is the principal determinant of Earth's climate state, the radiative "control knob" that sets global mean temperature (Lacis et al., 2010, 2013). Degree of control is shown by comparison of $CO_2$ amount with Antarctic temperature for the past 800,000 years (Fig. 28a). Control should be even tighter for global temperature than for Antarctic temperature, because of regional anomalies such as Antarctic temperature overshoot at terminations (Masson-Delmotte et al., 2006, 2010), but global data are not available.[16]

The $CO_2$ dial must be turned to ~260 ppm to achieve a Holocene-level interglacial. $CO_2 \sim$ 250 ppm was sufficient for quasi-interglacials in the period 800-450 ky b2k, with sea level 10-25 m lower than in the Holocene (Fig. S18 of Hansen et al., 2008). Interglacials with $CO_2 \sim$ 280 ppm, i.e., the Eemian and Holsteinian (~400 ky b2k), were warmer than the Holocene and had sea level at least several meters higher than today.

$CO_2$ and albedo change are closely congruent over the last 800,000 years (Fig. S18 of Hansen et al., 2008). GHG and albedo forcings, which are both amplifying feedbacks that boost each other, are each of amplitude ~3 W/m². So why do we say that $CO_2$ is the control knob?

First, $CO_2$, in addition to being a slow climate feedback, changes independently of climate. Natural $CO_2$ change includes an increase to ~1000 ppm about 50 million years ago (Zachos et al., 2001) as a result of plate tectonics, specifically volcanic emissions associated with movement of the Indian plate across the Tethys Ocean and collision with Asia (Kent and Muttoni, 2008). Humankind, mainly by burning fossil fuels, also moves the $CO_2$ control knob.

---

[16] The tight fit of $CO_2$ and Antarctic temperature (Fig. 28a) implies an equilibrium Antarctic sensitivity 20°C for $2 \times CO_2$ (4 W/m²) forcing (200 $\rightarrow$ 300 ppm forcing is ~2.3 W/m², Table 1 of Hansen et al., 2000), thus 10°C global climate sensitivity (Antarctic temperature change is ~ twice global change) with $CO_2$ taken as the ultimate control knob, i.e., if snow/ice area and other GHGs are taken to be slaves to $CO_2$-driven climate change. This implies a conventional climate sensitivity of 4°C for $2 \times CO_2$, as GHG and albedo forcings are similar for glacial-to-interglacial climate change and non-$CO_2$ GHGs account for ~20% of the GHG forcing. The inferred sensitivity is reduced to 2.5-3°C for $2 \times CO_2$ if, as some studies suggest, global mean glacial-interglacial temperature change is only about one-third of the Antarctic temperature change (Palaeosens, 2012; Hansen et al., 2013b).

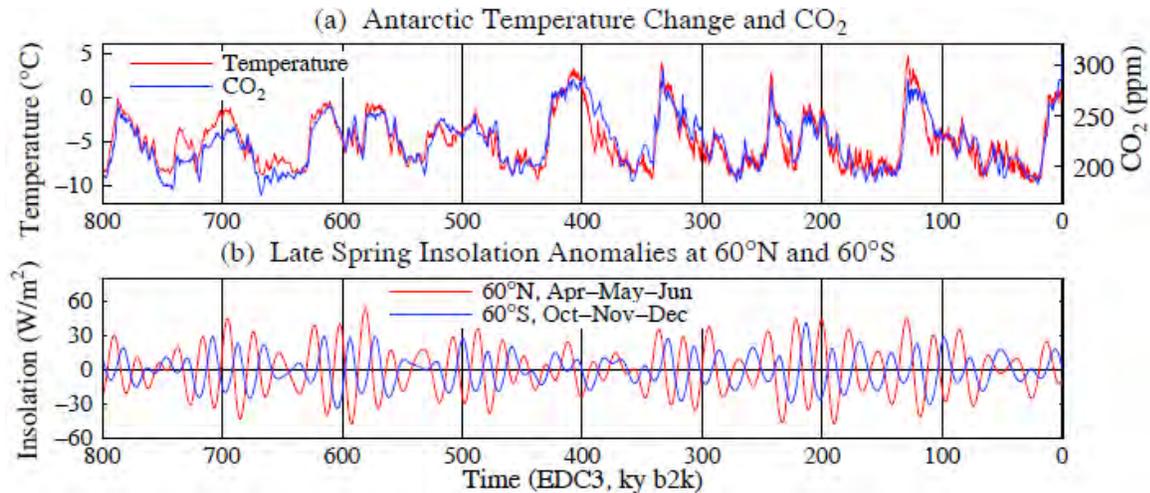

**Fig. 28.** (a) Antarctic (Dome C) temperature relative to last 10 ky (Jouzel et al., 2007) on AICC2012 time scale and $CO_2$ amount (Luthi et al., 2008). Temperature scale is such that standard deviation of T and $CO_2$ are equal, yielding $\Delta T$ (°C) = 0.114 $\Delta CO_2$ (ppm), (b) Late Spring insolation anomalies at 60°N and 60°S.

Second, $CO_2$ is more recalcitrant than snow and ice, i.e., its response time is longer. $CO_2$ inserted into the climate system, by humans or plate tectonics, remains in the climate system of order 100,000 years before full removal by weathering (Archer, 2005). Even $CO_2$ exchange between the atmosphere (where it affects climate) and ocean has a lag of order a millennium (Fig. 27). In contrast, correlations of paleo temperatures and sea level show that lag of sea level change behind temperature is of order a century, not a millennium (Grant et al., 2012).

We suggest that limitations on the speed of ice volume (thus sea level) changes in the paleo record are more a consequence of the pace of orbital changes and $CO_2$ changes, as opposed to being a result of lethargic ice physics. "Fast" changes of $CO_2$ have been identified, e.g., an increase of ~10 ppm in about a century at ~39.6 ky b2k (Ahn et al., 2012) and three increases of 10-15 ppm each within 1-2 centuries during the deglaciation following the last ice age (Marcott et al., 2014), but the magnitude of these $CO_2$ increases is not sufficient to provide a good empirical test of ice sheet sensitivity to the $CO_2$ forcing.

Supremacy of SMOC, the Southern Ocean meridional overturning circulation, in affecting the $CO_2$ control knob and thus glacial-interglacial change is contrary to the idea that the AMOC is a prime driver that flips global climate between quasi-stable glacial and interglacial states, yet AMOC retains a significant role. AMOC can affect $CO_2$ via the volume and residence time of NADW, but its largest effect is probably via its impact on the Southern Ocean. When AMOC is not shut down it cools the Southern Hemisphere, transferring heat from the Southern to the Northern Hemisphere at a rate ~1 petawatt, which is ~4 W/m² averaged over a hemisphere (Crowley, 1992). However, the Southern Ocean slowly warms when AMOC shuts (or slows) down; the response time is of the order of 1000 years because of the Southern Ocean's large thermal inertia (Stocker and Johnson, 2003). These mechanisms largely account for the nature of the "bipolar seesaw" (Broecker, 1998; Stocker, 1998; Stenni et al., 2011; Landais et al., 2015), including the lag between AMOC slowdown and Antarctic warming.

### 4.2.3 Dansgaard-Oeschger events and subsurface ocean warming

The magnitude and rapidity of Greenland climate change during Dansgaard-Oeschger events would deter prediction of human-made climate effects, if D-O events remained a mystery. Instead, however, enough is now understood about D-O events that they provide insight related to the vulnerability of ice shelves and ice sheets, including the role of subsurface ocean warming.

Broecker (2000) inferred from the rapidity of D-O warmings that a reduction of sea ice cover was probably involved. Li et al. (2005, 2010) modeling showed that removal of Nordic Sea ice cover is needed to yield the magnitude of observed Greenland warming. The spatial gradient of D-O warming, with smaller warming in northwest Greenland, agrees with that picture (Guillevic et al., 2013; Buizert et al., 2014). Such sea ice change is consistent with changes in deuterium excess in Greenland ice cores at D-O transitions, which indicate shifts of Greenland moisture source regions (Masson-Delmotte et al., 2005; Jouzel et al., 2007).

Fluckiger et al. (2006), Alvarez-Solas et al. (2010, 2011, 2013) and Marcott et al. (2011) noted modern and paleo data that point to ocean-ice shelf interaction as key to the ice discharge of accompanying Heinrich events, and they used a range of models to support this interpretation and overturn earlier suggestions of a central role for ice sheets via binge-purge oscillations (MacAyeal, 1993) or outburst flooding from subglacial reservoirs (Alley et al., 2006). Shaffer et al. (2004) and Petersen et al. (2013) conclude that subsurface ocean warming in the North Atlantic takes place during the stadial (cold) phase of all D-O events, and eventually this subsurface warming leads to ice shelf collapse or retreat, ice rafting, sea level rise, and sea ice changes. Rasmussen et al. (2003) examined ocean cores from the southeast Labrador Sea and found that for all 11 Heinrich events "…the icy surface water was overlying a relatively warm, poorly ventilated and nutrient rich intermediate water mass to a water depth of at least 1251 m." Collapse of a Greenland ice shelf fronting the Jakoshavn ice stream during the Younger Dryas cold event has been documented (Rinterknecht et al., 2014), apparently due to subsurface warming beneath the ice shelf leading to rapid discharge of icebergs.

Some D-O details are uncertain, e.g., the relation between changing sea ice cover and changing location of deep water formation (Rahmstorf, 1994) and whether an ice shelf between Greenland and Iceland contributed to the sea ice variability (Petersen et al., 2013). However, ocean-ice interactions emerge as key mechanisms, spurred by subsurface ocean warming, as ocean stratification slows but does not stop northward heat transport by AMOC.

We consider a specific D-O event for the sake of discussing mechanisms. D-O 22 cold phase, labeled C22 in ocean cores and coinciding with Heinrich H8 (Fig. 27), occurred as Northern Hemisphere insolation was rising (Fig. 27a). The North Atlantic surface was cooled by rapid ice discharge; sea level rose more than 40 m, a rate exceeding 1.6 m per century (Cutler et al., 2003). Ice discharge kept the North Atlantic highly stratified, slowing AMOC. Antarctic warming from a slowed AMOC increases almost linearly with the length of the D-O cold phase (Fig. 3, EPICA Community Members, 2006; Fig. 6, Capron et al., 2010) because of the Southern Ocean's large heat capacity (Stocker and Johnson, 2003). Antarctic warming, aided by the 2500-year D-O 22 event, spurred SMOC enough to raise atmospheric $CO_2$ 40 ppm (Fig. 27c).

As the Antarctic warmed, ocean heat transport to the North Atlantic would have increased, with most heat carried at depths below the surface layer. When the North Atlantic became warm enough at depth, stratification of cold fresh surface water eventually could not be maintained. The warming breakthrough may have included change of NADW formation location (Rahmstorf, 1994) or just large movement of the polar front. Surface warming east of Greenland removed most sea ice and Greenland warmed ~10°C (Capron et al., 2010). As the warm phase of D-O 21 began, AMOC was pumping heat from the Antarctic into the Nordic seas and Earth must have been slightly out of energy balance, cooling to space, so both Antarctica and Greenland slowly cooled. Once the North Atlantic had cooled enough, sea ice formed east of Greenland again, ice sheets and ice shelves grew, sea level fell, and the polar front moved southward.

Sea level rise associated with D-O events covers a wide range. Sea level increases as large as ~40 m were associated with large insolation forcings at 107 and 86 ky b2k (Fig. 27). However, rapid sea level change occurred even when forcing was weak. Roche et al. (2004) conclude from analyses of $\delta^{18}O$ that H4, at a time of little insolation forcing (~ 40 ky b2k, Fig. 27), produced 1.9 ±1.1 m sea level rise over 250 ±150 years. Sea level rise as great as 10-15 m occurred in conjunction with some other D-O events during 65-30 ky b2k (Lambeck and Chappell, 2001; Yokoyama et al., 2001; Chappell, 2002).

Questions about possible D-O periodicity and external forcing were raised by a seeming 1470-year periodicity (Schulz, 2002). However, improved dating indicates that such periodicity is an artifact of ice core chronologies and not statistically significant (Ditlevsen et al., 2007) and inspection of Fig. 27b reveals a broad range of time scales. Instead, the data imply a climate system that responds sensitively to even weak forcings and stochastic variability, both of which can spur amplifying feedbacks with a range of characteristic response times.

Two conclusions are especially germane. First, subsurface ocean warming is an effective mechanism for destabilizing ice shelves and thus the ice sheets buttressed by the ice shelves. Second, large rapid sea level rise can occur as a result of melting ice shelves.

However, ice shelves probably were more extensive during glacial times. So are today's ice sheets much more stable? The need to understand ice sheet vulnerability focuses attention on end-Eemian events, when ice sheets were comparable in size to today's ice sheets.

### 4.2.4 End-Eemian climate and sea level change

Termination II, ushering in the Eemian, was spurred by a late spring 60°N insolation anomaly peaking at +45 W/m$^2$ at 129.5 ky b2k (Fig. 27a), the largest anomaly in at least the past 425 ky (Fig. 3, Hansen et al., 2007b). $CO_2$ and albedo forcings were mutually reinforcing. $CO_2$ began to rise before Antarctic $\delta^{18}O$, as deglaciation and warming began in the Northern Hemisphere. Most of the total $CO_2$ rise was presumably from deep ocean ventilation in the Southern Ocean, aided by meltwater that slowed the AMOC and thus helped to warm the Southern Ocean.

The northern late spring insolation anomaly fell rapidly, becoming negative at 123.8 ky b2k (Fig. 27a), by which time summer insolation also began to fall (Fig. 26). Northern Hemisphere ice sheets must have increased intermittently while Southern Hemisphere ice was still declining, consistent with minor, growing ice rafting events C27, C27a, C27b and C26 and a sea level minimum during 125-121 ky b2k (Sec. 4.1.1). High Eemian climate variability in the Antarctic (Pol et al., 2014) was likely a result of the see-saw relation with North Atlantic events.

$CO_2$ (Fig. 27c) remained at ~270 ppm for almost 15 ky as the positive insolation anomaly on the Southern Ocean (Fig. 27a) kept the deep ocean ventilated. Sea level in the Red Sea analysis (Grant et al., 2012) shown in Fig. 27f seems to be in decline through the Eemian, but that must be a combination of dating and sea level error, as numerous sea level analyses cited in Sec. 4.1.1, our Supplement, and others (e.g., Chen et al., 1991; Stirling et al., 1998; Cutler et al., 2003), indicate high sea level throughout the Eemian and allow a possible late-Eemian maximum. Chen et al. (1991), using a U-series dating with 2σ uncertainty ±1.5 ky, found that the Eemian sea level high stand began between 132 and 129 ky b2k, lasted for 12 ky, and was followed by rapid sea level fall.

We assume that C26, the sharp cooling at 116.72 ky b2k in the NGRIP ice on the AICC2012 time scale, marks the end of fully interglacial Eemian conditions, described as 5e *sensu stricto* by Bauch and Erlenkeuser (2008). $\delta^{18}O$ in Antarctica was approaching a relative minimum (−46.7 per mil at EDML, see Fig. S20 for detail) and $CO_2$ was slowly declining at 263 ppm. In the next 300 years $\delta^{18}O$ increased to −45.2 and $CO_2$ increased by 13 ppm with lag ~1500 years, which we interpret as see-saw warming of the Southern Ocean in response to the C26-induced AMOC slowdown and resulting increased SMOC ventilation of $CO_2$.

Freshwater causing the C26 AMOC shutdown could not have been Greenland surface melt. Greenland was already 2000 years into a long cooling trend and the northern warm season insolation anomaly was in the deepest minimum of the last 150 ky (Fig. 26a). Instead C26 was one event in a series, preceded by C27b and followed by C25, each a result of subsurface North Atlantic warming that melted ice shelves, causing ice sheets to discharge ice. Chapman and Shackleton (1999) did not find IRD from C26 in the mid-Atlantic, but Carlson et al. (2008) found a sharp increase in sediments near the southern tip of Greenland that they identified with C26.

We suggest that the Southern Hemisphere was the source for brief late-Eemian sea level rise. The positive warm-season insolation anomaly on the Southern Ocean and AMOC slowdown due to C26 added to Southern Ocean heat, causing ice shelf melt, ice sheet discharge, and sea level rise. Rapid Antarctica ice loss would cool the Southern Ocean and increase sea ice cover, which may have left telltale evidence in ice cores. Indeed, Masson-Delmotte et al. (2011) suggest that abrupt changes of $\delta^{18}O$ in the EDML and TALDICE ice cores (those most proximal to the coast) indicate a change in moisture origin, likely due to increased sea ice. Further analysis of Antarctic data for the late Eemian might help pinpoint the melting and help assess vulnerability of Antarctic ice sheets to ocean warming, but this likely will require higher resolution models with more realistic sea ice distribution and seasonal change than our present model produces.

Terrestrial records in Northern Europe reveal rapid end-Eemian cooling. Sirocko et al. (2005) find cooling of 3°C in summer and 5-10°C in winter in southern Germany, annual layering in a dry Eifel maar lake revealing a 468 year period of aridity, dust storms, bushfires, and a decline of thermophilous trees. Similar cooling is found at other German sites and La Grande Pile in France (Kuhl and Litt, 2003). Authors in both cases interpret the changes as due to a southward shift of the polar front in the North Atlantic corresponding to C26. Cooling of this magnitude in northern Europe and increased aridity are found by Brayshaw et al. (2009) and Jackson et al. (2015) in simulations with high resolution climate models forced by AMOC shutdown.

While reiterating dating uncertainties, we note that the cool period with reduced NADW formation identified in recent high resolution ocean core studies for Eirik Drift site MD03-2664 (Fig. 25) near Greenland (Irvali et al., 2012; Galaasen et al., 2014) at ~117 ky b2k has length similar to the 468-year cold stormy period found in a German lake core (Sirocko et al., 2005). The Eirik core data show a brief return to near-Eemian conditions and then a slow decline, similar to the oscillation in the NGRIP ice core at 116.72 ky b2k on the AICC2012 time scale.

The principal site of NADW formation may have moved from the GIN Seas to just south of Greenland at end of the Eemian. Southward shift of NADW formation and the polar front is consistent with the sudden, large end-Eemian cooling in the North Atlantic and northern Europe, while cooling in Southern European was delayed by a few millennia (Brauer et al., 2007). Thus end-Eemian mid-latitude climate was characterized by an increased meridional temperature gradient, an important ingredient for strengthening storms.

## 5  Modern data

Observations help check our underlying assumption of nonlinear meltwater growth and basic simulated climate effects. As these data are updated, and as more extensive observations of the ocean and ice processes are obtained, a clearer picture should emerge over the next several years.

### 5.1  Ice sheet mass loss and sea level rise

The fundamental question we raise is whether ice sheet melt in response to rapid global warming will be nonlinear and better characterized by a doubling time for its rate of change or whether more linear processes dominate. Hansen (2005, 2007) argued on heuristic grounds that ice sheet disintegration is likely to be nonlinear if climate forcings continue to grow, and that sea level rise of several meters is possible on a time scale of the order of a century. Given current ice sheet melt rates, a 20-year doubling rate produces multi-meter sea level rise in a century, while 10-year and 40-year doubling times require about 50 years and 200 years, respectively.

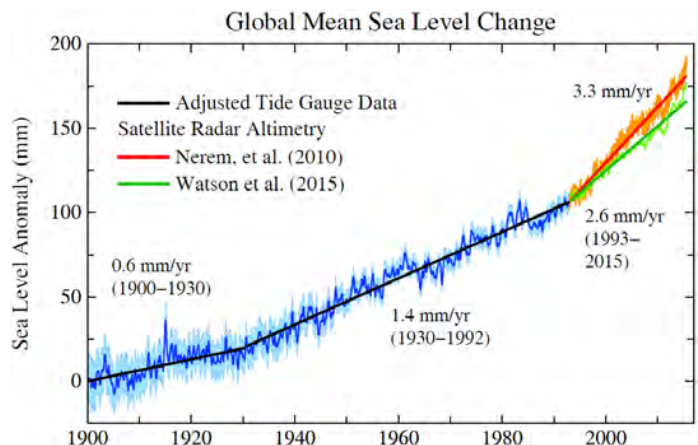

**Fig. 29.** Estimated sea level change (mm) since 1900. Data through 1992 are the tide gauge record of Church and White (2011) with the change rate multiplied by 0.78, so as to yield a mean 1901-1990 change rate 1.2 mm/year (Hay et al., 2015). The two estimates for the satellite era (1993-2015) are from Nerem et al. (2010, updated at http://sealevel.colorado.edu) and Watson et al. (2015).

Church et al. (2013) increased estimates of sea level rise compared to prior IPCC reports, but scenarios they discuss are close to linear responses to the assumed rising climate forcing. The most extreme climate forcing (RCP8.5, 936 ppm $CO_2$ in 2100 and GHG forcing 8.5 W/m²) is estimated to produce 0.74 m sea level rise in 2100 relative to the 1986-2005 mean sea level, with the "likely" range of uncertainty 0.52-0.98 m. Church et al. (2013) also discuss semi-empirical estimates of sea level rise, which yield ~0.7-1.5 m for the RCP8.5 scenario, but express low confidence in the latter, thus giving preference to the model-based estimate of 0.52-0.98 m. We note that Sec. 4.4.4.2 on ice sheet processes in the IPCC chapter on cryosphere observations (Vaughan et al., 2013) contains valuable discussion of nonlinear ice sheet processes that could accelerate ice sheet mass loss, but which are not fully included in current ice sheet models.

Empirical analyses are needed if we doubt the realism of ice sheet models, but semi-empirical analyses lumping multiple processes together may yield a result that is too linear. Sea level rises as a warming ocean expands, as water storage on continents changes (e.g., in aquifers and behind dams), and as glaciers, small ice caps, and the Greenland and Antarctic ice sheets melt. We must isolate the ice sheet contribution, because only the ice sheets threaten multi-meter sea level rise.

Hay et al. (2015) reanalyzed tide-gauge data for 1901-1990 including isostatic adjustment at each station, finding global sea level rise 1.2 ± 0.2 mm/year. Prior tide gauge analyses of 1.6-1.9 mm/year were inconsistent with estimates for each process, which did not add up to such a large value (Church et al., 2013). This estimate of 1.2 ± 0.2 mm/year for 1900-1990 compares with estimated sea level rise of ~0.2 m in the prior two millennia or ~0.1 mm/year (Kemp et al., 2011) and several estimates of ~3 mm/year for the satellite era (1993-present). Nerem et al. (2010) find sea level increase of 3.3 mm/yr in the satellite era, while Watson et al. (2015), based in part on calibration to tide gauge data, suggest alternative rates of 2.9 mm/yr or 2.6 mm/yr.

Accepting the analyses of Hay et al. (2015) for 1901-1990 and estimates of 2.5-3.5 mm/year for the satellite era leads to a picture of a rising sea level rate (Fig. 29) that differs from the perception of near-linear sea level rise created by Fig. 13.3 in the IPCC report (Church et al., 2013). We do not argue for the details in Fig. 29 or suggest any change-points for the rate of sea level rise, but the data do reveal a substantial increase in the rate of sea level rise.

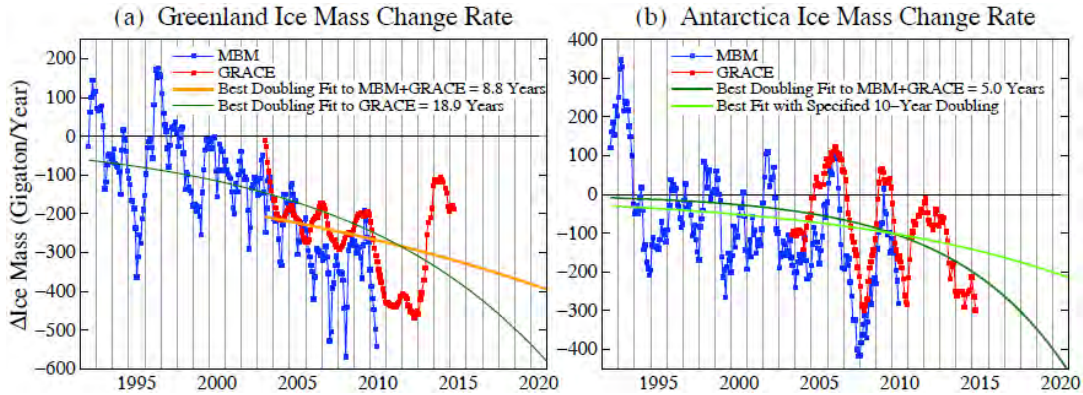

**Fig. 30.** Greenland and Antarctic ice mass change. GRACE data is extension of Velicogna et al. (2014) gravity data. MBM (mass budget method) data are from Rignot et al. (2011). Red curves are gravity data for Greenland and Antarctica only; small Arctic ice caps and ice shelf melt add to freshwater input.

The majority of sea level rise in the 20[th] century was from the several processes other than Greenland and Antarctica mass loss (Church et al., 2013), so the time scale for ice sheet mass loss may differ from the time scale for past sea level change. A direct measure of ice sheet mass loss is obtained from satellite gravity measurements by Velicogna et al. (2014), who find Greenland's mass loss in 2003-2013 of $280 \pm 58$ Gt/yr[17] accelerating by $25.4 \pm 1.2$ Gt/yr$^2$, and Antarctic mass loss $67 \pm 44$ Gt/year accelerating by $11 \pm 4$ Gt/year/year. The Velicogna et al. (2014) data are updated in Fig. 30. The reduced mass loss rate of Greenland in 2013-14 makes it difficult to infer an empirical growth rate for mass loss, as discussed below.

Reliability of mass loss inferred from gravity data is supported by comparison to surface mass balance studies (Fig. 30). Mass loss accelerations over 1992-2011 obtained via the mass budget method (Rignot et al., 2011) for Greenland ($21.9 \pm 1$ Gt/year/year) and Antarctica ($14.5 \pm 2$ Gt/year/year) are similar or larger than results from gravity analysis. A third approach, based on satellite radar altimetry, is consistent with the other two for mass loss from Greenland and West Antarctica (Shepherd et al., 2012), including the Amundsen Sea sector, which is the dominant contributor to Antarctic ice mass loss (Sutterley et al., 2014). Differences among techniques exist in East Antarctica, but mass changes there are small (Shepherd et al., 2012).

Best fit exponential doubling times for Greenland are 4.8 years (MBM 1992-2010 data), 18.9 years (GRACE 2003-2015 data) and 8.8 years (MBM + GRACE 1992-2015 data); in the latter case only GRACE data are used after 2002. The best fit to MBM + GRACE is shown in Fig. 30. The equivalent results for Antarctica are 5.3 years, 3.2 years and 5.0 years. Clearly the data records are too short to infer a doubling rate, let alone to confirm that mass loss is exponential. Thus we also show a 10-year doubling growth curve for Antarctic ice mass loss (Fig. 30b). The recent reduction of mass loss from Greenland illustrates how sensitive the empirical result is to record brevity, but the curves should become more informative over the next several years.

Additional insight is provided by the regional breakdown of the mass change data as achieved in the Velicogna et al. (2014) analysis. The regional data suggest that the Antarctic situation may be more threatening than indicated by the continental mass loss rate. This net mass loss combines mass loss via ice streams with regions of net snow accumulation. Queen Maud Land, e.g., is gaining $63 \pm 6$ Gt/year, accelerating $15 \pm 1$ Gt/year/year, but this mass gain may be temporary. Our simulations with increasing freshwater input indicate that circum-Antarctic cooling and sea ice increase eventually may limit precipitation reaching the continent, and recent SST and sea ice data have a tendency consistent with that expectation (Sec. 5.2).

---

[17] For comparison, our assumed freshwater injection of 360 Gt/year in 2011 with 10 year doubling yields an average mass loss 292 Gt/year for 2003-2013. Further, Velicogna et al. (2014) find an ice mass loss of $74 \pm 7$ Gt/yr from nearby Canadian glaciers and ice caps with acceleration $10 \pm 2$ Gt/yr$^2$, and there is an unknown freshwater input from melting ice shelves. Thus our assumed Northern Hemisphere meltwater was conservative.

Amundsen Sea glaciers are a gateway to West Antarctic ice, which has potential for several meters of sea level. Mass loss of the Amundsen Sea sector was $116 \pm 6$ Gt/year in 2003-2013, growing $13 \pm 2$ Gt/year/year (Velicogna et al., 2014; Rignot et al., 2014; Sutterley et al., 2014).

Totten Glacier in East Antarctica fronts the Aurora Subglacial Basin, which has the potential for ~6.7 m of sea level (Greenbaum et al., 2015). Williams et al. (2011) find that warm modified Circumpolar Deep Water is penetrating the continental shelf near Totten beneath colder surface layers. Details of how warmer water reaches the ice shelf are uncertain (Khazendar et al., 2013), but, as in West Antarctica, the inland sloping trough connecting the ocean with the main ice shelf cavity (Greenbaum et al., 2015) makes Totten Glacier susceptible to unstable retreat (Goldberg et al., 2009). Cook Glacier, further east in East Antarctica, also rests on a submarine inland-sloping bed and fronts ice equivalent to 3-4 meters of sea level. The Velicogna et al. (2014) analysis of gravity data for 2003-2013 finds the Totten sector of East Antarctica losing $17 \pm 4$ Gt/year, with the loss accelerating by $4 \pm 1$ Gt/year/year, and the Victoria/Wilkes sector including Cook Glacier losing $16 \pm 5$ Gt/year, with a small deceleration ($2 \pm 1$ Gt/year/year).

Greenland ice melt is subject to multiple feedbacks, some of which are largely absent on Antarctica, so it is not certain whether Greenland ice is less or more vulnerable than Antarctic ice. On the one hand, some differences make the Greenland ice sheet seem less vulnerable. Greenland does not have as much unstable ice volume sitting behind retrograde beds. Also, although surface cooling due to freshwater injection leads to subsurface ocean warming (Fig. 12), freshwater injection may also reduce poleward transport of heat by the Atlantic Ocean if the AMOC slows down, and North Atlantic cooling may affect summer surface melt on Greenland.

On the other hand, the Greenland ice sheet is subject to forcings and feedbacks that are less important on Antarctica. Greenland experiences extensive summer surface melt (Tedesco et al., 2011; Box et al., 2012), which makes surface albedo changes more important on Greenland. Greenland mass loss is thus more affected by snow darkening via dust, black carbon and biological substances including algae (Benning et al., 2014; Yasunari et al., 2015), which are in part an imposed climate forcing but in some cases also a substantial amplifying feedback. Soot from forest fires occurs naturally, but the magnitude of fire events is increasing Flannigan et al., 2013; Jolly et al., 2014) and may have contributed to widespread Greenland melt events in recent years (Keegan et al., 2014). Pigmented algae can substantially reduce spring and summer ice albedo and may be an important feedback in a warming world (Benning et al., 2014). Other amplifying feedbacks for Greenland include the ice surface elevation feedback, cryo-hydrologic warming in which percolating water alters thermal regime and weakens the ice sheet on decadal time scales (Colgan et al., 2015) and ocean-mediated melting of ice shelves and glacier fronts (Rignot et al., 2010). Increasing ice sheet surface melt and increasing ice stream mass discharge are both contributing to the observed growing mass loss rate of the Greenland ice sheet, as discussed in our response AC7962 on the ACPD website. Such mutually reinforcing processes provide expectation of nonlinear mass loss increase, if the climate forcing continues to increase.

Interpretation of Greenland mass loss is made difficult by its high variability. Large 2010-2012 mass loss was related to unusual summer high pressure over Greenland (Fettweis, 2013; Bellflamme et al., 2015), which produced a persistent "atmospheric river" of warm air of continental origin (Neff et al., 2014). However, weather patterns were much less favorable for surface melt in 2013 and 2014, and Greenland mass loss was much reduced (Fig. 30a).

We conclude that empirical data are too brief to imply a characteristic time for ice sheet mass loss or to confirm our hypothesis that continued high fossil fuel emissions leading to $CO_2 \sim 600$-900 ppm will cause exponential ice mass loss up to several meters of sea level. The empirical data are consistent with a doubling time of the order of a decade, but they cannot exclude slower responses. Our expectation of nonlinear behavior is based in part on recognition of how multiple amplifying feedbacks feed upon each other (Hansen et al., 1984; Pollard et al., 2015) and thus can result in large rapid change.

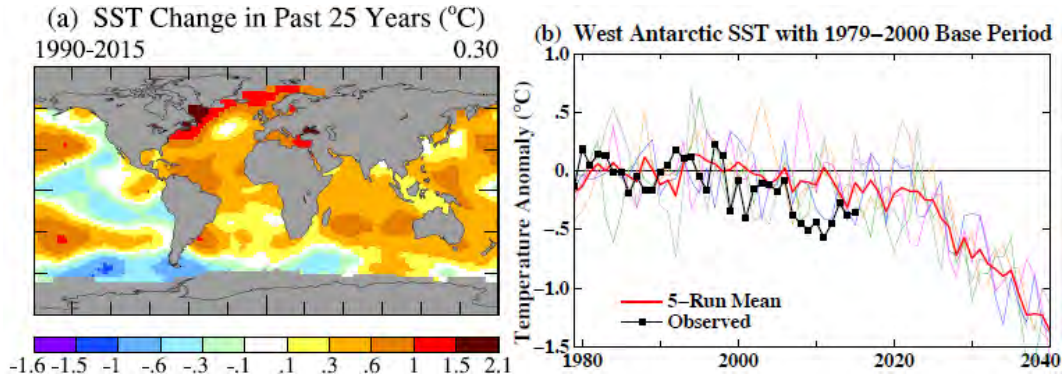

**Fig. 31.** (a) Observed 1990-2015 SST change based on local linear trends, and (b) SST anomaly relative to 1979-2000 for area south of 56°S between the dateline and 50°W. Base period excludes data prior to 1979 to avoid use of Southern Ocean climatology that artificially reduces variability (Huang et al., 2015).

## 5.2 Sea surface temperature and sea ice

The fundamental difference between climate forecasts of our model and CMIP simulations employed in IPCC assessments should appear in sea surface temperature (SST), which is well monitored. The Southern Ocean warms steadily in CMIP5 models (Fig. 12.11, Collins et al., 2013) that have little or no freshwater injection. In contrast, ice melt causes Southern Ocean cooling in our model, especially in the Western Hemisphere (Fig. 16). The model's cooling is largest in the Western Hemisphere because the specified freshwater injection (Fig. 14), based on data of Rignot et al. (2013) and Depoorter et al. (2013), is largest there. The cooling pattern is very strong by 2055-60 (Fig. 16), when freshwater injection reaches 3.8 Sv years on the North Atlantic and 7.6 Sv years on the Southern Ocean, amounts that yield 1 m global sea level rise with one-third from Northern Hemisphere ice. SST observations already show a cooling trend in the Southern Ocean off West Antarctica (Fig. 31) and are suggestive that the real world may be more sensitive than the model, but additional years of data are needed to confirm that.

Our model also differs from models in the predicted sense of Southern Hemisphere sea ice change. Freshwater effects dominate over direct effects of GHGs in our model, and thus sea ice cover grows. D. Thompson et al. (2011) suggest that $O_3$ depletion may account for observed Antarctic sea ice growth, but Sigmond and Fyfe (2014) found that all CMIP5 models yield decreasing sea ice in response to observed changes of $O_3$ and other GHGs. Ferreira et al. (2015) show that $O_3$ depletion yields a short time scale sea ice increase that is soon overtaken by warming and sea ice decrease with realistic GHG forcing. We suggest that these models are missing the dominant driver of change on the Southern Ocean: freshwater input.

Our modeled SMOC has begun to slow already (Fig. 32a), consistent with tracer observations in the Weddell Sea by Huhn et al. (2013), which reveal a 15-21% reduction in the ventilation of Weddell Sea Bottom Water and Weddell Sea Deep Water in1984-2008. Delayed growth of sea ice in the model (Fig. 32b) may be in part related to the model's muted vertical stratification, as we will discuss, and the model's general difficulty in producing Southern Hemisphere sea ice.

We infer that observed cooling in the western part of the Southern Ocean, growing Southern Ocean sea ice, and slowdown of at least the Weddell Sea component of SMOC are early responses to increasing freshwater injection. Although observed sea ice increase is smaller in 2015 than in the previous few years (data are updated daily at http://nsidc.org/data/seaice_index), but Hansen and Sato (2015) note a negative correlation between sea ice area and El Ninos, so we expect that sea ice growth may resume after the present strong El Nino fades.

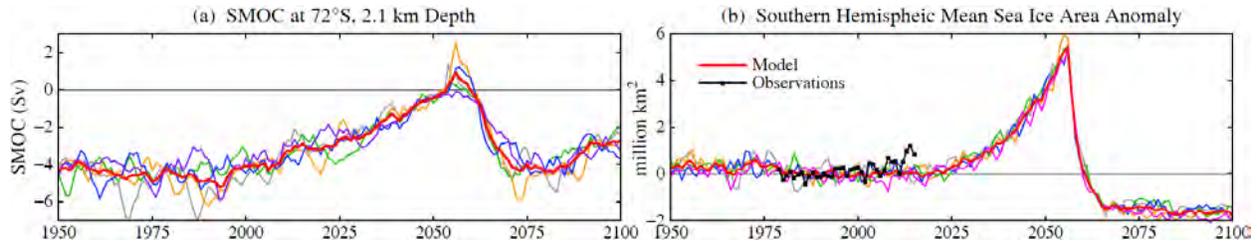

**Fig. 32.** (a) Global meridional overturning circulation (Sv) at 72°S. Freshwater injection near Antarctica is 720 Gt/year in 2011, increasing with 10-year doubling time, and half as much around Greenland. SMOC diagnostic includes only the mean (Eulerian) term. (b) Annual-mean Southern Hemisphere sea ice area anomaly ($10^6$ km$^2$) in five runs (relative to 1979-2000). Observations include 2015.

Let's compare North Atlantic and Southern Ocean responses to freshwater forcing. Modeled AMOC response does not become significant until ~2040 (Fig. 33). Even discounting the decade lead in Southern Hemisphere forcing (720 Gt/year in 2011, double that in the North Atlantic), SMOC still responds quicker, albeit gradually, to freshwater forcing (Fig. 32).

Observations suggest that the real world may be responding more quickly than the model to freshwater forcing in the North Atlantic. Rahmstorf et al. (2015) develop an AMOC index based on SST in the "global warming hole" southeast of Greenland (Drijfhout et al., 2012), and they use the AMOC index to conclude that an AMOC slowdown unprecedented in the prior 1000 years occurred in the late 20[th] century. That slowdown seems to have been a response to the "Great Salinity Anomaly", which is thought to have resulted from an estimated ~2000 km$^3$ anomalous sea ice export from the Arctic (Dickson et al., 1988). Although the AMOC partially recovered in the early 21[st] century, further slowdown has returned in the past several years, judging from a measurement array (Robson et al., 2014) as well as from the AMOC index (Rahmstorf et al., 2015). The recent AMOC slowdown could be related to ice melt, as Greenland (Fig. 30) and neighboring ice caps contributed more than 1000 km$^3$ meltwater in 2011-2012. The model (Fig. 33), in contrast, does not reach substantial AMOC reduction until ~2040, when the annual freshwater injection on the North Atlantic is ~7500 km$^3$/year and the cumulative injection is ~ 1.2 Sv years.

A useful calibration of AMOC sensitivity to freshwater forcing is provided by the 8.2 ky b2k glacial Lake Agassiz freshwater outburst (Kleiven et al., 2008) that occurred with the demise of the Hudson Bay ice dome. Freshwater injected onto the North Atlantic was ~2.5-5 Sv years (Clarke et al., 2004). Proxy temperature records (see LeGrande et al., 2006) suggest that real world cooling reached about 6°C on Greenland, 3-4°C in the East Norwegian Sea and 1.7°C in Germany. The duration of the 8.2 ky b2k event was 160 years (Rasmussen et al., 2014). The LeGrande et al. (2006) model, which has the same atmospheric model as our present model but does not include the basic improvements in the ocean described in Sec. 3.1, produced results not inconsistent with real world changes, but the modeled temperature response seemed to be on the low side (Fig. 1, LeGrande et al., 2006). The mean decrease of the modeled AMOC was 40% with AMOC recovering within 20-30 years, but secondary and tertiary slowdowns in some of the model runs extending as long as in the observed 8.2 ky b2k event (160 years). Although this model response is within the range suggested by paleo data, it is on the weak side.

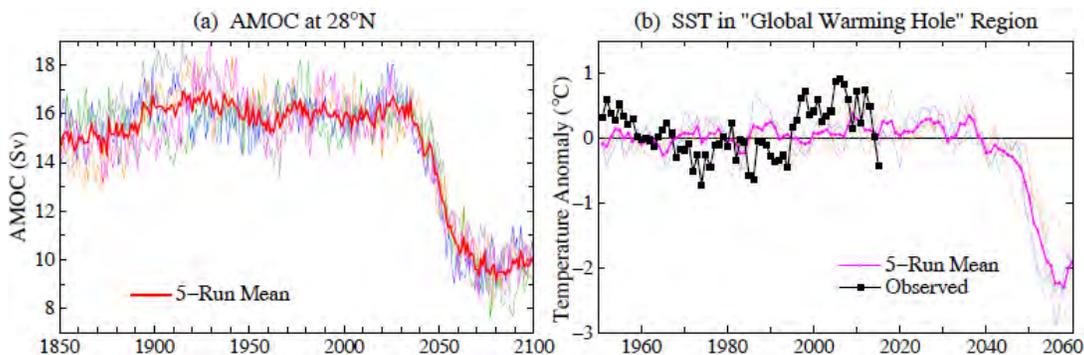

**Fig. 33.** (a) AMOC (Sv) at 28°N in simulations with the forcings of Sec. 4.2 (i.e., including freshwater injection of 720 Gt/year in 2011 around Antarctica, increasing with a 10-year doubling time, and half that amount around Greenland), and (b) SST (°C) in North Atlantic region (44-60°N, 10-50°W).

This model check based on the 8.2 ky b2k event does not prove that the model has correct sensitivity for today's weaker forcing. We suspect that the model is less sensitive than the real world because the model has difficulty maintaining vertical stratification, which could result from coarse vertical resolution, excess parameterized small scale mixing, or numerical noise. Excessive mixing could also explain a too-long climate response time, as discussed in connection with Fig. 4. Hansen et al. (2011) showed that surface temperature is probably too sluggish in response to a climate forcing, not only in the GISS model, but in several other models. Hofmann and Rahmstorf (2009) suggest another reason for models being too insensitive to freshwater forcing: a bias in ocean model development spurred by desire for a stable AMOC. Below we suggest studies that are needed to investigate the model sensitivity issue.

### 5.3 Southern Ocean internal processes

Although the ocean surface is observed in detail on a daily basis, our main interest is in implications for long-term processes in the ocean below. Paleoclimate data discussed in Sec. 4 reveal that the Southern Ocean, as gateway to the global deep ocean, exerts a powerful control over glacial/interglacial climate.

The Southern Ocean has significant control on release of ocean heat to space. In an extreme case, polynyas form in the dead of Antarctic winter, as upwelling warm water melts the sea ice and raises the air temperature by tens of degrees, increasing thermal radiation to space, thus serving as a valve that releases ocean heat. Today, as surface meltwater stabilizes the vertical water column, that valve is being partially closed. de Lavergne et al. (2014) relate the absence of large open ocean polynyas in recent decades to surface freshening. Release of heat to the atmosphere and space, which occurs without the need for large open ocean polynyas, is slowed by increasing sea ice cover in response to increasing ice shelf melt (Bintanja et al., 2013).

Internal Southern Ocean effects of ocean surface freshening and cooling seem to be well underway. Schmidtko et al. (2014) and Roemmich et al. (2015) document changes in the Southern Ocean in recent decades, especially warming of Circumpolar Deep Water (CDW), which they and others (Jacobs et al., 2011; Rignot et al., 2013) note is the likely cause of increased ice shelf melt. Observations of ocean surface freshening and freshening of the water column (Rintoul, 2007; Jacobs and Giulivi, 2010) and deep ocean warming (Johnson et al., 2007; Purkey and Johnson, 2013) leave little doubt that these processes are occurring.

### 6 Summary and Implications

Via a combination of climate modeling, paleoclimate analyses, and modern observations we have identified climate feedback processes that help explain paleoclimate change and may be of critical importance in projections of human-made climate change. Here we summarize our interpretation of these processes, their effect on past climate change, and their impact on climate projections. We then discuss key observations and modeling studies needed to assess the validity of these interpretations. We argue that these feedback processes may be understated in our model, and perhaps other models, because of an excessively diffusive ocean model. Thus there is urgency to obtain better understanding of these processes and models.

## 6.1 Ocean stratification and ocean warming

Global ocean circulation (Fig. 16) is altered by the effect of low density freshwater from melting of Greenland or Antarctica ice sheets. While the effects of shutdown of NADW have been the subject of intensive research for a quarter of a century, we present evidence that models have understated the threat and imminence of slowdown and shutdown AMOC and SMOC. Below we suggest modeling and observations that would help verify the reality of stratification effects on polar oceans and improve assessment of likely near-term and far-term impacts.

Almost counter-intuitively, regional cooling from ice melt produces an amplifying feedback that accelerates ice melt by placing a lid on the polar ocean that limits heat loss to the atmosphere and space, warming the ocean at the depth of ice shelves. The regional surface cooling increases Earth's energy imbalance, thus pumping into the ocean energy required for ice melt.[18]

## 6.2 Southern Ocean, $CO_2$ control knob, and ice sheet time scale

Our climate simulations and analysis of paleoclimate oscillations indicate that the Southern Ocean has the leading role in global climate change, with the North Atlantic a supporting actor. The Southern Ocean dominates by controlling ventilation of the deep ocean $CO_2$ reservoir.

$CO_2$ is the control knob that regulates global temperature. On short time scales, i.e., fixed surface climate, $CO_2$ sets atmospheric temperature because $CO_2$ is stable, thus the ephemeral radiative constituents, $H_2O$ and clouds, adjust to $CO_2$ amount (Lacis et al., 2010, 2013).

On millennial time scales both $CO_2$ and surface albedo (determined by ice and snow cover) are variable and contribute about equally to global temperature change (Hansen et al., 2008). However, here too $CO_2$ is the more stable constituent with time scale for change $\sim 10^3$ years, while surface albedo is more ephemeral judging from the difficulty of finding any lag of more than order $10^2$ years between sea level and polar temperature (Grant et al., 2012).

Here we must clarify that ice and snow cover are both a consequence of global temperature change, generally responding to the $CO_2$ control knob, but also a mechanism for global climate change. Specifically, regional or hemispheric snow and ice respond to seasonal insolation anomalies (as well as to $CO_2$ amount), thus affecting hemispheric and global climate, but to achieve large global change the albedo driven climate change needs to affect the $CO_2$ amount.

We also note that Southern Ocean ventilation is not the only mechanism affecting airborne $CO_2$ amount. Terrestrial sources, dust fertilization of the ocean, and other factors play roles, but deep ocean ventilation seems to be the dominant mechanism on glacial-interglacial time scales.

The most important practical implication of this "control knob" analysis is realization that the time scale for ice sheet change in Earth's natural history has been set by $CO_2$, not by ice physics. With the rapid large increase of $CO_2$ expected this century, we have no assurance that large ice sheet response will not occur on the century time scale or even faster.

## 6.3 Heinrich and Dansgaard-Oeschger events

Heinrich and Dansgaard-Oeschger events demonstrate the key role of subsurface ocean warming in melting ice shelves and destabilizing ice sheets, and they show that melting ice shelves can result in large rapid sea level rise. A cold lens of fresh meltwater on the ocean surface may make surface climate uncomfortable for humans, but it abets the provision of warmth at depths needed to accelerate ice melt.

---

[18] Planetary energy imbalance induced by meltwater cooling helps provide the energy required by ice heat of fusion. Ice melt to raise sea level 1 m requires a 10-year Earth energy imbalance 0.9 W/m² (Table S1, Hansen et al., 2005b).

## 6.4 End-Eemian climate events

We presented evidence for a rapid sea level rise of several meters late in the Eemian, as well as evidence of extreme storms in the Bahamas and Bermuda that must have occurred when sea level was near its maximum. This evidence is consistent with the fact that the North Atlantic was cooling in the late Eemian, while the tropics were unusually warm, the latter being consistent with the small obliquity of Earth's spin axis at that time.

Giant boulders of mid-Pleistocene limestone placed atop an Eemian substrate in north Eleuthera, which must have been deposited by waves, are emblematic of stormy end-Eemian conditions. Although others have suggested the boulders may have been emplaced by a tsunami, we argue that the most straightforward interpretation of all evidence favors storm emplacement. In any case, there is abundant evidence for strong late Eemian storminess and high sea level.

A late Eemian shutdown of the AMOC would have caused the most extreme North Atlantic temperature gradients. AMOC shutdown in turn would have added to Southern Ocean warmth, which may have been a major factor in the Antarctic ice sheet collapse that is required to account for the several meters of rapid late Eemian sea level rise.

Confirmation of the exact sequence of late Eemian events does not require absolute dating, but it probably requires finding markers that allow accurate correlation of high resolution ocean cores with ice cores, as has proved possible for correlating Antarctic and Greenland ice cores. Such accurate relative dating would make it easier to interpret the significance of abrupt changes in two Antarctic ice cores at about End-Eemian time (Masson-Delmotte, 2011), which may indicate rapid large change in Antarctic sea ice cover.

Understanding end-Eemian storminess is important in part because the combination of strong storms with sea level rise poses a special threat. However, sea level rise itself is the single greatest global concern, and it is now broadly accepted that late Eemian sea level reached +6-9 m, implicating a substantial contribution from Antarctica, at a time when Earth was little warmer than today (Dutton et al., 2015; Supplement to our present paper).

## 6.5 The Anthropocene

The Anthropocene (Crutzen and Stoermer, 2000), the era in which humans contributed to global climate change, is usually assumed to have begun in the past few centuries. Ruddiman (2003) suggests that it began earlier, as deforestation began to affect $CO_2$ about 8000 years ago. Southern Ocean feedbacks considered in our present paper are relevant to that discussion.

Ruddiman (2003) assumed that 40 ppm of human-made $CO_2$ was needed to explain a 20 ppm $CO_2$ increase in the Holocene (Fig. 27c), because $CO_2$ decreased ~20 ppm, on average, during several prior interglacials. Such a large human source should have left an imprint on $\delta^{13}CO_2$ that is not observed in ice core $CO_2$ (Elsig, et al., 2009). Ruddiman (2013) suggests that $^{13}C$ was taken up in peat formation, but the required peat formation would be large and no persuasive evidence has been presented to support such a dominant role for peat in the glacial carbon cycle.

We suggest that Ruddiman's hypothesis may be right, but the required human-made carbon source is much smaller than he assumed. Decline of $CO_2$ in interglacial periods is a climate feedback, a result of declining Southern Ocean temperature, which slows the ventilation of the deep ocean and exhalation of deep ocean $CO_2$. Human-made $CO_2$ forcing needed to avoid Antarctic cooling and atmospheric $CO_2$ decline is only the amount needed to counteract the weak natural forcing trend, not the larger feedback-driven $CO_2$ declines in prior interglacials, because the feedback does not occur if the natural forcings are counteracted.

The warm season insolation anomaly on the Southern Ocean was positive and growing 8 ky ago (Fig. 27a). Thus the human-made $CO_2$ contribution required to make the Southern Ocean a $CO_2$ source sufficient to yield the observed $CO_2$ growth (Fig. 27c) is unlikely to have been larger than ~10 ppm, but quantification requires carbon cycle modeling beyond present capabilities.

However, the modest requirement on the human $CO_2$ source and the low $\delta^{13}C$ content of deep-ocean $CO_2$ make the Ruddiman hypothesis more plausible and likely.

## 6.6 The Hyper-Anthropocene

A fundamentally different climate phase, a Hyper-Anthropocene, began in the latter half of the 18[th] century as improvements of the steam engine ushered in the industrial revolution (Hills, 1989) and exponential growth of fossil fuel use. Human-made climate forcings now overwhelm natural forcings. $CO_2$, at 400 ppm in 2015, is off the scale in Fig. 27c. $CO_2$ climate forcing is a reasonable approximation of the net human forcing, because forcing by other GHGs tends to offset negative human forcings, mainly aerosols (Myhre et al., 2013). Most of the $CO_2$ growth occurred in the past several decades, and three-quarters of the ~1°C global warming since 1850 (update of Hansen et al., 2010, available at http://www.columbia.edu/~mhs119/Temperature/) has occurred since 1975. Climate response to this $CO_2$ level, so far, is only partial.

Our analysis paints a very different picture than IPCC (2013) for continuation of this Hyper-Anthropocene phase, if GHG emissions continue to grow. In that case, we conclude that multi-meter sea level rise would become practically unavoidable, probably within 50-150 years. Full shutdown of the North Atlantic Overturning Circulation would be likely within the next several decades in such a climate forcing scenario. Social disruption and economic consequences of such large sea level rise, and the attendant increases of storms and climate extremes, could be devastating. It is not difficult to imagine that conflicts arising from forced migrations and economic collapse might make the planet ungovernable, threatening the fabric of civilization.

Our study, albeit with a coarse-resolution model and simplifying assumptions, raises fundamental questions that point toward specific modeling and measurement needs.

## 6.7 Modeling priorities

Predictions from our modeling are shown vividly in Fig. 16, which shows simulated climate four decades in the future. However, we concluded that the basic features there are already beginning to evolve in the real world, that our model underestimates sensitivity to freshwater forcing and the stratification feedback, and that the surface climate effects are likely to emerge sooner than models suggest, if GHG climate forcing continues to grow.

This interpretation arises from evidence of excessive small scale mixing in our ocean model and some other models, which reduces the stratification feedback effect of freshwater injection. Our climate model, with ~3°C equilibrium sensitivity for 2×$CO_2$, achieves only ~60% of its equilibrium response in 100 years (Fig. 4). Hansen et al. (2011) conclude that such a slow response is inconsistent with Earth's measured energy imbalance; if the ocean were that diffusive it would be soaking up heat faster than the measured planetary energy imbalance ~0.6 W/m$^2$ (Hansen et al., 2011; von Schuckmann et al., 2016). Hansen (2008) found the response time of climate models of three other modeling centers to be as slow or slower than the GISS model, implying that the oceans in those models were also too diffusive and thus their climate response times too long. The climate response time is fundamental to interpretation of climate change and the impact of excessive small scale mixing, if such exists, is so important that we suggest that all models participating in future CMIP studies should be asked to calculate and report their climate response function, R (Fig. 4). An added merit of that information is the fact that R permits easy calculation of the global temperature response for any climate forcing (Eq. 1).

It may be possible to quickly resolve or at least clarify this modeling issue. A fundamental difficulty with ocean modeling is that the scale of eddies and jet-like flows is much smaller than comparable features in the atmosphere, which is the reason for the Gent and McWilliams (1990) parameterization of eddy mixing in coarse resolution models. However, computers at large modeling centers today allow simulations with ocean resolution as fine as ~0.1°, which can resolve eddies and minimize need for parameterizations. Winton et al. (2014) used a GFDL model [one of the models Hansen (2008) found to have a response function, R, similar to that of our model] with 0.1° ocean resolution for a 1%/year increasing $CO_2$ experiment, finding ~25% increase in transient global warming, which is about the increment needed to to increase R (100 years) to ~0.75, consistent with Earth's measured energy imbalance (Hansen et al., 2011). The increased surface response implies that small-scale mixing that limits stratification is reduced. Sabe et al. (2015) show that this model with 0.1° ocean resolution yields 3-4°C warming along the United States East Coast at doubled $CO_2$ and cooling ($\sim -1$°C) southeast of Greenland, both temperature changes a result of AMOC slowdown that reduces poleward transport of heat.

The model results are striking because similar temperature patterns seem to be emerging in observations (Figs. 31, S24). Annual and decadal variability limit interpretation, but given the AMOC sensitivity revealed in paleoclimate data, we infer that stratification effects are beginning to appear in the North Atlantic due to the combination of ice melt and GHG forcing. Eddy-resolving ocean models are just beginning be employed and analyzed (Bryan et al., 2014), but there needs to be an added focus in CMIP runs to include freshwater from ice melt. CMIP5 simulations led to IPCC estimates of AMOC weakening in 2100 (Collins et al., 2013) of only 11% for the weakest forcing scenario and 34% for the strongest forcing ($CO_2$ = 936 ppm), but the CMIP5 runs do not include ice melt. This moderate change on century time scale may be a figment of (1) excluding ice melt, and (2) understated stratification, as can be checked with improved high-resolution models that include realistic meltwater injection. Reliable projections of AMOC and North Atlantic climate will not flow simply from new high resolution model runs, as Winton et al. (2014) note that ocean models have other tuning parameters that can sensitively affect AMOC stability (Hofmann and Rahmstorf, 2009), which is reason for a broad comparative study with the full set of CMIP models.

High resolution ocean models are also needed to realistically portray deepwater formation around Antarctica, penetration of warm waters into ice shelf environments, and, eventually, ocean-ice sheet feedbacks. More detailed models should also include the cooling effect of ice phase change (heat of fusion) more precisely, perhaps including iceberg tracks. However, there is merit in also having a coarser resolution version of major models with basically the same model physics. Coarser resolution allows long simulations, facilitating analysis of the equilibrium response, paleoclimate studies, and extensive testing of physical processes.

## 6.8 Measurement priorities

A principal issue is whether ice melt will increase exponentially, as we hypothesize if GHGs continue to grow rapidly. Continuous gravity measurements, coupled with surface mass balance and physical process studies on the ice sheets, are needed to obtain and understand regional ice mass loss on both Greenland and Antarctica. Ocean-ice shelf interactions need to be monitored, especially in Antarctica, but some Greenland ice is also vulnerable to thermal forcing by a warming ocean via submarine glacial valleys (Morlighem et al., 2014; Khan et al., 2014).

Summer weather variability makes mass loss in the Greenland melt season highly variable, but continued warming of North American continental air masses likely will spur multiple amplifying feedbacks. These feedbacks need to be monitored and quantified because their combination can lead to rapid meltwater increase. In addition to feedbacks discussed in Sec. 5.1, Machguth et al. (2016) note that meltwater injection to the ocean will increase as surface melt and refreeze limits the ability of firn to store meltwater. Meltwater in the past several years is already of the magnitude of the "Great Salinity Anomaly" (Dickson et al., 1988) that Rahmstorf et al. (2015) conclude produced significant AMOC slowdown in the late 20[th] century.

Continued global measurements of SST from satellites, calibrated with buoy and ship data (Huang et al., 2015) will reveal whether coolings in the Southern Ocean and southeast of Greenland are growing. Internal ocean temperature, salinity and current measurements by the ARGO float program (von Schuckmann et al., 2016), including planned extensions into the deep ocean and under sea ice, are crucial for several reasons. ARGO provides global measurements of ocean quantities that are needed to understand observed surface changes. If climate models are less sensitive to surface forcings than the real world, as we have concluded, the ARGO data will help analyze the reasons for model shortcomings. In addition, ARGO measurements of the rate of ocean heat content change are the essential data for accurate determination of Earth's energy imbalance, which determines the amount of global warming that is still "in the pipeline" and the changes of atmospheric composition that would be needed to restore energy balance, the fundamental requirement for approximately stabilizing climate.

## 6.9 Practical implications

The United Nations Framework Convention on Climate Change (UNFCCC, 1992) states:

> *"The ultimate objective of this Convention and any related legal instruments that the Conference of the Parties may adopt is to achieve, in accordance with the relevant provisions of the Convention, stabilization of greenhouse gas concentrations in the atmosphere at a level that would prevent dangerous anthropogenic interference with the climate system. Such a level should be achieved within a time frame sufficient to allow ecosystems to adapt naturally to climate change, to ensure that food production is not threatened and to enable economic development to proceed in a sustainable manner."*

"Dangerous" is not defined by the Convention. The public understands the word.

Our present paper has several implications with regard to the concerns that the Framework Convention is meant to address:

First, our conclusions suggest that a target of limiting global warming to 2°C, which has sometimes been discussed, does not provide safety. In other words, 2°C global warming is dangerous in the sense that the public would understand. We cannot be certain that multimeter sea level rise will occur if we allow global warming of 2°C. However, we know the warming would remain present for many centuries, if we allow it to occur (Solomon et al., 2010), a period exceeding the ice sheet response time implied by paleoclimate data. Sea level reached +6-9 m in the Eemian, a time that we have concluded was probably no more than a few tenths of a degree warmer than today. We observe accelerating mass losses from the Greenland and Antarctic ice sheets, and we have identified amplifying feedbacks that will increase the rates of change. We also observe changes occurring in the North Atlantic and Southern Oceans, changes that we can attribute to ongoing warming and ice melt, which imply that this human-driven climate change seems poised to affect these most powerful overturning ocean circulation systems, systems that we know have had huge effects on the planetary environment in the past. We conclude that, in the sense of the word dangerous that the public appreciates, 2°C global warming is dangerous.

Second, our study suggests that global surface air temperature, although an important diagnostic, is a flawed metric of planetary "health", because faster ice melt has a cooling effect for a substantial period. Earth's energy imbalance is in some sense a more fundamental climate diagnostic. Stabilizing climate, to first order, requires restoring planetary energy balance. The Framework Convention never mentions temperature – instead it mentions stabilization of greenhouse gas concentrations at a level to avoid danger. It has been shown that the dominant climate forcing, $CO_2$, must be reduced to no more than 350 ppm to restore planetary energy balance (Hansen et al., 2008) and keep climate near the Holocene level, if other forcings remain unchanged. Rapid phasedown of fossil fuel emissions is the crucial need, because of the millennial time scale of this carbon in the climate system. Improved understanding of the carbon cycle is needed to determine the most effective complementary actions. It may be feasible to restore planetary energy balance via improved agricultural and forestry practices and other actions to draw down atmospheric $CO_2$ amount, if fossil fuel emissions are rapidly phased out.

Third, we not only see evidence of changes beginning to happen in the climate system, as discussed above, we have associated these changes with amplifying feedback processes. We understand that in a system that is out of equilibrium, a system in which the equilibrium is difficult to restore rapidly, a system in which major components such as the ocean and ice sheets have great inertia but are beginning to change, the existence of such amplifying feedbacks presents a situation of great concern. There is a possibility, a real danger, that we will hand young people and future generations a climate system that is practically out of their control.

We conclude that the message our climate science delivers to the public is this: we have a global emergency. Fossil fuel $CO_2$ emissions should be reduced as rapidly as practical.

**Acknowledgments.** This paper is dedicated to Wally Broecker, the "father of global warming", whose inquisitive mind has stimulated much of the world's research aimed at understanding global climate. Completion of this study was made possible by a generous gift from The Durst Family to the Climate Science, Awareness and Solutions program at the Columbia University Earth Institute. That program was initiated in 2013 primarily via support from the Grantham Foundation for Protection of the Environment, Jim and Krisann Miller, and Gerry Lenfest and sustained via their continuing support. Other substantial support is provided by the Flora Family Foundation, Elisabeth Mannschott, Alexander Totic and Hugh Perrine. Concepts about "greenhouse, icehouse, madhouse" conditions during MIS 5e in Bermuda and the Bahamas were fostered by A. Conrad Neumann, while John T. Hollin understood nearly half a century ago the importance of West Antarctica's contributions to rapid climate, ice surge, and sea-level changes. We are grateful to numerous friends and colleagues who are passionate about the geology and natural history of Bermuda and the Bahamas. We thank Anders Carlson, Elsa Cortijo, Nil Irvali, Kurt Lambeck, Scott Lehman, and Ulysses Ninnemann for their kind provision of data and related information, the editors of ACP for development of effective publication mechanisms, and referees and commenters for many helpful suggestions on the ACPD version of the paper. Support for climate simulations was provided by the NASA High-End Computing (HEC) Program through the NASA Center for Climate Simulation (NCCS) at Goddard Space Flight Center.

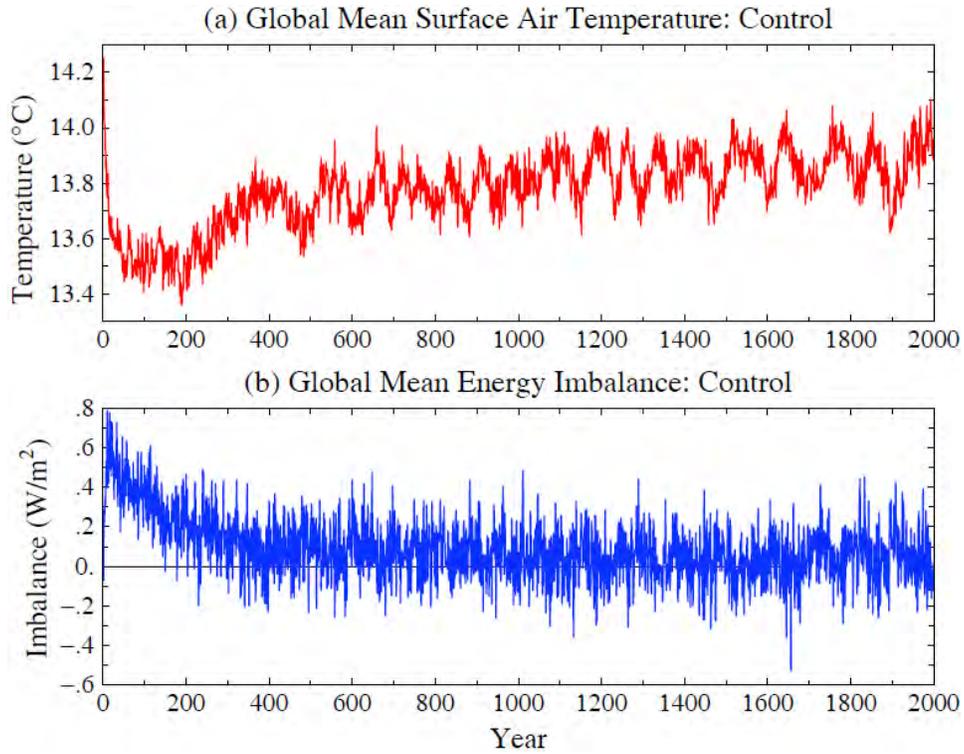

**Fig. S1.** Surface air temperature (°C) and planetary energy imbalance (W/m²) in the control run.

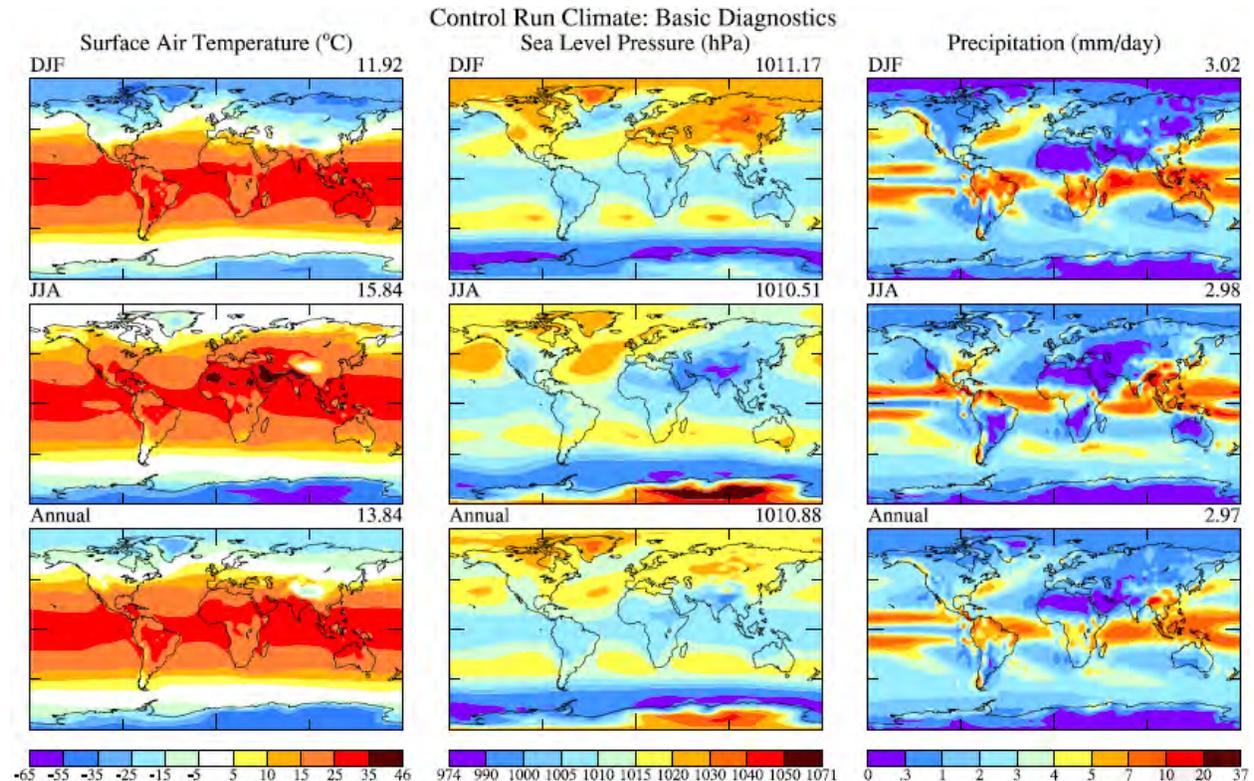

**Fig. S2.** Surface air temperature (°C), sea level pressure (hPa) and precipitation (mm/day) in Dec-Jan-Feb (upper row), JJA (middle row) and annual mean (lower row) in the climate model control run.

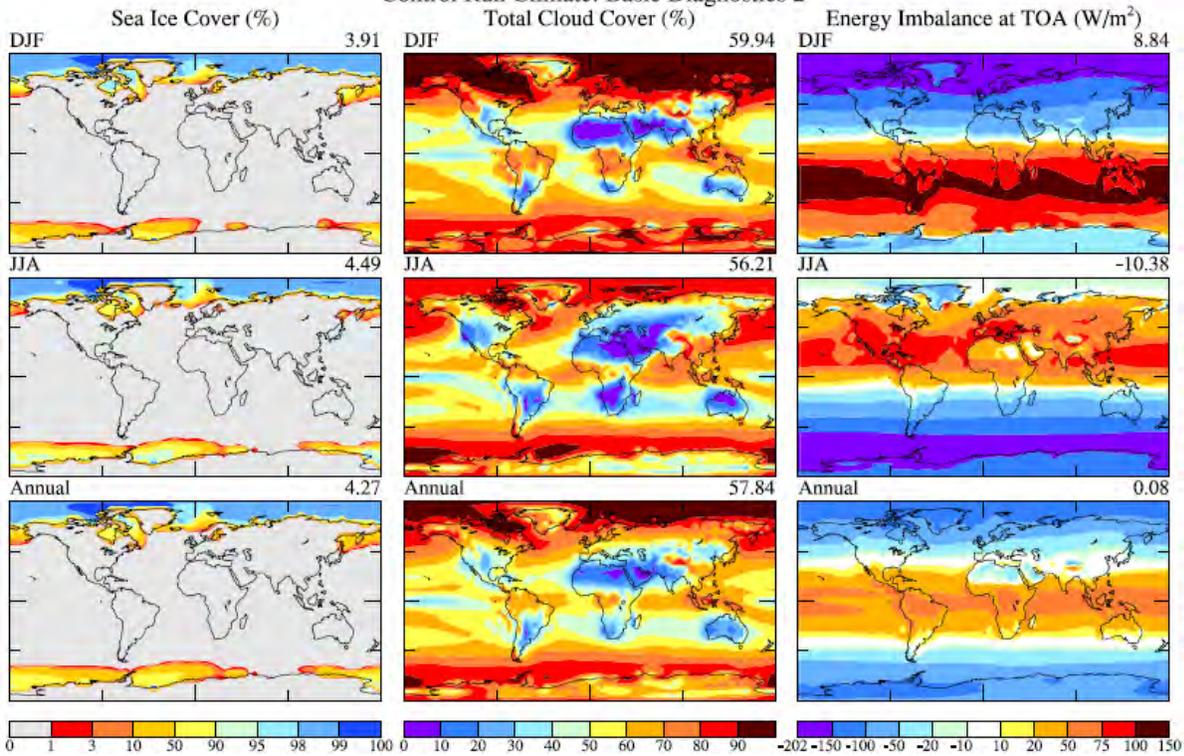

**Fig. S3.** Sea ice cover (%), cloud cover (%) and top of atmosphere energy imbalance (W/m²) in Dec-Jan-Feb (upper row), JJA (middle row) and annual mean (lower row) in climate model control run.

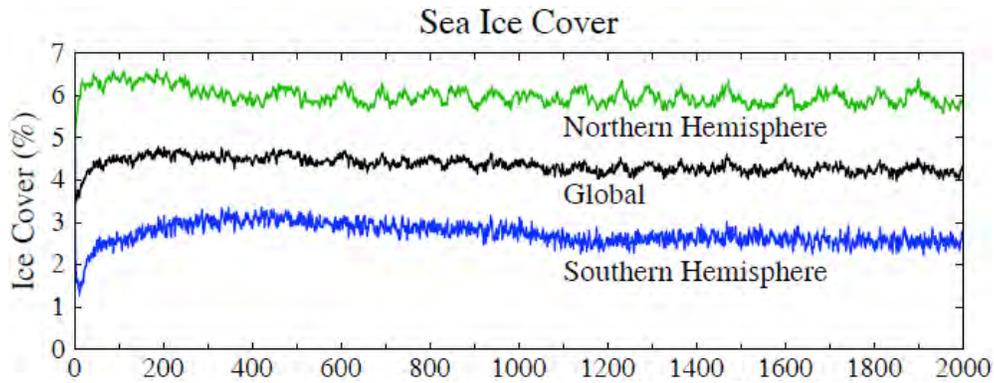

**Fig. S4.** Hemispheric and global sea ice cover (%) versus time in the control run.

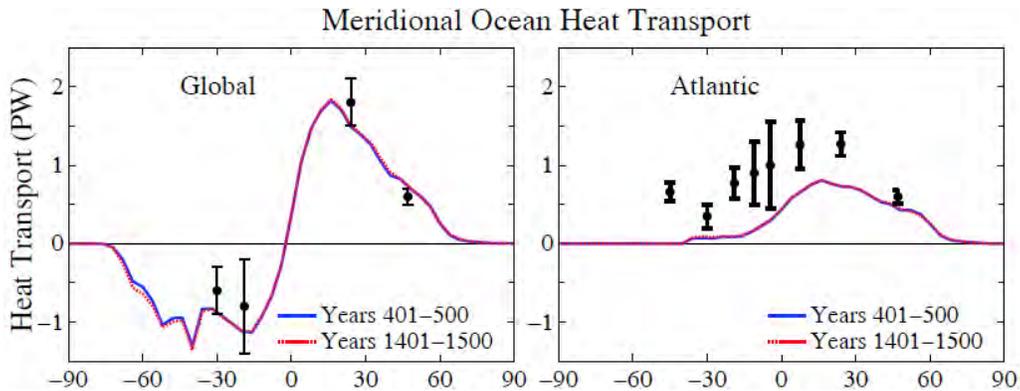

**Fig. S5.** Poleward transport of heat (PW) by the ocean in 5th and 15th centuries of the control run. Observational estimates (black dots with error bars) are from Ganachaud and Wunsch (2003).

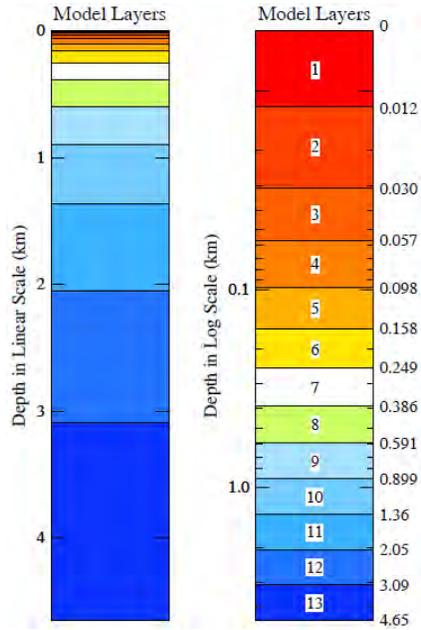

**Fig. S6.** Layer depths in ocean model.

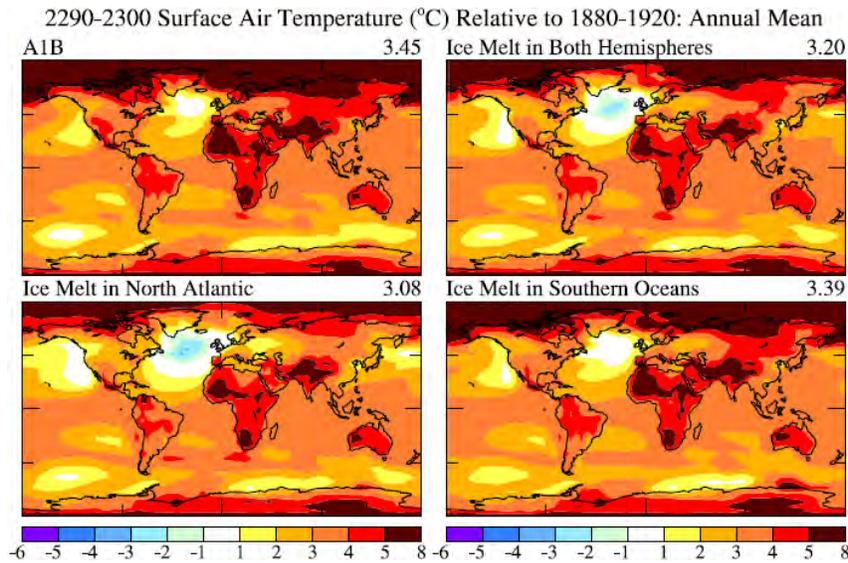

**Fig. S7.** Surface air temperature change (°C) relative to 1880-1920 in 2290-2300 for the four climate forcing scenarios shown in Fig. 8.

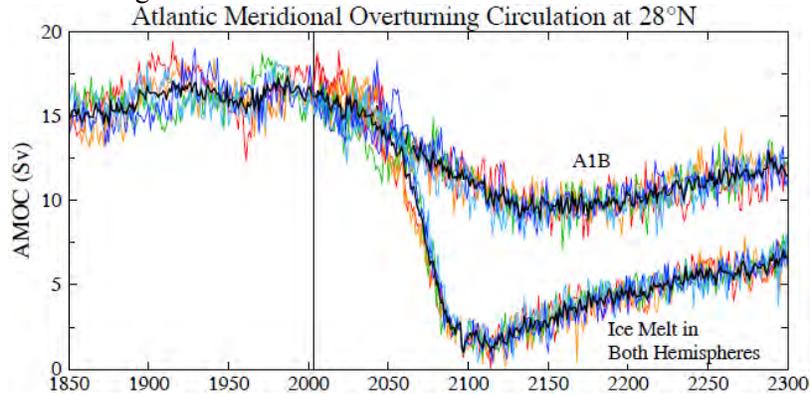

**Fig. S8.** AMOC strength (Sv) at 28N in five ensemble members and their mean (heavy black line) for the A1B GHG scenario and for that scenario plus ice melt in both hemispheres with 10-year doubling time reaching a maximum 5 m contribution to sea level.

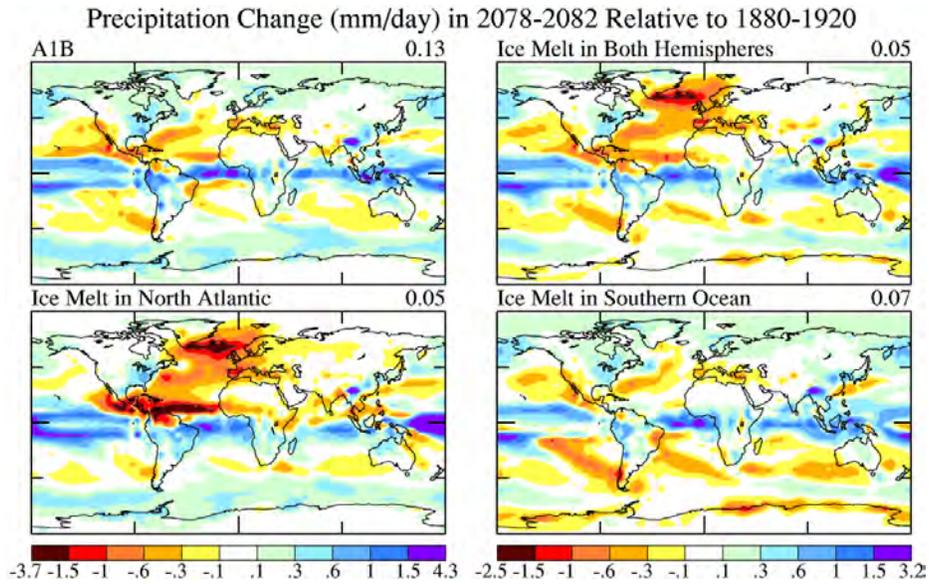

**Fig. S9.** Precipitation change (mm/day) in 2078-2082 for the same four scenarios as in Figs. 6 and 8.

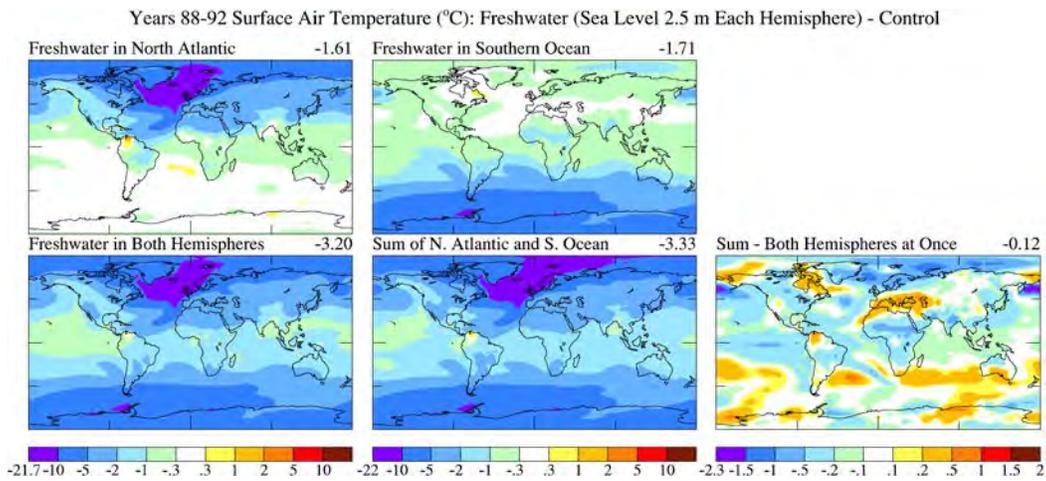

**Fig. S10.** Surface air temperature change (°C) in pure freshwater experiments at time of peak cooling (years 88-92) in three experiments with 2.5 m freshwater in each hemisphere. The sum of responses to the hemispheric forcings is compared with the response to forcing in both hemispheres in the bottom row.

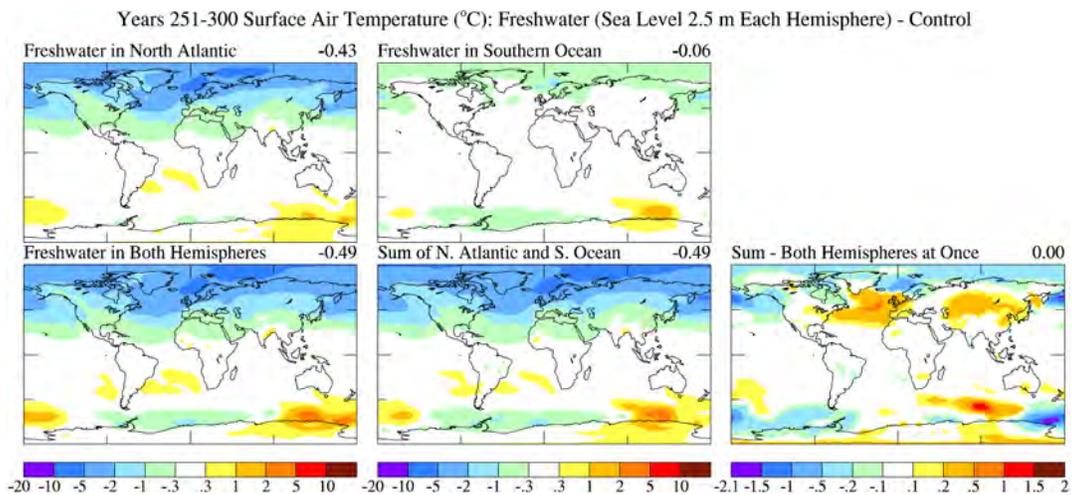

**Fig. S11.** Same as Fig. S10, but for years 251-300.

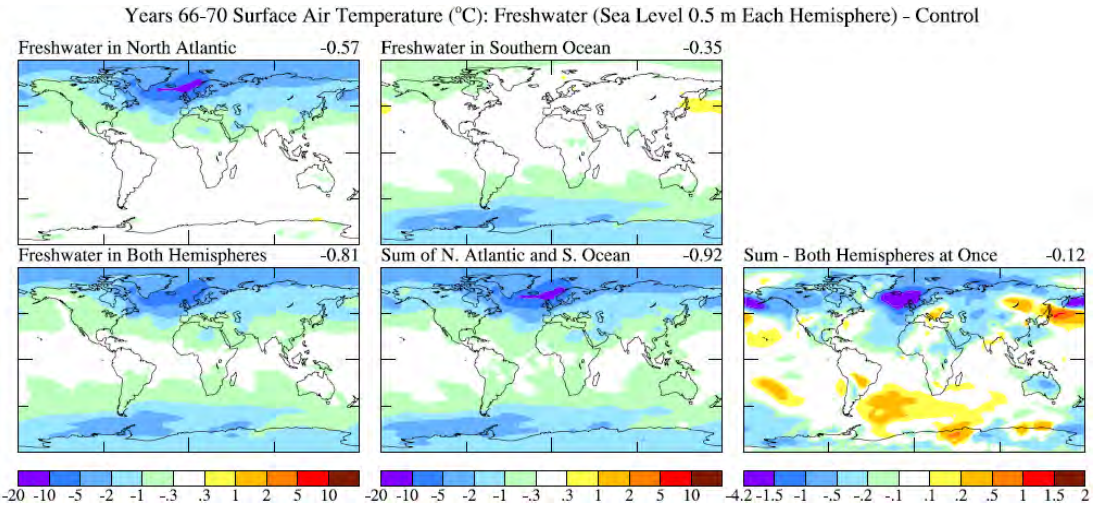

**Fig. S12.** Same as Fig. S10, but for hemispheric freshwater inputs of 0.5 m at years 66-70.

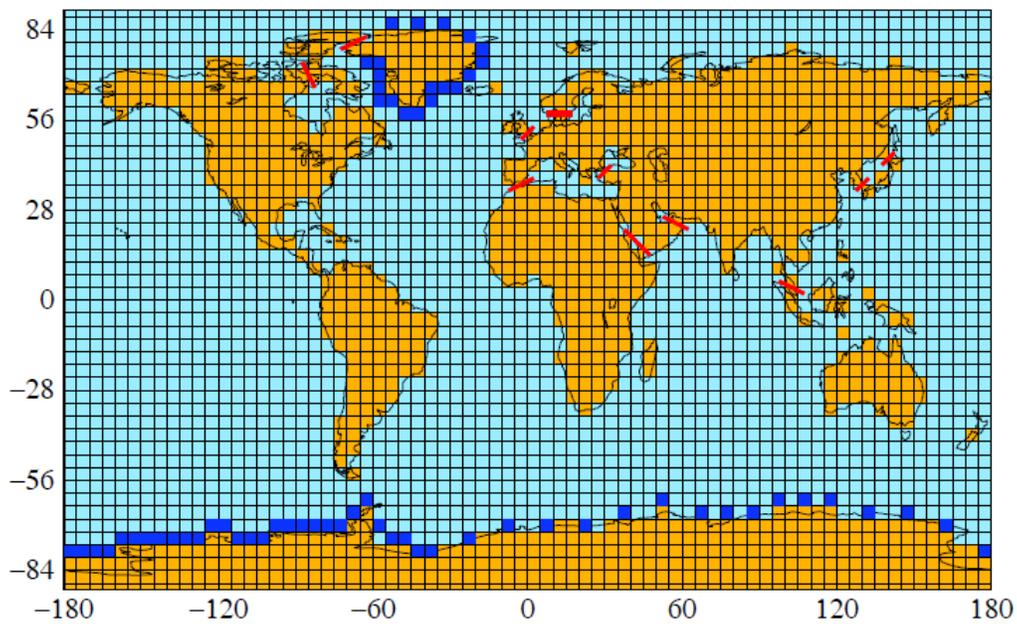

**Fig. S13.** Climate model grid. Dark blue gridoxes are locations of freshwater insertion. Red lines mark the 12 straights connecting ocean gridboxes.

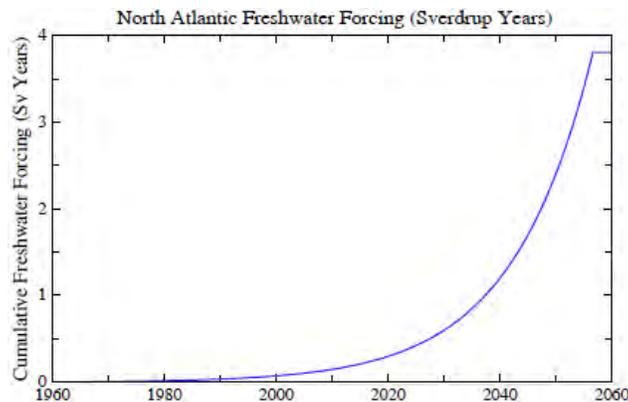

**Fig. S14.** Freshwater forcing (Sv years) in the North Atlantic in modified forcings scenario, i.e., the runs that have 360 Gt freshwater injection in 2011 with freshwater at earlier and later times based on 10-year doubling. Freshwater injection onto the Southern Ocean is double the North Atlantic rate.

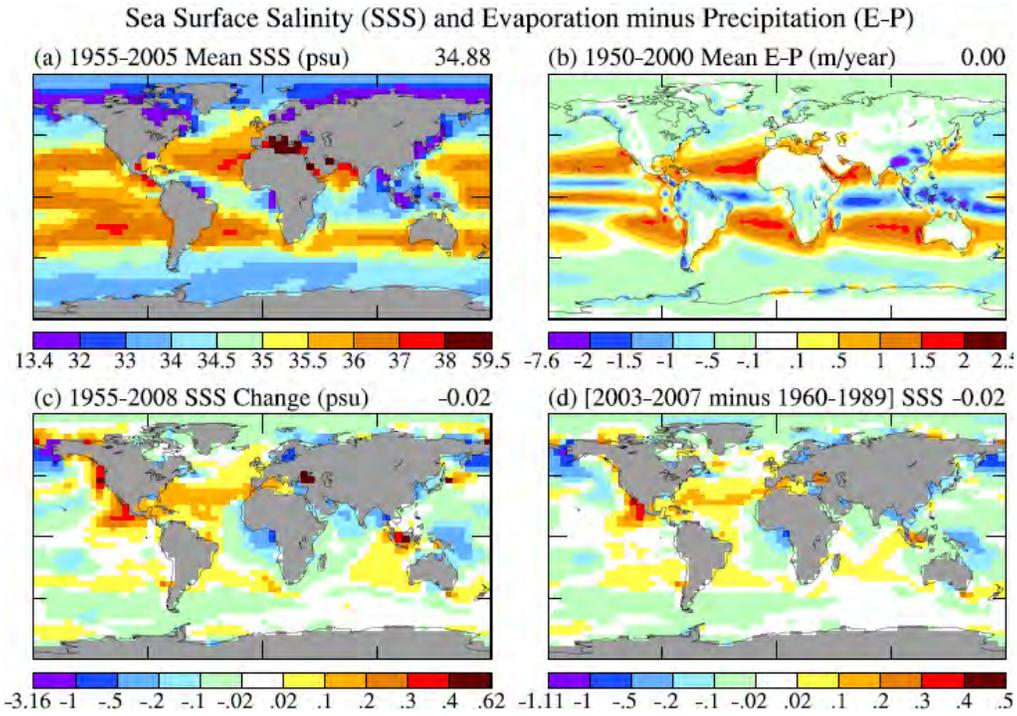

**Fig. S15.** (a) Simulated sea surface salinity (psu), (b) evaporation minus precipitation (m/yr), and (c,d) salinity change (m/yr), periods being chosen to allow comparison with observations, as discussed in text.

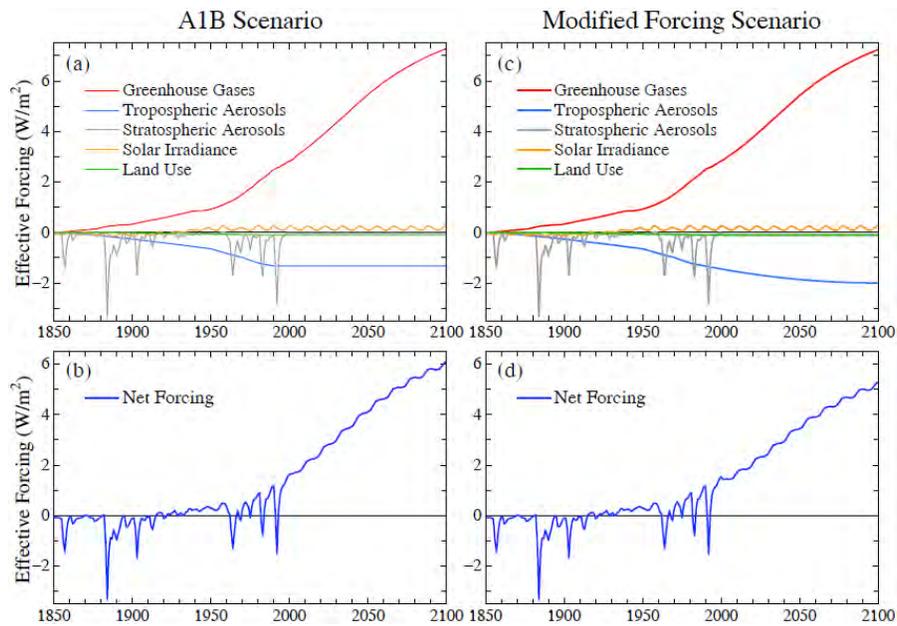

**Fig. S16.** Effective global climate forcings (W/m²) in our climate simulations relative to values in 1850.

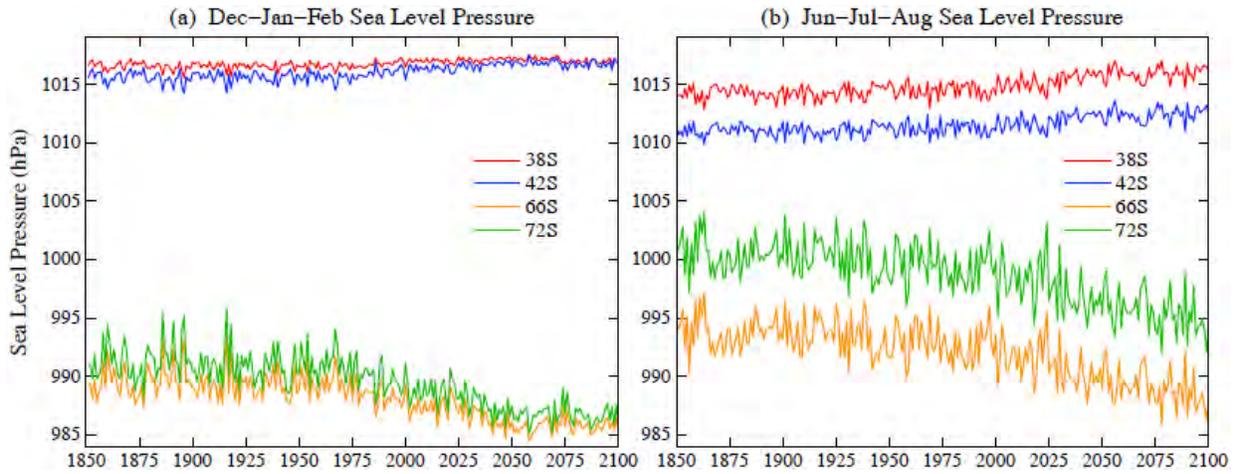

**Fig. S17.** Sea level pressure (hPa) at four latitudes in (a) Dec-Jan-Feb and (b) Jun-Jul-Aug. Model is driven by "modified" forcings including ice melt reaching the equivalent of 1 m sea level by mid-century.

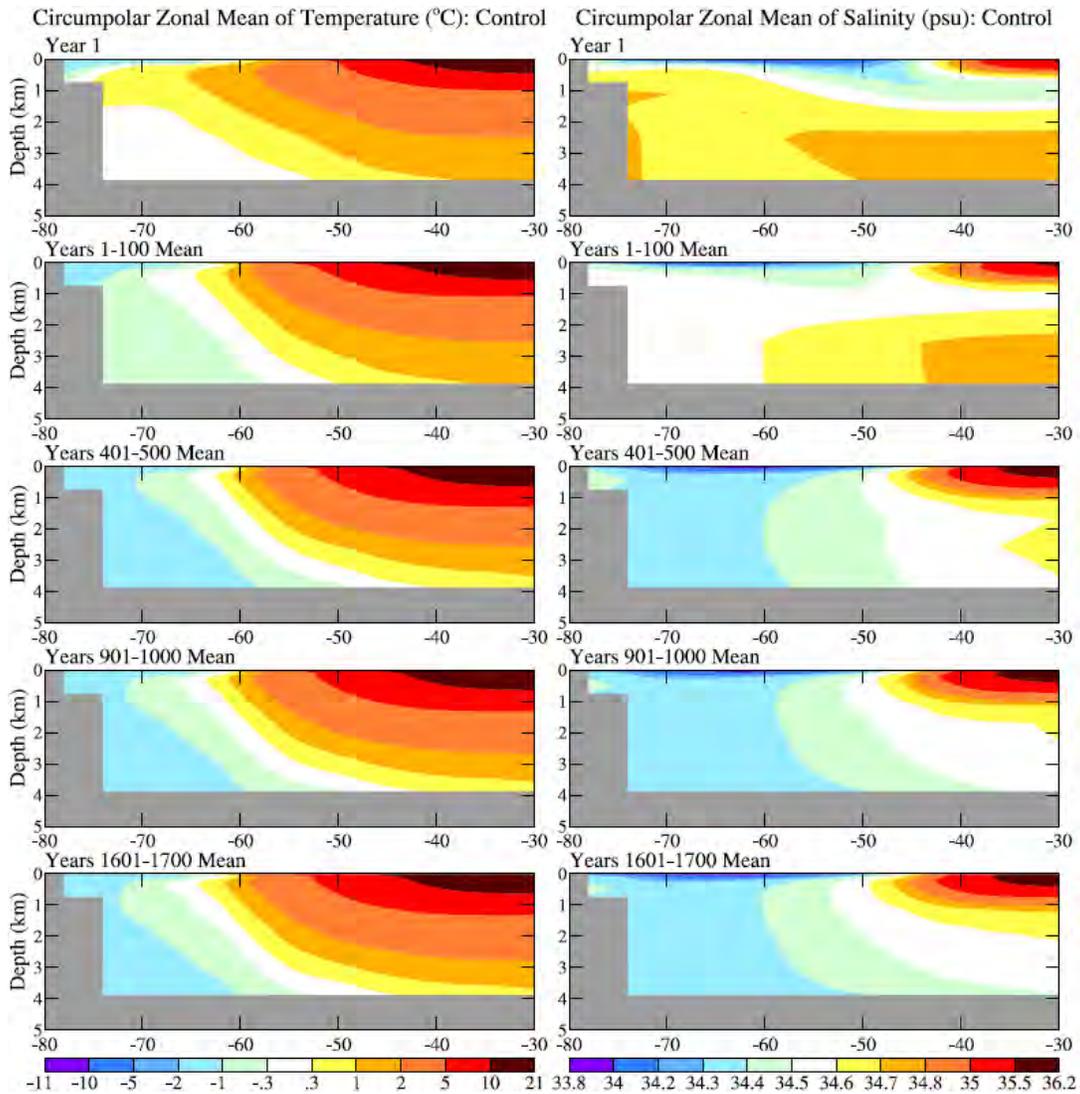

**Fig. S18.** Ocean temperature (°C) and salinity (psu) in the control run.

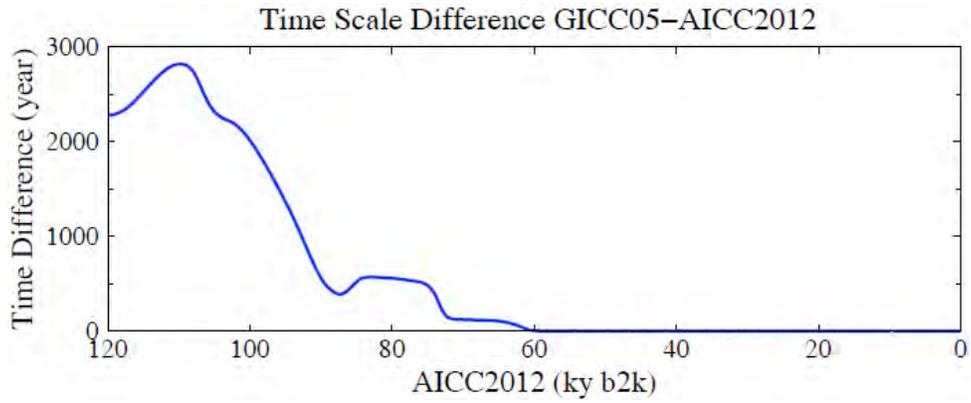

**Fig. S19.** Difference (years) between the GICC2005modelext and AICC2012 time scales (Bazin et al., 2013; Veres et al., 2013; Rasmussen et al., 2014; Seierstad et al., 2014).

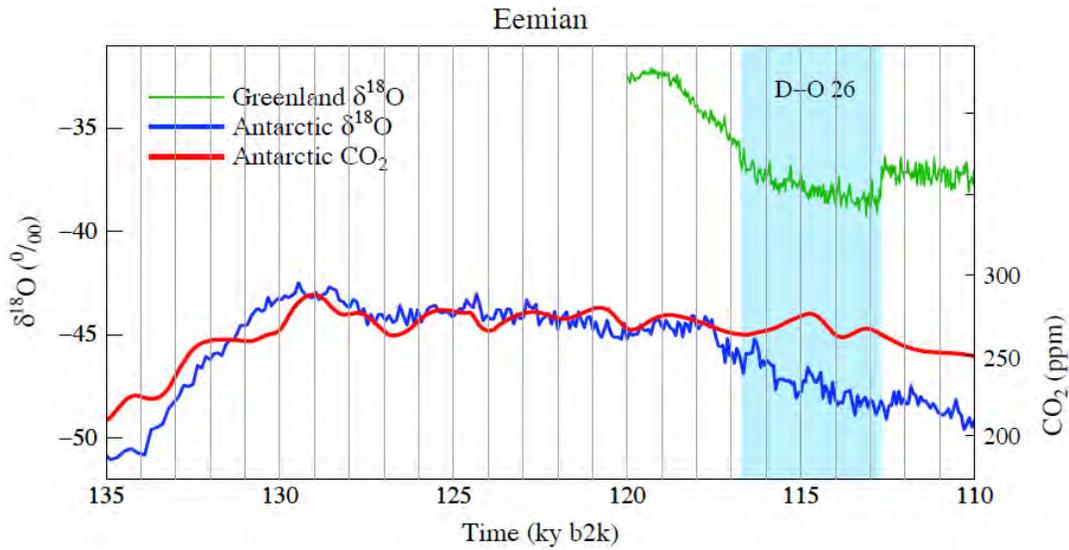

**Fig. S20.** Expansion of data from Fig. 27b,c. $CO_2$ increases during D-O 26 lag Antarctic temperature rises by 1500-2000 years.

Change in 2078-2082 Relative to 1880-1920

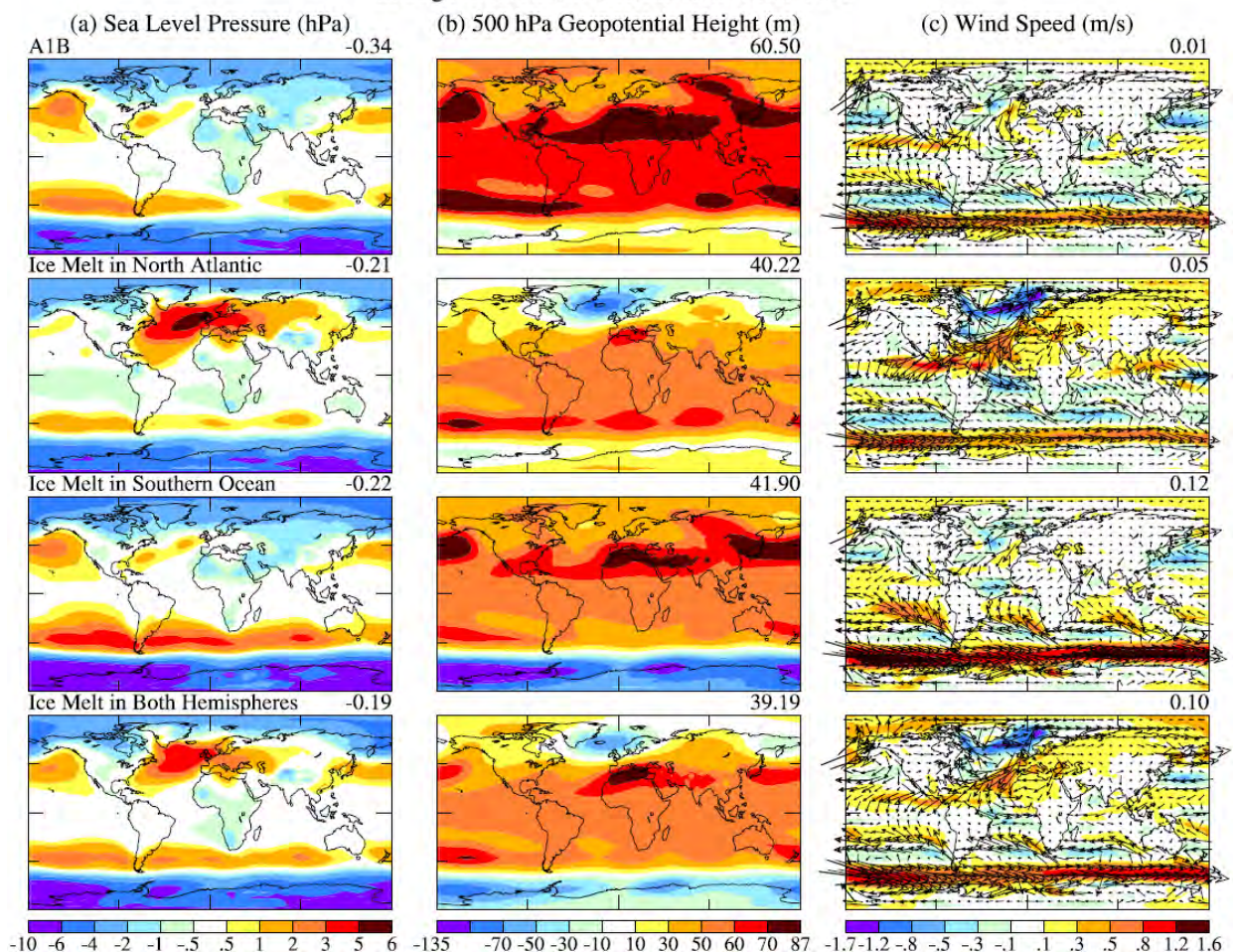

**Fig. S21.** Change in 2078-2082, relative to 1880-1920, of the annual mean **(a)** sea level pressure (hPa), **(b)** 500 hPa geopotential height (m), and **(c)** wind speed (m/s), for the same four scenarios as in Fig. 6. Numbers in upper right corners are the global mean change.

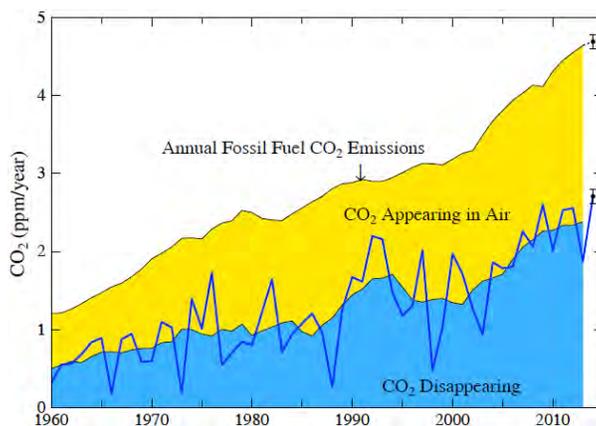

**Fig. S22.** Top curve: global fossil fuel $CO_2$ emissions (ppm/year). Measured $CO_2$ increase in air is the yellow area. The 7-year mean of $CO_2$ being absorbed by the ocean, soil and biosphere is blue (5- and 3-year means at the end; dark blue line is annual). 2014 global emissions estimate as 101% ±2% of 2013 emissions. $CO_2$ emissions from Boden et al. (2013) and atmospheric $CO_2$ from P. Tans (www.esrl.noaa.gov/gmd/ccgg/trends) and R. Keeling (www.scrippsco2.ucsd.edu/).

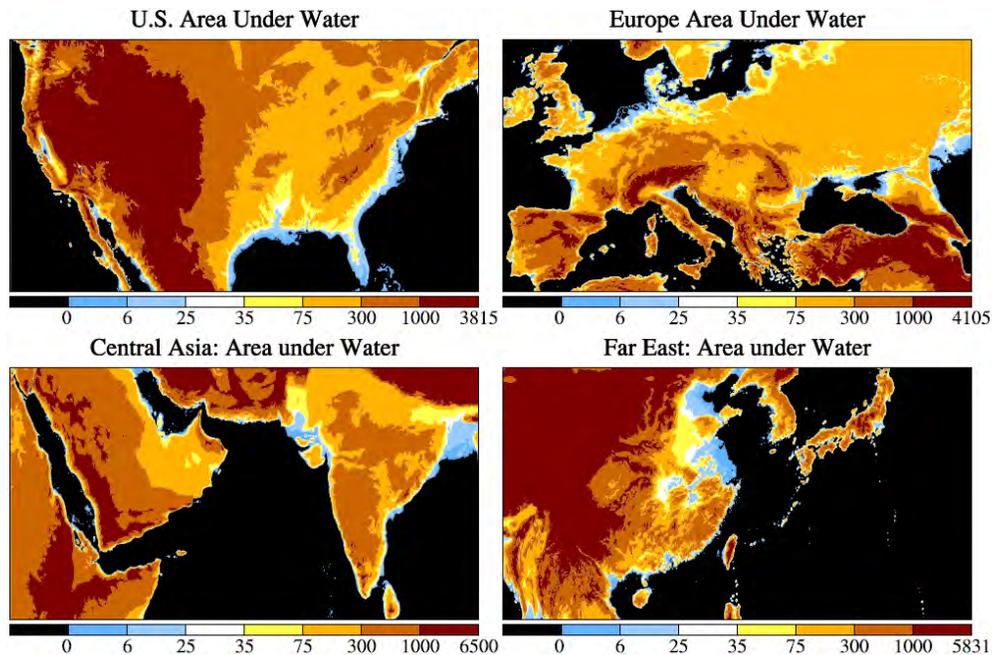

**Fig. S23.** Areas (light and dark blue) that nominally would be under water for 6 and 25 m sea level rise.

## Annual Mean Surface Temperature Anomaly (°C)

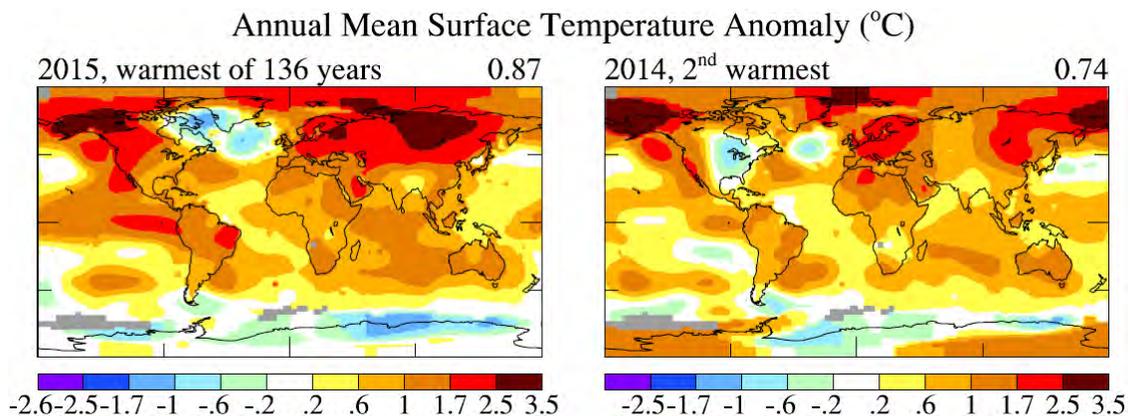

**Fig. S24.** Observed surface temperature relative to 1951-1980 mean (update of Hansen et al., 2010; maps and other graphs are updated monthly at http://www.columbia.edu/~mhs119/Temperature/).

**Supplement S2: Eemian sea level: Evidence for early double peaks and late peak highstand**

In Bermuda, Land et al. (1967) were among the first to recognize both a complex Eemian sea level record, and a much higher peak highstand late in the interglacial. Land et al., (1967, Fig. 5 and p.1005) stated: *"Later in the same (MIS 5e) interglacial period the sea rose again, at least to +11 m (east of Spencer's Point)."* Hearty (2002) later surveyed the same Spencer's Point deposits to a more precise +9.2 m ("+" indicates above today's sea level).

In the Mediterranean, a 'double 5e' Eutyrrhenian (Eemian) was a prominent stratigraphic sea level feature described in the 1980s (e.g., Hearty, 1986). Aharon et al. (1980) described a double-5e sea level history from the Papua New Guinea and suggested the higher, late rise was the result of West Antarctic ice collapse. In South Carolina, Hollin and Hearty (1990) similarly documented a double 5e sea level with a rapid late rise several meters higher than the early sea stand. Evidence of a rapid but brief, late rise was further described in Bermuda and the Bahamas in the 1990s (e.g., Hearty and Kindler, 1995). Neumann and Hearty (1996) estimated only a few hundred years to rise to and incise a +6 m notch in the Bahamas. Rapid rise to and brevity at these higher levels is inferred from the prevalence of notches and rubble benches in the Bahamas, in contrast to broad terraces and reefs formed earlier at the +2-3 m level. Additional geological details of these carbonate platform sea level records were contained in a number of interim papers, and summarized in Hearty et al. (2007). Most recently, Godefroid and Kindler (2015) added: *"The MIS 5e record is remarkable. In particular, beach deposits and an intertidal notch at +11 m above msl strongly suggest that sea-level peaked at a much higher elevation than previously assessed, implying pronounced melting of polar ice."*

In the Bahamas, less than 5% of documented Eemian exposures contain coral reefs, and no Eemian *in situ* exposed reefs are known from Bermuda, so U/Th coral dating is not the primary geochronological method available in these areas. Regardless, many of these sparsely distributed reef deposits in the Bahamas have been U/Th dated (e.g., Chen et al., 1991; Hearty et al., 2007; W. Thompson et al., 2011) and correlated with the diagnostic oolites. The geochronological age of Quaternary deposits is based on 275 whole rock and 507 land snail amino acid racemization (AAR) age estimates from U/Th and [14]C calibrated age models (Hearty and Kaufman, 2000, 2009). Of key importance, the Eemian-MIS 5e in the Bahamas is defined by its position in the stratigraphic sequence of the rocks, the oolitic and pristine aragonitic sedimentology, a unique landsnail fauna (Garrett and Gould, 1984), and numerous additional diagnostic characteristics (e.g., Hearty and Neumann, 2001, p. 1883). There is little disagreement among researchers of the defining characteristics of MIS 5e in the Bahamas.

*What gives carbonate platforms such as Bermuda and the Bahamas the unique quality of preserving such a detailed geologic record?* Because carbonate sediments, particularly ooids, respond and cement quickly, the highly mobile sediments that mantle flat-topped carbonate platforms effectively record and preserve rock evidence of short-lived energetic events such as storms and rapid sea level changes. Corals and coral reefs respond too slowly and cannot record such brief changes. Likewise, similar short-term events are not preserved on coasts dominated by siliciclastic or volcanic sediments (e.g., US East Coast and much of Caribbean region) due to the instability and slowness of cementation ($>10^6$ yr) of non-carbonate sediments.

In a global multidisciplinary review of MIS 5e, Hearty et al. (2007) assembled shoreline stratigraphy, field information, and geochronological data from 15 sites to construct a composite curve of Eemian sea level change. Their reconstruction has sea level rising in the early Eemian to +2-3 m. Mid-Eemian sea level may have fallen a few meters to a level near today's sea level.

Sea level then rose rapidly in the late Eemian to +6-9 m, cutting multiple bioerosional notches in older limestone in the Bahamas and elsewhere.

Along the northeast Yucatan Peninsula, Mexico, Blanchon et al. (2009) used a sequence of coral reef crests to investigate coral reef "back-stepping", i.e., the fact that coral reef building moves shoreward as sea level rises with a higher temporal precision than possible with U-series dating alone. They documented an early +3 m sea level jump by 2-3 m to +6 m within an "ecological" period, i.e., within several decades, in the late Eemian about 121 ky b2k based on U/Th ages. W. Thompson et al. (2011) reexamined the Eemian using corrected U-series coral reef data from the Bahamas and interpreted a mid-Eemian sea level at +4 m at 123 ky b2k, a maximum at +6 m at 119 ky b2k, and at 0 m at some time in between. Note that no known coral reef crests are higher than +2-3 m across the entire archipelago (Hearty and Neumann, 2001; p. 1883).

In Western Australia, O'Leary et al. (2013) assembled one of the most comprehensive Eemian sea level studies that includes: 1) 28 "far field" study sites along the 1400 km coastline; 2) application of a multi-disciplinary approach using geomorphology, stratigraphy, and sedimentology; 3) high-precision U/Th dating and screening of over 100 in situ corals; and 4) incorporation of GIA correction regionally yielding a more precise eustatic sea level history. The O'Leary et al. (2013) analyses suggest that sea level was relatively stable at 3-4 m in most of the early-mid Eemian, followed by a brief but rapid (<1000 yr) late-Eemian sea level rise to about +9 m. U-series dating of the corals has the sea level rise begin at 119 ky b2k and peak sea level at 118.1 ± 1.4 ky b2k.

The far field *eustatic* sea level changes documented across Western Australia (O'Leary et al., 2013) agree closely with the relative sea level shifts from near and mid field Bermuda and the Bahamas (Hearty et al., 2007). Nearly all global sites in the Hearty et al. (2007) study showed the same *relative* changes: early prolonged stability, a minor mid regression, then finally rapid upward shifts of 3 to 5 m late in the Eemian. Such rapid sea level changes require ice sheet growth and melting, regional glacio-isostatic adjustment (GIA), or both.

**Supplement S3: Ocean wave splash near the location of Eleuthera boulders**

Cox et al. (2012) discuss the inadequacy of hydrodynamic modeling to realistically describe movement of boulders by large storms. Specifically, they found that storms in the North Atlantic had thrown boulders as large as 80 tons to a height 11 m AHWM (above high water mark) on the shore on Ireland's Aran Islands, the specific storm on 5 January 1991 being driven by a low pressure system that recorded a minimum 946 mb (equivalent to a category 3 hurricane). Winds gusted to 80 knots and the closest weather station to the Aran Islands recorded gale force winds for 23 hours and sustained winds of 40 knots for five hours. The storm waves built on swell previously developed by strong winds during the prior two weeks.

Cox et al. (2012) note that existing hydrodynamic modeling equations would not lift the boulders, and they cite two reasons to disregard those equations. First, they note that wave height measurements frequently reveal waves twice the SWH (significant wave height) of wave models. Second, existing wave equations do not include the effects of reflection from cliff and shoreline, and the attendant wave amplification. Cox et al. note that wave heights at shoreline cliffs can be much greater than the equilibrium height of approaching deep-water waves. The waves steepen as they shoal, impact the coast, reflect back, meet advancing wave crests causing a mixture of constructive and destructive interference, with intermittent production of very large individual waves capable of quarrying and transporting large blocks and boulders.

These considerations help explain why megaboulders (as large as ~1000 tons) on Eleuthera are only found just south of the Glass Window Bridge at the apex of an embayment that funnels the waves before they encounter a steep shoreline cliff (Figs. 1-3 of Hearty, 1998; see also Hearty, 1997). The special effect of the location and shoreline cliff is shown in a photo (Fig. 1). Despite relatively calm conditions on Eleuthera, as indicated by the waters in the photo, immediately southwest of the narrow Eleuthera island, the northeast side of Eleuthera was being battered by large waves generated in the North Atlantic by the 1991 "Perfect Storm". The Perfect Storm originated as an extratropical low east of Nova Scotia that tracked first toward the southeast and then west, sweeping up remnants of Hurricane Grace, which deepened the low. The storm eventually reached a peak intensity with sustained winds of 75 mph (120 km/h), a category 1 hurricane, making landfall on Nova Scotia on 2 November. The shoreline cliffs immediately south of the Glass Window Bridge, facing slightly east of due north (Fig. 3 Hearty, 1998), were battered by the deep long-period waves generated by the storm in the North Atlantic.

Irregularity of ocean spash in this setting probably helps account for how an unsuspecting bread truck driver, seduced by the relative calm and fair weather (Fig. 1), was swept off the road by one of the bursts as water swept across the road. The truck was thrown/washed well into the shallow waters on the Caribbean-facing side of the island – the driver escaped in these relatively calm waters to the southwest, but his rusted out truck frame remains there today.

Further confirmation of the ability of storm waves to lift large boulders was provided recently by May et al. (2015). Despite the fact that this storm did not have the "advantage" of being stationary for the long period required to develop deep powerful waves, the typoon produced longshore transport of a 180 ton block and lifted boulders of up to ~24 tons to elevations as high as 10 m. May et al. (2015) conclude that these observed facts "…demand a careful re-evaluation of storm-related transport where it, based on the boulder's sheer size, has previously been ascribed to tsunamis."

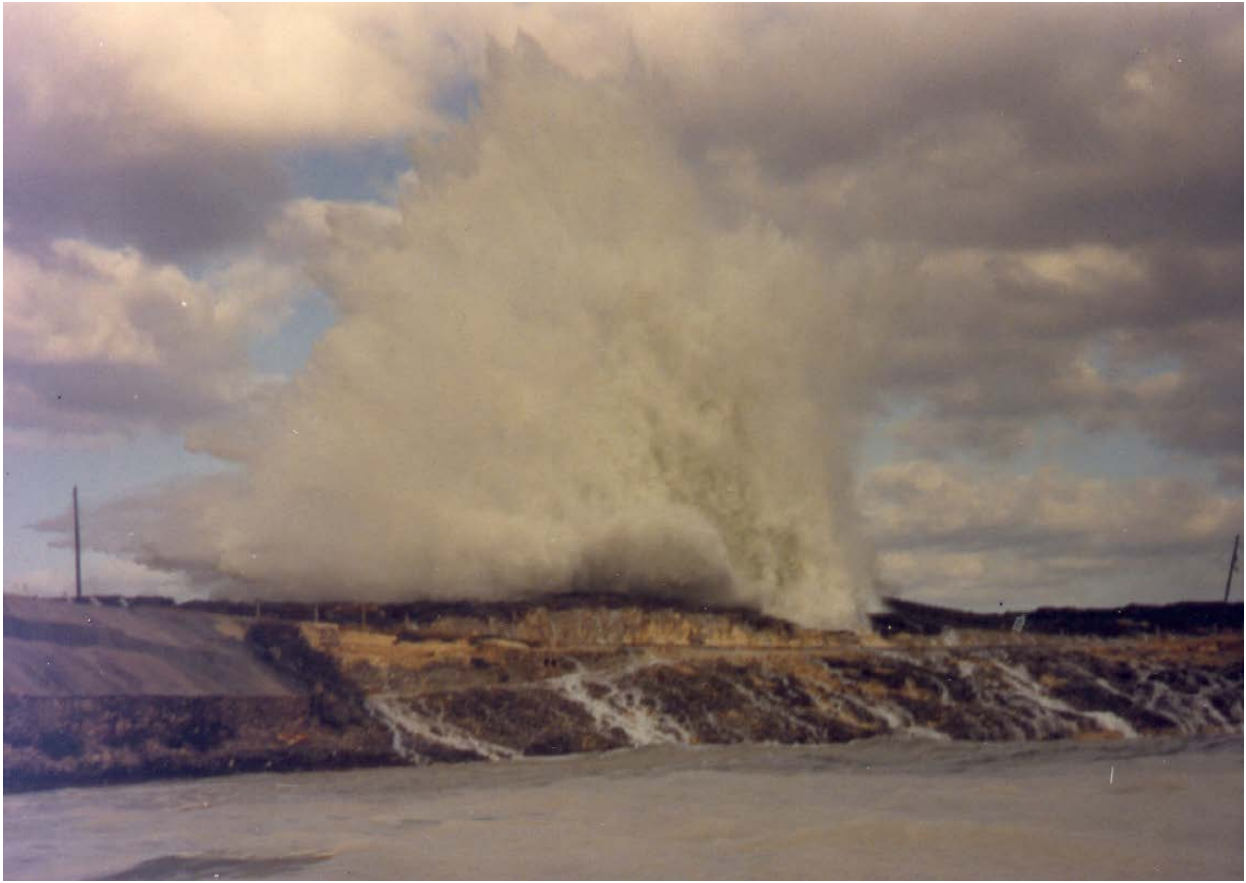

**Photograph S1.** Photo taken 31 October 1991 from a few hundred meters offshore of the southern protected bank-side at the narrow part of Eleuthera near the Glass Window Bridge, looking northeast (see text). The telephone pole on the left and the 15-20 m cliff provide scale.